# Exploration of unknown indoor regions by a swarm of energy-constrained drones

Rappel, Ori
Faculty of Aerospace Engineering
Technion, Haifa 32000, Israel

Ben-Asher, Joseph
Faculty of Aerospace Engineering
Technion, Haifa 32000, Israel

Bruckstein, M. Alfred
Computer Science Department Technion,
Haifa 32000, Israel

**Several distributed algorithms are presented for the exploration of unknown indoor regions by a swarm of flying, energy constrained agents. The agents – identical, autonomous, anonymous and oblivious – uniformly cover the region and thus explore it using predefined action rules based on locally sensed information and the agent's energy level. While flying drones have many advantages in search and rescue scenarios, their main drawback is a high power consumption during flight combined with limited, on-board energy. Furthermore, in these scenarios agent size is severely limited and consequently so are the total weight and capabilities of the agents. The region is modeled as a connected sub-set of a regular grid composed of square cells that the agents enter, over time, via entry points. Some of the agents may settle in unoccupied cells as the exploration progresses. Settled agents conserve energy and become "virtual pheromones" for the exploration and coverage process, beacons that subsequently aid the remaining, and still exploring, mobile agents. The termination of the coverage process is based on a backward propagating information diffusion scheme. Various algorithmical alternatives are discussed and upper bounds derived and compared to experimental results. Finally, an optimal entry rate that minimizes the total energy consumption is derived for the case of a linear regions.**

# I. Introduction

In many real-life scenarios, one is interested in deploying agents over an unknown region in order to explore it and possibly create an environment capable of detecting disturbances, localized activities and various events of interest. Often such scenarios are characterized by no a-priori knowledge on the region to be explored and covered, movement on the ground may be hampered by rubble and debris, agents lacking a global reference frame and limited communication capabilities with other agents and/or with the deploying station. For example, operating in a post-earthquake, collapsed building, rescue workers would obviously benefit from having a map of the free space inside the wreckage and a sensor network that can detect survivors. In general, unknown indoor environments present the major challenges of difficult ground movement, lack of a global reference frame (due to the impossibility of receiving the GPS signal) and severe degradation of RF communication due to multi-path and absorption effects. Sensing ranges, in most real-life scenarios, are also severely limited due to irregularities in the terrain. These irregularities might also hamper and slow down the movement of ground-based exploration and rescue robots.

In our previous work [1] we presented the concept of exploring the region with a swarm of flying drones that eventually uniformly cover the region. The agents – identical, autonomous, anonymous and oblivious - enter the region one after another from entry points and explore the region in a distributed manner according to some predefined action rules based on locally sensed information. While flying drones (MAV's) have many advantageous in search and rescue scenarios, their main drawback is a high power consumption during flight combined with limited, on-board energy (i.e., battery). This is especially critical in the aforementioned scenarios in which agent size is severely limited and consequently so are the battery weight and agent capabilities. While the algorithms in [1] were designed with the energy constraint in mind it was not part of the action rules. In this work we extend the discussion to energy constrained drones and answer three questions. First, what is the upper bound on the energy consumption of the swarm as a whole and of an individual agent. Second, is there an optimal entry rate that will minimize the energy consumption. Third, can the available energy be integrated into the action rules an agent while keeping the basic concepts described in [1].

The simplest solution to the energy constraint and the one implemented in our algorithms is to shorten the flight time as much as possible. In addition, the payload capacity of small MAV's is limited and the deployment of dispensable beacons that aid the exploration process is not relevant. Therefore, as a meta-rule, some of the agents settle and become beacons that can guide the exploration by the remaining mobile agents. The resulting sensing and covering infrastructure, effectively a directed acyclic graph of the region (or a directed forest if more than entry point exists) composed of settled agents, may be used to implicitly generate a map and an ad-hoc sensor network to aid and guide subsequent operations.

*Exploration* is commonly defined in robotics research as the process of visiting of every part of a region by at least one agent at least once. In the context of this article, *uniform filling* (dispersal) refers to uniformly distributing agents in an unknown (but bounded) environment hence a uniformly filled region is obviously an explored region. *Deployment* is the process through which the agents enter the region, and the combination of deployment with the filling process must result in uniform coverage of the region of interest with agents Note, some researchers refer by "uniform filling/ dispersal" to the combined deployment and uniform distribution of agents.

The algorithms described in this paper are based on two basic meta-concepts. The first concept, mentioned above, is the use of agents in the Dual Role of (mobile) Explorers and of Physical Beacons or Pheromones that guide exploring agents while also creating the uniform coverage. An agent entering the region starts as a mobile agent and hence may move, once at every time step, to one of its neighboring cells in the region. The agents can also decide to become stationary and settle (i.e., become a beacon) if the proper conditions are met. The change from a mobile agent to a stationary one is considered irreversible. Once settled, an agent serves as a "beacon" to other, mobile agents thereby guiding their movement. The conditions for a change of the state and the actions taken in each state differentiate between the algorithms described herein. From the energy perspective the Dual Role meta-concept implements a "settle as quickly as possible" since once an agent settles its energy consumption decreases significantly. The second meta-concept, termed Backward Propagating Closure, realizes the requirement for a deterministic indication of termination and at the same time reduces the energy consumption by effectively funneling the drones to empty regions. Settled agents in completely covered parts of the region signal that information to the mobile agents thereby reducing the energy consumption as well shortening the termination time.

The energy constraint can be mitigated using *open-loop behavior* and/or *closed-loop control*. Open-loop behavior refers to agent behaviors (i.e., local action rules) that are designed to minimize the energy consumption of an agent ([1], [2]). Closed-loop behavior refers to conditioning the online (real-time) behavior of a drone on its energy level. In this paper algorithms presented in [1] are modified to take into account the current energy level thus combining the two methods. Once the energy level of an agent falls beneath a threshold it either lands (if mobile) or transitions to a Low Energy state (if already settled). Due to the energy constraint the full exploration and coverage of a region may be impossible. In these situations, it is critical that both the entry of additional agents be halted and a deterministic indication of process termination be given to an external observer at the entry point/s. Two approaches to achieving these objectives are investigated. The first gives precedence to terminating the process as fast as possible while the other gives precedence to achieving maximum exploration. As will be shown in the paper, the first is more robust with respect to the energy consumption model of the energy while the second is greedier with respect to the covered area.

This research has three important contributions. First, distributed algorithms for the exploration and uniform coverage of unknown region by a group of energy constrained flying drones. Upper bounds are derived, and compared to experimental results, for the covered area, termination time and the number of agents that will be run out of energy. In addition, a criteria for the prevention of energy-constrained uniform coverage is presented. Second, the exploration of a linear graph is thoroughly analyzed and upper bounds on the energy consumption are derived both for the individual agent and the whole swarm. Lastly, expressions for the entry rate that minimizes the total energy consumption in a linear region are derived.

In the following section, related work is a reviewed. Next, definitions and assumptions pertinent to all the algorithms are presented. In section IV the algorithms are described while in section V the experimental results are presented and discussed. Analytical expressions for the case of a linear graph are derived in section VI followed by conclusions in section VII.

## II. Related Work

Numerous papers have been published on coverage and/or exploration of unknown regions using swarms of agents. The large number of possible assumptions on the capabilities of the agents, agent activation schemes, deployment schemes and types of topologies resulted in rich and diverse research topics. A review of the common assumptions and required outcomes is given in Das in [3, p. 403]. A more extensive overview of the literature can be found in Flocchini et al [4]. Relatively limited research, however, has been done on the exploration of graphs by multiple, energy constrained agents. Therefore, we first review relevant works on uniform coverage by a swarm of agents followed by works on exploration by energy constrained agents and finally works on exploration by a swarm of energy constrained agents.

The problem of uniform filling was formulated by Hsiang et al. [5] for the case of ground robots. In their paper the authors describe several distributed algorithms based on the 'follow-the-leader" concept and use both Depth First Search and Breadth First Search strategies to guide the filling process. The agents enter the region via a door and the objective of the algorithms is to minimize the filling time defined as the time until all the cells are filled with an agent (termed the "make span"). The agents however are not energy limited and the discussion of the energy consumption centers on the total number of steps.

The most common assumption regarding the initial location of the agents is that they are located outside the region, as is the case in this paper. However, some researchers assume the initial location of the agents is inside the region based on some arbitrary spatial distribution (e.g. [6], [7]). When the initial location is outside the region a deployment process (i.e., the rules that according to the agents enter the region ) is commonly defined [2] [8], [9], [10].

Additional ways to divide the body of work is according to the way the region is modeled. The region to be filled may be a part of the two-dimensional plane (i.e., continuous) [2], [11], [12] or modeled as graph (usually as grid graph) [6]. Moreover, the region may be restricted to be simply-connected which means none of the obstacles is completely surrounded by empty cells [13], [6] however a more realistic and common assumption is for the region to be non-simply connected [12], [14], [15], [16]. In this research we assume the region of interest can be modeled as a non-simply connected grid graph since the presence of pillars and/or piles of rubble cannot be ruled out.

The problem of energy-constrained region exploration was introduced by Betke et al in [17]. In the paper two algorithms for the single explorer problem with a limited energy are presented and evaluated. In both the agent is assumed to travel on a grid and to return to the starting point to "refuel". The research in the field of energy constrained exploration by a muti-agent system focused on ground-based robots and dealt with three general, complementary problems. The first, given a region **R** (in many cases a tree) and agents with an energy bound of **E** what is the minimal effort to explore the region. The second, given a region **R** and *k* agents, what is the required energy by each agent in order to explore the region. The third, and the one dealt with in this work, is given **k** agents with an energy bound of **E** what is the maximal area that can be explored. Works that treat the three types of problems are described below. In all the papers the algorithms are decentralized, and the agents are initially located at the entry point and return to it at the of the process unless explicitly stated otherwise. We focus on the case in which there are no charging stations in the regions although the effect of charging stations – stationary or mobile – has also attracted interest.

In [18], [19] and [20] Das et al focus on the first type of problems and present an exploration strategy of a tree by a group of energy constrained agents based on dividing a depth first search into bounded-length segments. The objective is to minimize the number of agents used ( [18], [19]) or the number the total number of edges traveled ([20] ).

The second type of problems is investigated in [21] and the agents are assumed synchronous, to have a communication range of 1 and capable of storing information in the various node The authors describe an online algorithm for the exploration of an n-node tree by a group of k mobile with the objective of minimizing the maximal distance traveled by each robot. The results are compared with optimal trajectories calculated offline. In [22] the third type of problem is discussed. The agents are assumed to have unlimited communication range and the region type is limited to a tree. Kwok et al. [23] also treat the third-type of problems however the objective is maximal covered area and not maximal explored area. The authors describe energy-aware coverage algorithms based on Voronoi partitioning and Loyld's algorithm with the cost function dependent upon the energy of the agents.

In many of the papers the performance of the algorithms is assessed using competitive analysis, i.e., comparison of the online algorithm to an optimal offline algorithm which has a complete map of the region. Chang et al. [13] however introduce a new metric to evaluate algorithms called the energy-time-product, that describes in one number the trade-off between the (total) energy consumption and the process time. In the paper the authors assume ant-like agents that communicate by leaving marks on explored nodes and investigate the effect of the deployment (i.e., dispatching) method on above metric.

While most research assume the agents are ground robots, the advantages of airborne robots in exploration (i.e., increased field of view) have been recognized by some researchers in the last decade. The idea of using airborne drones as a beacon was first mentioned by Stirling et al. [2], [8] and later by Aznar et al. [24], [25], with the objective of exploring (not uniformly filling) a region. In [2], [8],[9], [24], [25] each drone has a unique Id with the exploration strategies based on either depth first search [2], [8] or on a potential field generated by virtual pheromones in [24], [25]. In Rappel et al. [26] the authors present a DFS-based uniform coverage strategy by airborne agents however the issue of collision avoidance, critical when using airborne agents, is problematic since the number of mobile agents in each cell is not limited.

Energy is the prime constraint on the performance of swarms composed of flying agents. That is why most papers that assume flying agents also discuss the energy consumption and use it as one of the performance metrics. Stirling et al. [2] proposed an energy model that takes into account the energy consumption of a drone before its entry into the region, during movement and when operating as a beacon. Based on this model the total energy consumption is computed and sensitivity to different parameters analyzed. Aznar et al. [24] predict the energy consumption using recurrence equations.

A settled agent can signal other agents using various mediums – RF, acoustic, and visual amongst others. Possibly the best solution for the scenarios described above is visual as it is both localized and secure from interference by neighboring agents. First suggested by Peleg [27] as a means of signaling without explicit communication, the idea has been since extended for the lights to be used only as a memory [28]. Das et al. [29] investigate the additional capabilities imparted to agents with lights. Prior to this work the idea was used by Poudel et al. [30] that present a uniform scattering algorithm that is linear with time for synchronous robots with lights, a compass and a visibility range of 2. It is important to note that comparison of the performance metrics (e.g., cover time) of the various approaches, especially the issue of energy consumption, is problematic due to the differing assumptions about the agent's capabilities, the modeling of the region and the time scheme.

In summary, there are numerous papers on the exploration and coverage of unknown regions by multi-agent systems - with and without an energy constraint – using ground robots. The existing literature on exploration using flying drones discusses the energy consumption but does not treat the problem of energy constrained flying drones. This is of prime importance due to the limited energy capacity of flying drones combined with the severe consequences of running of out of energy in midflight. To the best knowledge of the authors, this research is the first to present some relevant results on the topic.

## III. Preliminaries

### A. Problem Formulation

In the *coverage problem*, a swarm of drones seeks to enter and uniformly cover an a-priori unknown region **R** in a decentralized manner such that in the terminal configuration every cell is occupied by a drone. The unknown region, **R**, to be covered is a connected region that is modeled as an orthogonal grid graph and access to the region is via well-defined *entry point/s*. It is an arbitrary, possibly non-simply connected (i.e., obstacles may be surrounded by empty cells) grid composed of equally sized cells forming both the empty region to be explored and topological features such as walls and obstacles. Each cell of **R** can contain a single mobile agent and/or a single settled agent. The agents enter the region in a prescribed order from an "agent source" that has a sufficient number of drones to cover the region via the entry point/s.

The region **R** can also be represented by graph $\mathbb{G}(V, E)$ with the vertices, *V*, representing the centers of the empty cells and the edges, *E*, representing the connections between empty, adjacent cells. In the rest of this paper, the terms cell and vertex are interchangeable. Unless stated otherwise, $n = |\mathbf{V}|$ i.e. the cardinality of the graph is equal to the number of empty cells in the region and $m = |\mathbf{E}|$. A sample, non-simply connected region with one entry point, eight rooms and a double corridor is depicted in Figure 1.

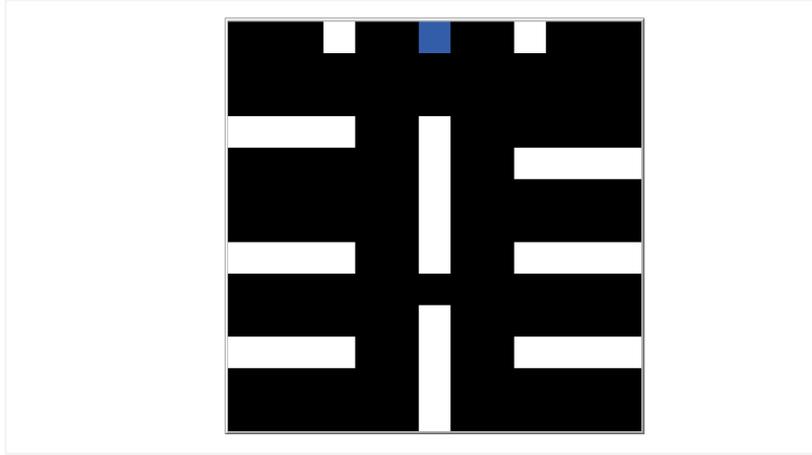

Figure 1  Typical Region. The empty cells are marked in black, the obstacles in white and the entry point in blue

### B. Timing model

Time, $t$, is discretized such that $t = 0, 1, 2, 3, \ldots, k, \ldots$. And each step is further sub-divided into M equal sub time-steps, $dt$, such that $dt = 1/M$. Each agent wakes up once in every time step at a randomly selected sub time-step as given by Definition 1 and Eq. (1). This is the most probable in real–life, decentralized multi-agent systems in which each agent has its own internal timer that controls when it wakes up. For sufficiently large values of M the probability that two agents will wake up in the same instance is negligible - see Figure 2. Consequently, the time between successive action cycles of a specific agent is bounded by (0,2) and the number of activations of agent $a_j$ between two successive action cycles of agent $a_i$ is in the range [0:2]. Such time models are called fair 2-bounded schedulers in the literature [31].

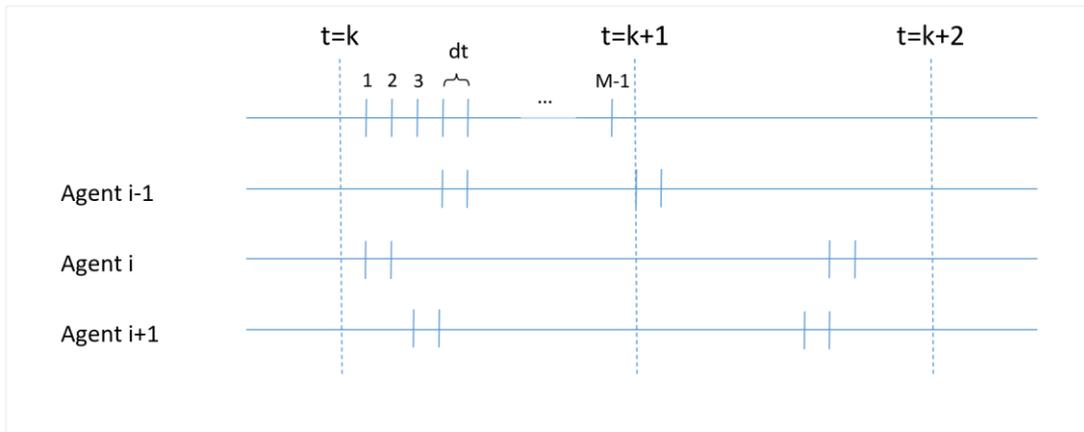

Figure 2  An illustration of the asynchronous time model. The top row shows the subdivision of each time step into sub time steps of size $dt$. Each agent wakes up exactly once per time step $t$ at one of these sub-steps

**Definition 1**: The wake-up time of agent $a_i$ at time step $k$ is denoted by $t_{i,k}$ and defined as:

$$t_{i,k} = t_k + \frac{U(0, M-1)}{M} \tag{1}$$

with $U(0, M-1)$ denoting the uniform distribution over the integers $\{0, 2, \ldots M-1\}$.

## C. The Agents

The agents are assumed to be *miniature air vehicles* (MAV's) capable of motion both in the horizontal plane (i.e., parallel to the to the surface) and in the vertical plane (i.e., perpendicular to the surface). In the horizontal plane the agents move in one of four principal directions (i.e., "Manhattan Walk") whereas in the vertical plane the agents can only move straight down (i.e., parallel to the gravity vector). Hence, at every time step, every agent is either in flight and searching for a place to land, or has already landed somewhere in **R**. The former type of agent is called *mobile* whereas the latter is called *settled*. We assume that agents can sense whether a neighboring agent is mobile or settled. Agents always start as mobile agents and may settle down at some time step. Settled agents remain in place forever, providing useful information to the mobile agents. Each agent can sense the presence of agents in any adjacent cell at Manhattan distance 1 from it (i.e., the sensing range of agents is 1). While mobile agents can sense both mobile and settled agents, the settled agents can only sense neighboring settled agents. Figure 3 depicts the sensing volume and possible directions of movement.

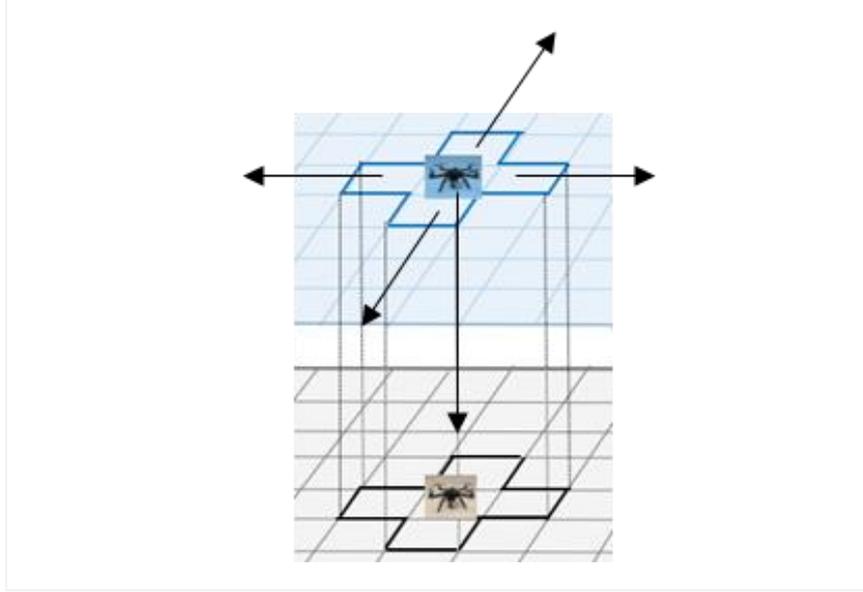

Figure 3  Sensing and moving region of an agent.  The ground, occupied by a settled agent is marked in grey while the plane in which the flying agents move is marked in blue. The mobile agent, a drone, is shown both in the air (dark blue) and settled on the ground (brown). The arrows denote the possible directions of movement.

**Definition 2**: The set of all entities – agents and/or walls - sensed by agent $a_i$ at cell (or node) $u \in V$ at time $t$ is denoted by $\xi_u(i,t)$. This set is also called "the neighbors of $a_i$ at time $t$ at cell $u$". For brevity, we omit the reference to both a specific time and agent whenever possible. As $\mathbb{G}(V,E)$ is a grid graph and the agents perform a Manhattan-type walk, the positions of the neighbors of agent $a_i$ are enumerated systematically according to Figure 4 with a mobile agent being in the cell enumerated as "5" and a settled agent in the cell enumerated as "0". Furthermore, cell "0" is the cell referred to when discussing the cell directly underneath a mobile agent.

The set $\xi_u(i,t)$ is hence the 10-tuple $\xi_u(i,t) = \{\xi_0, \xi_1, \dots, \xi_9\}$ where $\xi_j$ has a value from the set $\{-1, 0, S_j\}$ corresponding to {wall, empty, state of agent j}. The set $\xi_u^G$ refers to the sub-set of sensed entities on the ground, $\xi_u^G = \{\xi_j : j = 0:4\}$ while the set $\xi_u^A$ refers to the sub-set of mobile neighboring entities in the air, $\xi_u^A = \{\xi_j : j = 5:9\}$.

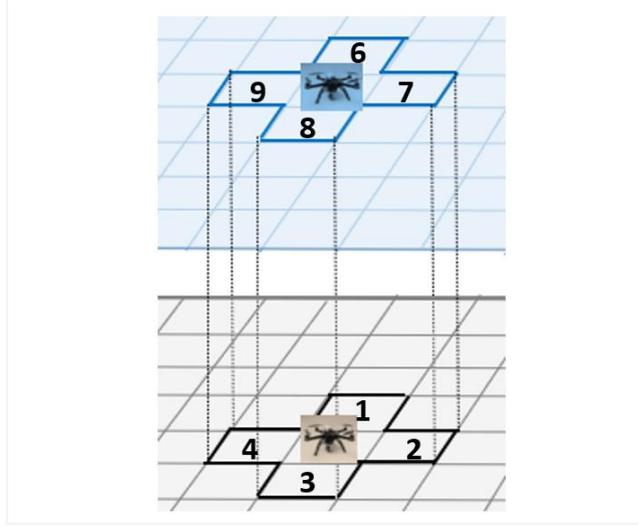

Figure 4  Neighboring region enumeration

The agents are identical (i.e., act according to the same action rules), anonymous (i.e., have no identifiers) and autonomous (i.e., have no central controller). The agents may also utilize a small, limited memory– depending on the algorithm. At every wake-up, the agent does a Look, Compute and Action (LCA) action cycle [32] in which it senses its surroundings, makes a decision and implements the action. An agent that wakes up at time $t$ senses the configuration of the agents at the previous sub time step $t - dt$. All actions are assumed to be instantaneous (i.e., atomic). The agent may do several types of actions. The first is "move" from one cell to another; the second is "settle in place", the third is "settle in a neighboring cell" which is composed of a moving to the relevant cell and settling in place and the fourth is "project" in which a settled agent becomes a beacon and projects (e.g., using lights) some information.

The four basic states of an agent are "in the agent source", "mobile", "settled" or "failed". We assume that agents in the agent source are not operating and thus don't consume any energy while the energy consumption of failed agents is irrelevant. The total energy consumption of an agent $a_i$ until time $t$ may thus be expressed as

$$E_i(t) = E_m^i(t) + E_s^i(t) \tag{2}$$

in which the components of the energy consumption when mobile and settled are denoted by $E_m$ and $E_s$ respectively. Assuming constant power consumption in the mobile and settled states we can re-write Eq. (2) as:

$$E_i = P_m t_m^i + P_s t_s^i \tag{3}$$

In order to simplify the discussion, every time step in which the agent is initially mobile increases $t_m^i$ by 1 resulting in $t_m^i, t_s^i \in \mathbb{Z}$.

**Definition 3:** The power ratio, denoted by α, is the ratio between the power consumption when settled and the power consumption when mobile, and is expressed as

$$\alpha = \frac{P_s}{P_m} \tag{4}$$

Division of Eq. (3) by $P_m$, using the definitions for α and assuming $P_m = 1$ yields:

$$E_i = t_m^i + \alpha t_s^i \qquad (5)$$

**Definition 4**: The energy available to an agent when it enters the region is $E_0$.
Assuming $P_m \gg P_s$ yields

$$E_i \cong t_m^i \qquad (6)$$

meaning the energy consumption is approximately linear with respect to the time an agent is mobile. While this approximation simplifies the discussion, its effect is investigated in the latter part of this work.

### D. Agent entry model

In this research the agents are initially outside the region and enter it one-at-a-time. We use the constant rate entry model in which the time interval between successive entry-attempts, denoted as $\Delta T$, is constant – see Figure 5. When $\Delta T > 1$, the time window in which agents may enter the region is limited to the first time step and in the other $\Delta T - 1$ time steps only the agents inside the region are active. In other words, no agent will enter the region during the interval $\Delta T$ if entry does not occur during the first time step. Whether an agent actually enters or not depends on the occupancy of the entry point. Since movement of the mobile agents move out of the entry point/s depend on the filling strategy, the actual entry rate depends on progression of the filling process and vis-versa. An agent entering the environment at a given time step remains there without waking for the rest of the current time step and activates only at the subsequent time step. Thus, on entering the region the agent consumes one unit of energy. The effect of the time interval on the covering process and specifically the total energy is investigated both analytically and experimentally in this work.

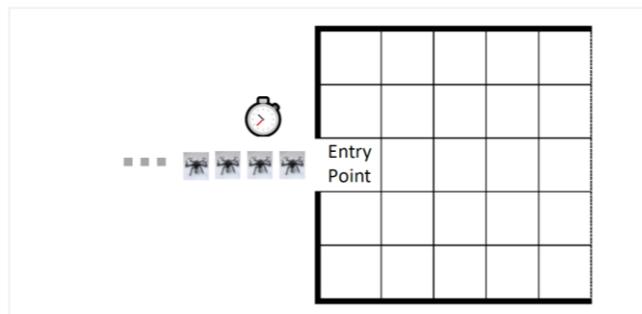

Figure 5  The Constant Rate Agent Entry model. Agents, located outside the region, may enter every $\Delta T$ time steps

### E. Performance Metrics

We are interested in the performance of the swarm as a whole as it solves the coverage problem and therefore define three metrics that measure the overall performance – termination time, total energy used, and number of agents used. Both the total energy used and number of agents used are measures of the effort to cover the region while the termination time measures its result. In addition to these three aggregative metrics, we are also interested in the maximal energy consumption of any of the agents since it directly relates to the energy limit of the agents. Furthermore, when the agents have insufficient energy to explore and cover the whole region it is of interest to know the area that was covered as well as the cost. In this research that cost is described by the number of depleted agents (i.e., that ran out of energy).

**Definition 5:** The *termination time*, $T_C(\mathbf{R})$, of a coverage process done by a multi-agent system with respect to a region $R$ is the first time step that the process of covering $R$ with agents terminates.

**Definition 6:** The set of agents participating in the coverage process at time step $t = k$ is defined by $A(t) = \{a_1, a_2, \dots, a_{N(t)}\}$ and the cardinality of $A(t)$, is denoted by $\mathbf{N(t)} = |\mathbf{A(t)}|$. Note, when referring to the number of agents without explicit mention of the time the meaning is the number of agents at $T_C$, $N = N(T_C)$.

**Definition 7:** The *total energy consumption* used by a multi-agent system with respect to a graph environment $\mathbf{R}$ is equal to $\sum_i E_i$ – the sum of all individual agents' energy consumptions. Total energy consumption is also denoted $E_{Total}(\mathbf{R})$.

**Definition 8:** The *maximum specific energy consumption* by any agent at the termination of the process, is denoted by $max\ (E_i)$, and defined by $max\ (E_i) = max(E_i: i = 1: N)$

**Definition 9:** The *Covered Area* is defined as number of cells in the region occupied by a settled agent at the time of process termination.

**Definition 10:** The *number of depleted agents*, abbreviated as $\mathbf{NDA,}$ is defined as the cardinality of the set $\{a_i \in A(T_C) | E_i = 0\}$

## IV.     The Single Layer, Energy Aware Coverage Algorithms

In all drones the energy capacity is limited and is the main constraint on the operational (flight) time. We describe in this section a couple of approaches and several algorithms that treat this constraint. In [1] two families of exploration and coverage algorithms are described however the energy capacity of the drones is not limited in either. One of the conclusions in [1] is that the *Single Layer Coverage* (SLC) algorithms significantly out-performs the *Dual Layer Coverage* (DLC) algorithms with respect to the termination time, number of agents and, most importantly, total energy used in the process. In this paper the energy constraint is integrated into the SLC algorithms, and the modified algorithms called the *Single Layer, Energy Aware Coverage* (abbreviated as SLEAC and pronounced "sleek") algorithms.

When the energy capacity is limited and the current energy level is not part of the local action rules, agents will enter the region, advance but run out of energy and crash before settling. Moreover, the knowledge that completely filling the region is impossible will not propagate to the user and agents will continue to enter the region ad-infinitum. Hence the local action rules of both moving and settled agents must be modified. We define two energy levels – $E_{critical}^{mobile}$ and $E_{critical}^{settled}$ – below which the behavior of the agent will change. When the energy of a mobile agent falls below $E_{critical}^{mobile}$ the agent will settle (in an occupied cell) and turn itself off in order to prevent multiple signals being projected from the same cell. $E_{critical}^{mobile}$ must be at least equal to one since the energy consumption of a mobile agent is one unit of energy per time step. Moreover, the agents can be mobile for at most $E_0 - E_{critical}^{mobile}$ time steps and can move to a maximal distance of $d_{max}$ cells from an entry point before settling.

$$d_{max} = E_0 - E_{critical}^{mobile} - 1^*  \qquad (7)$$

We define a new state of a settled agent – Low Energy. A settled agent will transition to this state when its energy level is below $E_{critical}^{settled}$ (i.e., due to locally sensed information). An agent in the Low Energy state will display Low Energy (as well as its step count). The state chart of an energy-aware agent is described in Figure 6. Note that mobile agents always transition to Beacon before transitioning to Low Energy and/or Closed Beacon in subsequent time steps.

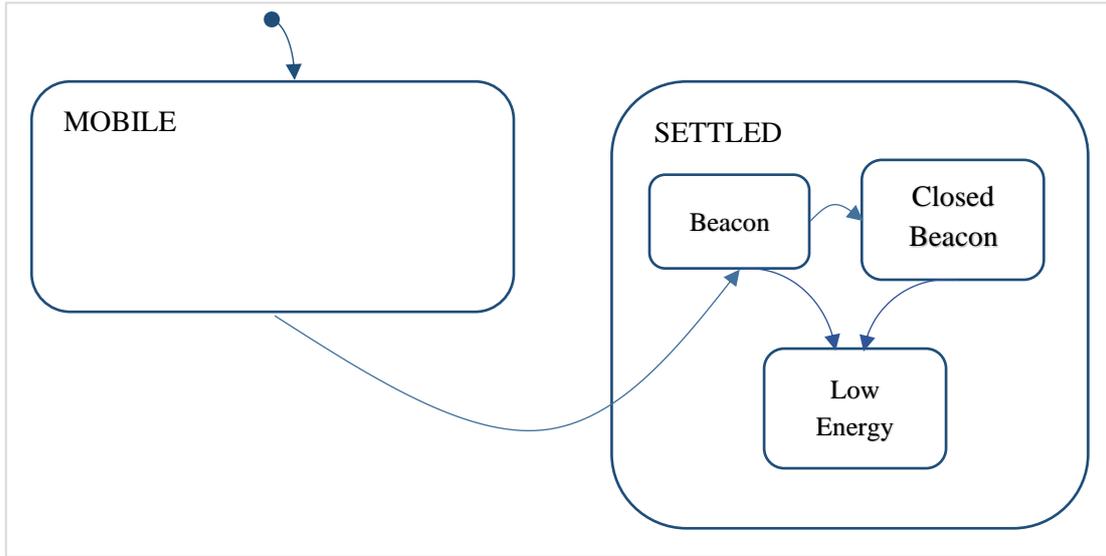

Figure 6  SLEAC algorithms - agent state diagram

There are two possible outcomes of running the SLEAC algorithms on a region $R$. The region may either be fully covered when $max(E_i) < E_{max}$ or not when $max(E_i) \geq E_{max}$ with the second case being the more relevant to the energy constraint problem. We thus modify the definition of the termination time in [1] to:

**Definition 11**: The ***termination time***, $T_C(R)$, *of a coverage process done by an energy limited, multi-agent system using a Single Layer, Energy Aware Coverage algorithm with respect to a region $R$ is the first time step at which either:*
*(i) external observer/s located at the entry point/s receive an indication of Low Energy, or*
*(ii) every cell in $R$ is occupied by one settled agent and external observer/s located at the entry point/s receive an indication of Closed Beacon.*

The Backward Propagating Closure (BPC) meta-concept, described at length in [1] is used to propagate information about the Closed Beacon state of agents to the entry point. This information is needed to generate a deterministic indication of termination to the user, guide the advancing agents to relevant parts of the region and guide the remaining mobile agents back to the entry point/s after process termination. In the SLEAC algorithms the BPC meta-concept is modified and used to propagate additional information about the Low Energy state of agents. Consequently, and a direct outcome of Definition 11, is that when the settled agent/s at the entry point transition to "Low Energy" the process will terminate, and additional agents will not enter the region.

---

* The additional unit of energy is subtracted from $E_0$ because the agent moves only in the time step following its entry.

Two alternative approaches to the termination of the process by transition of the settled agent/s at the entry points/s to Low Energy are investigated in this work. The objective of the first approach is to inform the external observer as fast as possible that there is an active energy constraint in the system and thus reduce the number of depleted drones. This is achieved by terminating the process as soon as possible after the first agent in the region transitions to Low Energy. Specifically, a settled agent will transition to Low Energy if any of its neighbors is in state Low Energy and regardless of its own energy level. The objective of the second approach is to explore as large part of the region as possible. This is achieved by delaying as much as possible the propagation of the Low Energy indication. Consequently, a settled agent will transition to Low Energy if all of its relevant neighbors changed their state to Low Energy†. We denote them Approach 1 and Approach 2. From the above objectives it is clear that transition of the entry point/s to Low Energy will be faster using Approach 1 however the covered area will be smaller compared to Approach 2 and vis-versa. Moreover, the number of additional agents that will enter the region after the 1st settled agent transitions to Low Energy will be smaller when using Approach 1‡.

The three SLEAC algorithms described in the following sub-sections are based the SLLG, SLUG and SLTT algorithms described in [1] and the behavior of the settled agents is according to either of the approaches described above. In each of the sub-sections the flowchart and the pseudocode are given with the modification to the basic algorithm in **Bold.** Reference to relevant parts of the flowcharts is made in the pseudocode.

### A.    Single Layer, Limited Gradient, Energy Aware (SLLG-EA) algorithm

In SLLG-EA, if the current energy level is beneath the pre-defined $E_{critical}^{mobile}$ the agent will settle and shut itself down. Otherwise, and as in SLLG, the mobile agents advance up a gradient of step counts projected by settled agents as long as its steepness is exactly one. When moving down the gradient the mobile agents can move over settled agents in the Closed beacon state. The behavior of the settled agents in SLLG-EA is significantly different than in SLLG. Using Approach 1 (shown in Figure 8) the settled agent first checks if its energy level is below the predefined $E_{critical}^{settled}$. If it is, the agent transitions to the Low Energy state. If not, the agent checks if any of its neighbors are in the Low Energy state in order to transition to Low Energy as soon as possible in accordance with the objective of Approach 1. If there are no neighboring agents in Low Energy the action rules are the same as in SLLG. Similarly, if Approach 2 (shown in Figure 9) is used the settled agent first checks its own energy level. If its energy level is not below $E_{critical}^{settled}$, transition to Low Energy will occur when the following three conditions are met simultaneously: (1) there are no neighboring empty cells; (2) there is at least one neighboring cell in the Low Energy state; and (3) all neighboring agents with a step count greater than own step count by exactly one are either in Low Energy or Closed Beacon state. The third condition treats scenarios in which some of the neighbors transitioned to Closed Beacon while others to Low

---

† In addition to transitioning when its energy level is below $E_{critical}^{settled}$.

‡ The relative merit of each approach also depends on the number of entry points into the region. This issue is not investigated in this paper.

Energy. This may occur, for example, to the settled agent in the middle of a "T-junction" with one branch in the Low Energy state and the other in the Closed Beacon state.

At every wake up time each agent behaves according to the flowcharts in Figure 7 - Figure 9 and pseudocode in Figure 10 - Figure 13.

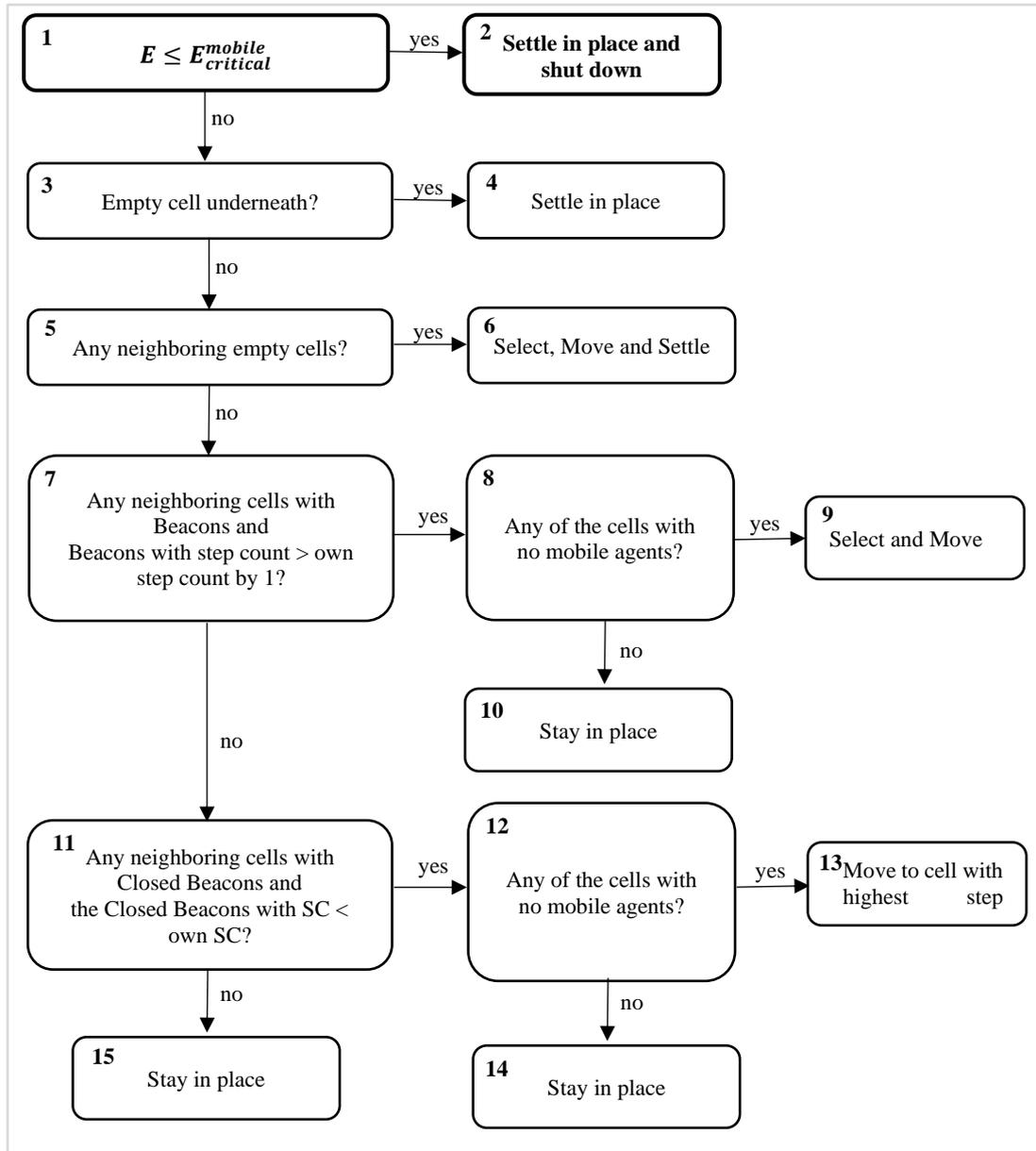

Figure 7      Single Layer, Limited Gradient-Energy Aware – mobile agent flowchart

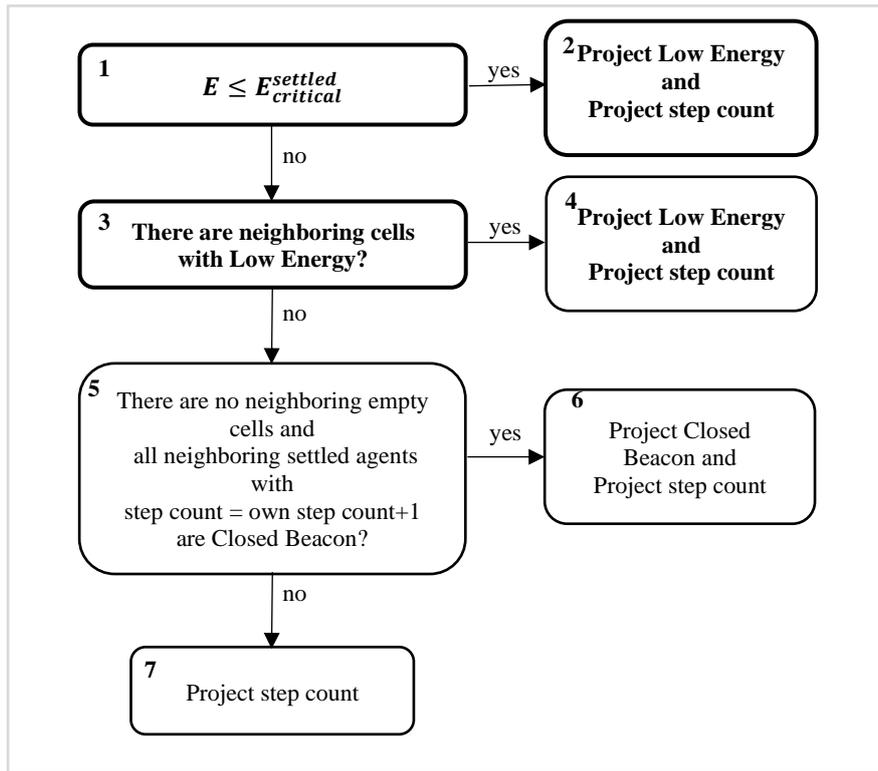

Figure 8　　　　　　　Single Layer, Limited Gradient-Energy Aware – Approach 1 - settled agent flowchart

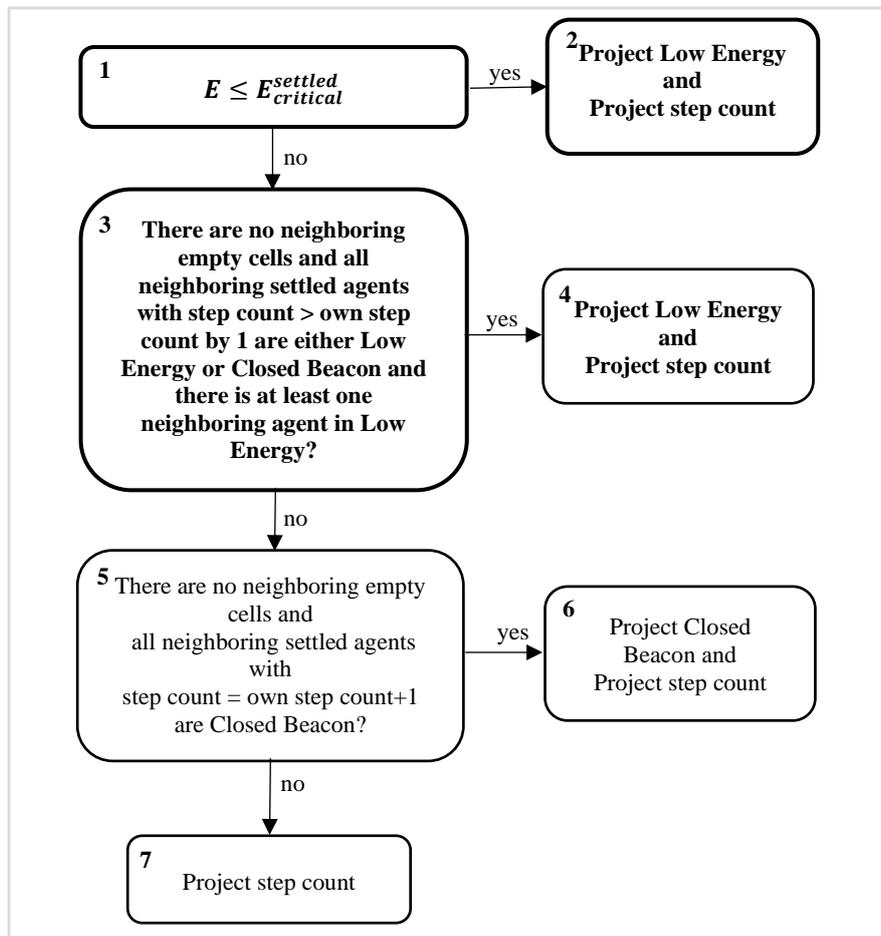

Figure 9　　　　　　　Single Layer, Limited Gradient-Energy Aware – Approach 2 - settled agent flowchart

**Initialization**: for all agents $a \in A$ located at $t=0$ at the agent source set $S_i = (s_1, s_2) = ("mobile", 0)$ and $E_i = E_0$

```
If s_1 = "mobile"
    If E ≤ E_critical^mobile                                /* if the current energy level is E_critical^mobile [1]
        u ← u;   shutdown                                   /* settle in place and shutdown [2]
    Elseif ∃(ξ^G) = 0 then                                  /* if there are empty, neighboring cells
        If ξ_0^G = 0 then                                   /* if the cell directly underneath is empty [3]
            u ← u;   s_1 ← "settled";   s_2 ← 1             /* set step count as 1 [4]
        Else                                                /* if not [5]
            ξ̃^G = {ξ_j^G = 0: j = 1:4}                      /* define set ξ̃^G as all empty neighboring
                                                            /* cells
            v ← random(|ξ̃^G|)                              /* select a cell from ξ̃^G
                                                            /* settle at selected empty cell, update
                                                            /* state, increase step count by 1 [6]
            u ← v;   s_1 ← "settled";   s_2 ← s_2 + 1
        End If
    Else
        possiblePos ← ∅                                     /* initialize possiblePos - the set of possible
                                                            /* locations
        destSC ← s_2 + 1                                    /* define the step count of the destination
                                                            /* cell
        relevantCells ← ξ^G with ξ^G.s_1 = "Beacon" and ξ^G.s_2 = destSC
                                                            /* the subset of ξ̃^G that is filled with agents in
                                                            /* Beacon state and have a SC eq. to destSC
        If |relevantCells| > 0                              /* if there are neighboring cells with Beacons
                                                            /* and step count > own step count by 1 [7]
            possiblePos ← relevantCells with ξ^A = 0        /* the subset of relevantCells with no mobile
                                                            /* agents
            If |possiblePos| > 0                            /* if there are cells to move-to [8]
                v = random(|possiblePos|)                   /* select a cell from possiblePos [9]
                u ← v;   s_1 ← "mobile";   s_2 ← destSC     /* move to selected cell and increase Step
                                                            /* Count by 1
            Else
                u ← u;   s_1 ← "mobile";   s_2 ← s_2        /* stay in place [10]
            End If
        Else
            destSC ← 0                                      /* initialize the step count of the destination
                                                            /* cell
            relevantCells ← ξ^G with ξ^G.s_1 = "Closed Beacons and ξ^G.s_2 < s_2
                                                            /* search for cells that are Closed Beacons
                                                            /* and have a step count lower than own
                                                            /* step count
```

Figure 10  Single Layer, Limited Gradient-Energy Aware pseudocode – mobile agent - part 1

```
            If |relevantCells| > 0                              /* [11]
            possiblePos ← relevantCells with ξ^A = 0    /*the subset of relevantCells with no
                                                         /*mobile agents
                If |possiblePos| > 0                     /*if there are cells to move-to [12]
                    destSC = max (possiblePos.s_2 < s_2)  /* find in possiblePos the cell/s with the
                                                         /* highest step count [13]
                    possiblePos ← possiblePos with possiblePos.s_2 = destSC and ξ^A = 0
                                                         /* find all the cells in possiblePos with Step
                                                         /*Count eq. to destSC and no mobile agents
                    v = random(|possiblePos|)            /*select a cell from possiblePos [13]
                    u ← v;   s_1 ← "mobile";   s_2 ← destSC
                                                         /*move to selected cell and set step count
                                                         /* to destSC
                Else
                    u ← u;   s_1 ← "mobile";   s_2 ← s_2   /*stay in place [14]
                End If

            Else                                         /* [15]
                u ← u;   s_1 ← "mobile";   s_2 ← s_2   /*stay in place
            End If
        End If
    End If
Else                                                     /* if the agent is not mobile
    Do SLLG-EA settled agent logic as defined in Figure 12 or Figure 13
End
```

Figure 11    Single Layer, Limited Gradient-Energy Aware pseudocode – mobile agent - part 2

```
If s_1 ≠ "mobile"                                    /* if the agent is not mobile
    If E < E_critical^settled                        /* If the current energy < E_critical^settled [1]
        s_1 ← "Low Energy"                           /* state is "Low Energy" [2]
        Project (s_1)                                /* project state and step count
        Project (s_2)
    Elseif ∃ξ_j^G.s_1 = Low Energy, j = 1:4          /* There are neighboring agents in state
                                                     /* "Low Energy" [3]
        s_1 ← "Low Energy"                           /* state is "Low Energy" [4]
        Project (s_1)                                /* project state and step count
        Project (s_2)
    Else ∄ξ_j^G = 0, j = 1:4                         /* If there are no empty cells [5]
        ξ̌^G ← ξ^G with ξ^G.s_2 = ξ_0^G.s_2 + 1       /* ξ̌^G - the set of neighboring settled agents
                                                     /* with step count > own step count
        ξ̂^G ← ξ^G with ξ^G.s_1 = "Closed Beacon"     /* ξ̂^G - the set of neighboring Closed Beacon
        If ξ̌^G = ξ̂^G                                 /* [5]
            s_1 ← "Closed Beacon"                    /* state is "Closed Beacon" [6]
            Project (s_1)                            /* project state and step count
            Project (s_2)
        Else
            s_1 ← "Beacon"                           /* state is "Beacon"  [7]
            Project (s_2)
        End If
    End If
End If
```

Figure 12    Single Layer, Limited Gradient-Energy Aware – Approach 1 - settled agent pseudocode

```
If s₁ ≠ "mobile"                                              /* if the agent is not mobile
    If E < E_{critical}^{settled}                             /* If there current energy < E_{critical}^{settled} [1]
        s₁ ← "Low Energy"                                     /* state is "Low Energy" [2]
        Project (s₁)                                          /* project state and step count
        Project (s₂)
    Else                                                      /* If there are no empty cells [3]
        ξ̆^G ← ξ^G with ξ^G.s₂ = ξ₀^G.s₂ + 1                   /* ξ̆^G - the set of neighboring settled agents with
                                                                    step count > own step count
        ξ̂^G ← ξ^G with ξ^G.s₁ = Low Energy or Closed Beacon
                                                              /* ξ̂^G - the set of neighboring settled agents in Low
                                                              /* Energy or Closed Beacon
        If ξ̆^G = ξ̂^G and |ξ̂^G| > 0 and ∄ξ_j^G = 0, j = 1:4
                                                              /* [3]
            s₁ ← "Low Energy"                                 /* state is "Low Energy" [4]
            Project (s₁)                                      /* project state and step count
            Project (s₂)
        Else
            ξ̂^G ← ξ^G with ξ^G.s₁ = "Closed Beacon"   /* ξ̂^G - the set of neighboring Closed Beacon
            If ξ̆_u^G = ξ̂_u^G                                 /* [5]
                s₁ ← "Closed Beacon"                          /* state is "Closed Beacon" [6]
                Project (s₁)                                  /* project state and step count
                Project (s₂)
            Else
                s₁ ← "Beacon"                                 /* state is "Beacon" [7]
                Project (s₂)
            End If
        End If
    End If
End If
```

Figure 13  Single Layer, Limited Gradient-Energy Aware – Approach 2 - settled agent pseudocode

*Example*

An example of the energy constrained coverage process using SLLG-EA is described in Figure 14 for Approach 1 and in Figure 15 for Approach 2. The region is a $41 \times 41$ square with the entry point at the exact middle as shown in Figure 14(a). In addition, $E_0 = 15$, $E_{critical}^{mobile} = 1$ and $E_{critical}^{settled} = 1$ which means $d_{max}$, the maximum distance an agent can move from the entry point is 13. A mobile agent at the expansion frontier with $E = 2$ will move and settle in a neighboring empty cell and transition to state Beacon. The move and settle action will reduce its energy by one unit hence in its next wake-up it will sense its energy level is 1 and transition to Low Energy. This transition bounds the covered area because advance is only possible over settled agents in Beacon state. The power factor, $\alpha$, is zero hence once an agent settles its energy level remains unchanged and $\Delta T = 1$. To facilitate better understanding, colors are used to differentiate between the different types of cells. Empty cells are black, the entry point is blue, cells with agents in the Beacon state are colored yellow and similarly cells with Closed Beacon are brown and cells in Low Energy are gray. The mobile agents are the green "X".

Figure 14(b)-(f) show snapshots of the process at different times using Approach 1. In Figure 14(b) and (c) the energy constraint is not yet active whereas in Figure 14(d) the first settled agent transitions to Low Energy. Figure 14(e) and (f) show the fast propagation of the Low Energy indication to the entry point. Note that all cells neighboring a Low Energy settled agent also transition to Low Energy as described in the algorithm above.

Figure 15(a) shows the coverage at the first time step an agent transition to Low Energy using Approach 2 while Figure 15(b) shows the process several time steps later when additional agents at the boundary transitioned to Low Energy and the indication began propagating towards the entry point. As seen in Figure 15(c) – (e) the two process, exploration and propagation of the Low Energy indication, continue simultaneously. The potential blocking of advancing agents by agents moving down the gradient is seen in both Figure 15(d) – (e) (see blue oval in Figure 15(e)). It does not occur since the action rules in SLLG prevent retracing over Beacons, effectively separating the advancing agents from the retracing ones. Lastly, the funneling of the mobile agents to the still relevant part of the region is clearly in the sequence of snapshots.

As the exploration process continues the covered region takes the form of a square rotated by 45 degrees with respect to the region's boundary. Moreover, this shape first forms at the main diagonals (shown in Figure 15(b) and (d) as blue lines). At each time step, given several possible directions of advance, an agent will arbitrarily and with equal probability select one. Points along the main diagonals are reached by moving on an equal number of vertices on the horizontal and on the vertical axis and at whatever order. In contrast, points on the horizontal or vertical axis require moving continuously in the same direction hence repeatedly selecting, at every wake-up, the same direction. Thus, the probability an agent will move along a certain path depends on the vertices comprising that path. Consequently, the square shape results from the combination of the limit on the distance (due to the energy constraint) and the different likelihoods an agent will advance on one path versus another.

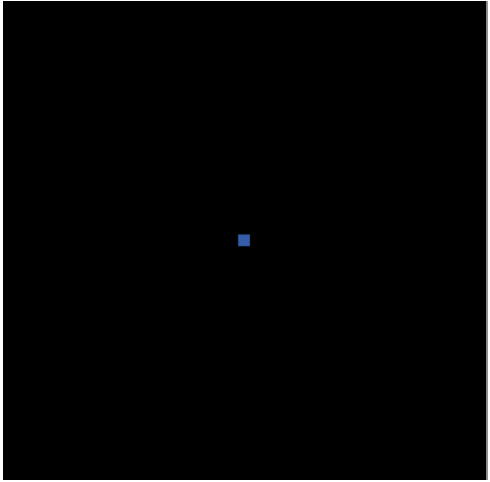

(a) Initial Configuration

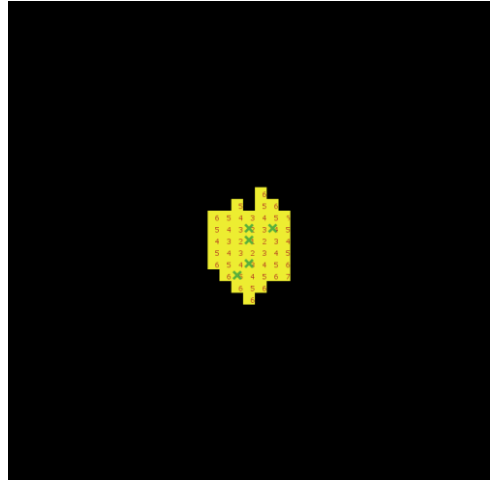

(b) t = 54

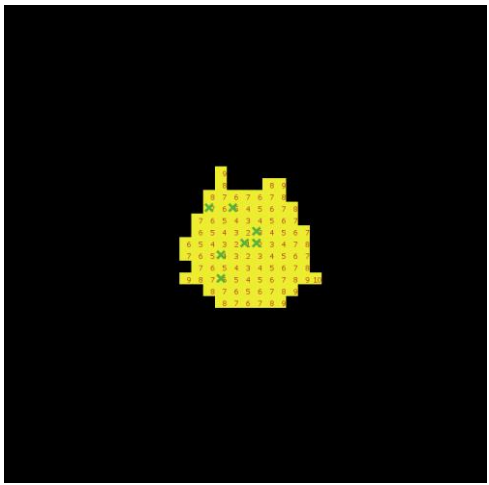

(c) t = 103

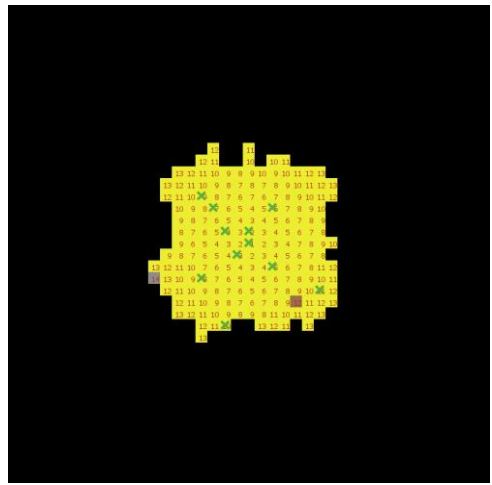

(d) t = 210

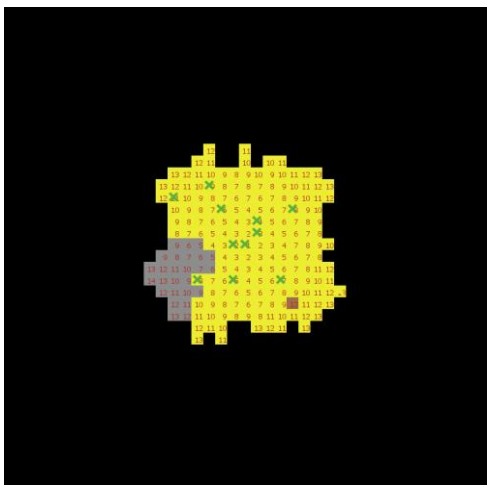

(e) t = 212

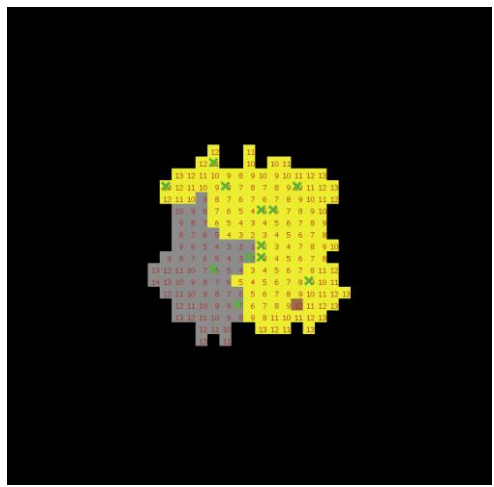

(f) t = 214

Figure 14 Single Layer, Limited Gradient – Energy Aware – Coverage process using Approach 1 of a square region. The entry point is in the middle and $E_0 = 15, E_{critical}^{mobile} = 1$ and $E_{critical}^{settled} = 1$. The snapshots of the process are ordered left to right, top to bottom

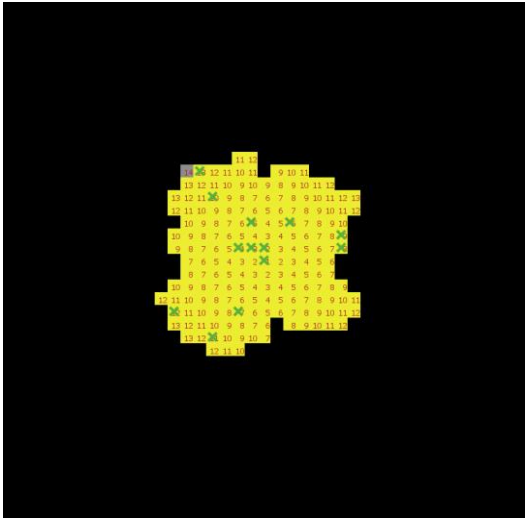

(a) t = 199

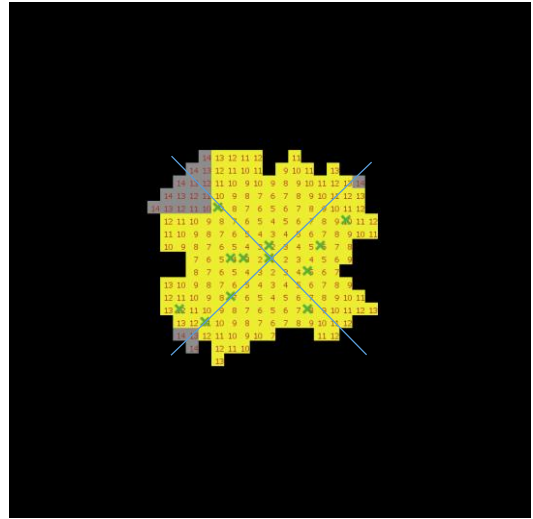

(b) t = 228

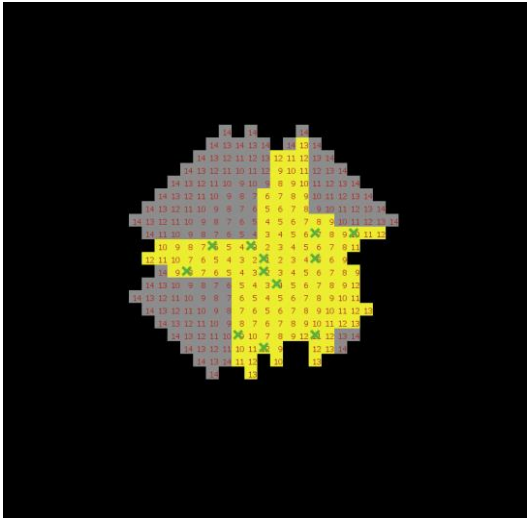

(c) t = 305

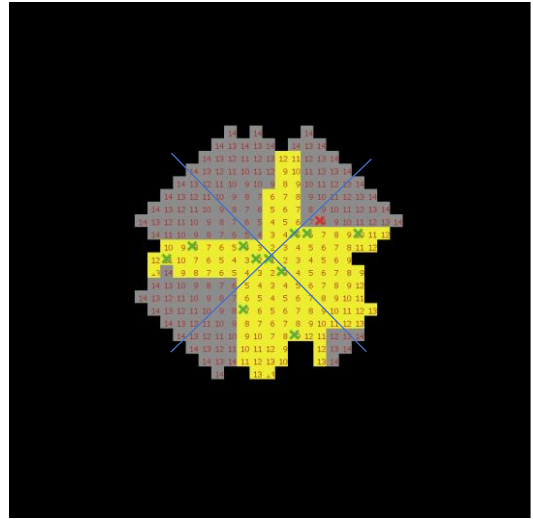

(d) t = 310

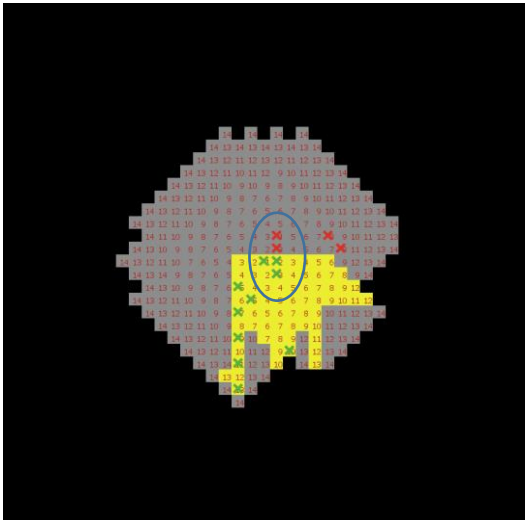

(e) t = 352

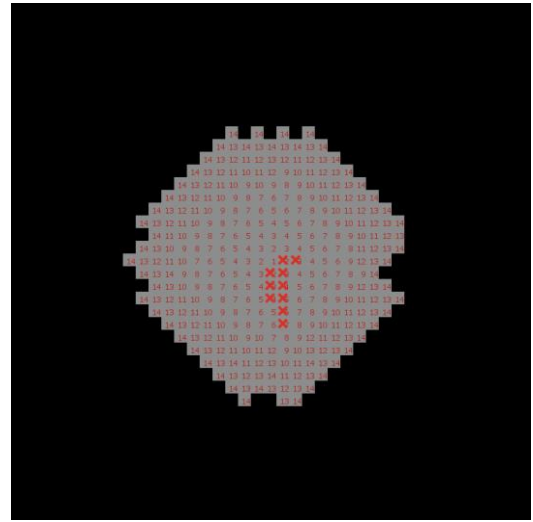

(f) t = 403

Figure 15 Single Layer, Limited Gradient – Energy Aware – Coverage process using Approach 2 of a square region. The entry point is in the middle and $E_0 = 15$, $E_{critical}^{mobile} = 1$ and $E_{critical}^{settled} = 1$. The snapshots of the process are ordered left to right, top to bottom

## B. Single Layer, Unlimited Gradient, Energy Aware (SLUG-EA) algorithm

The local action-rules in SLUG-EA are modified to take into consideration the instantaneous energy level in the same manner as in SLLG-EA described above. Thus, a mobile agent will settle and shut itself down when its energy level is lower then $E_{critical}^{mobile}$ and settled agents will transition to Low Energy if their energy level is below $E_{critical}^{settled}$. The specific conditions used to determine the transition to Low Energy in both approaches is shown in block 3 in Figure 18 and Figure 19. Otherwise, the mobile agents advance up a gradient of step counts projected by settled agents regardless of the steepness. When moving down the gradient the mobile agents can move over agents in either the Beacon or Closed beacon state.

The main difference between SLUG-EA and SLLG-EA is in the definition of the step count. In SLUG-EA the step count of a mobile agent $a_i$ at time step $t$ is the step count of the settled agent directly underneath (i.e., $sc_i = \xi_0.s_2$) rather than the distance from the entry point (in SLLG and SLLG-EA). Consequently, the step count of a settled agent using SLUG-EC, may be greater than the distance the agent moved from the entry point. In other words, the possible trajectories from the entry point to a certain cell may differ from one another by their energy consumption. This difference is critical when the energy is constrained and negligible when it is not. When the onboard energy is limited, an agent using SLUG-EC may settle with a step count equal to the maximal possible distance (for a given $E_0$ and $E_{critical}^{mobile}$) but without transiting to the Low Energy state. All following agents that may use a less efficient trajectory will be definition have $E_{critical}^{mobile}$ and hence settle and shut down. Figure 16 depicts an example of such a scenario. In the example the region is a $41 \times 41$ square with the entry point in the middle and $E_0 = 15, E_{critical}^{mobile} = 1$ and $E_{critical}^{settled} = 1$. The settled agent, marked by a surrounding blue oval, projects a step count of 14 but has $E = 3$. This is the result of the agent advancing along the dotted path prior to settling and becoming a Beacon. Since $\Delta T = 2$, following agents will move along the path with a steepness of one and reach said agent with $E = 1$ and consequently settle in place, shut down and never transition to Low Energy. To prevent such events, the transition of a settled agent to Low Energy in SLUG-EA also occurs when $step\ count = d_{max}$.

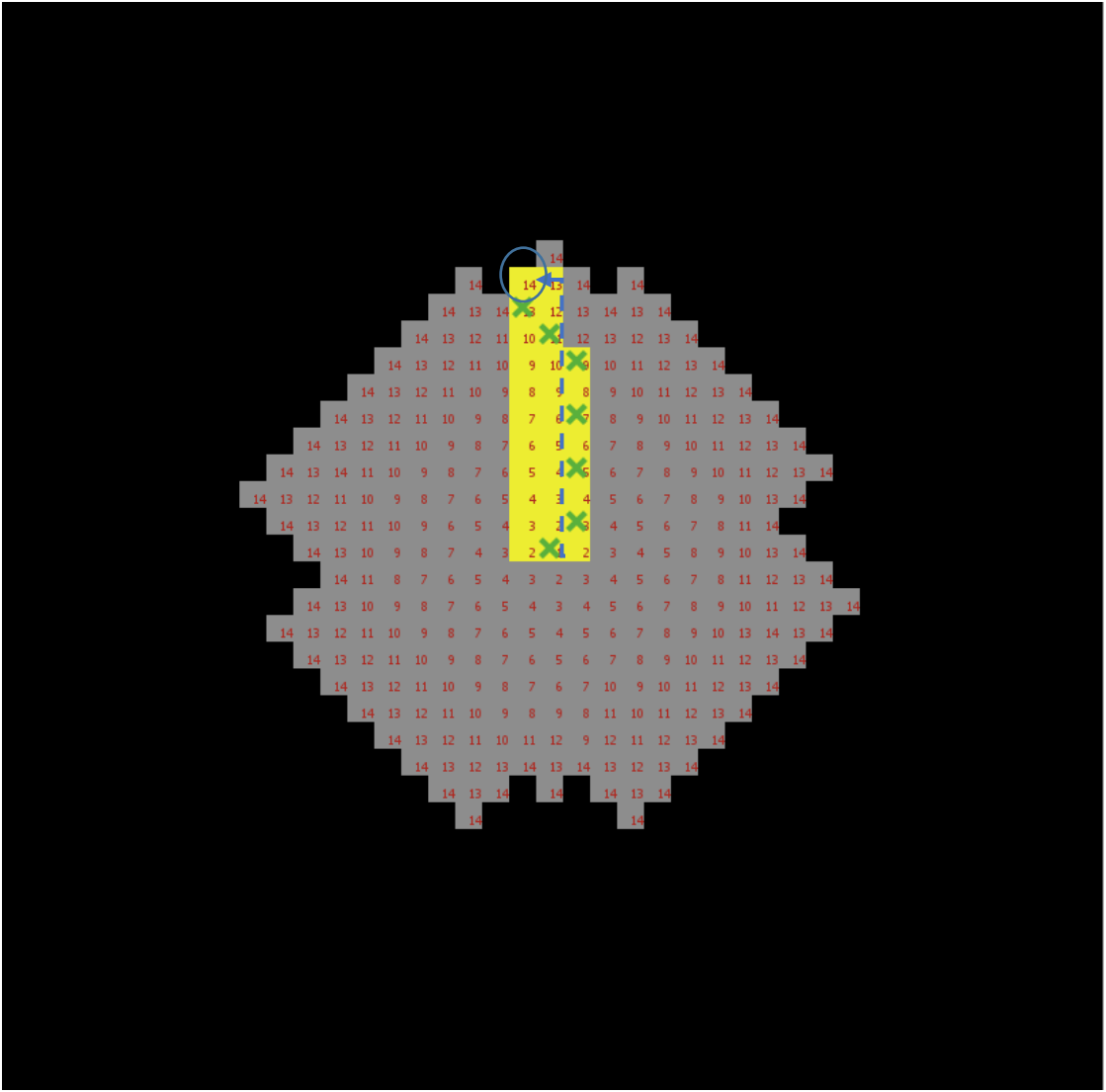

Figure 16 The problem of unlimited gradient

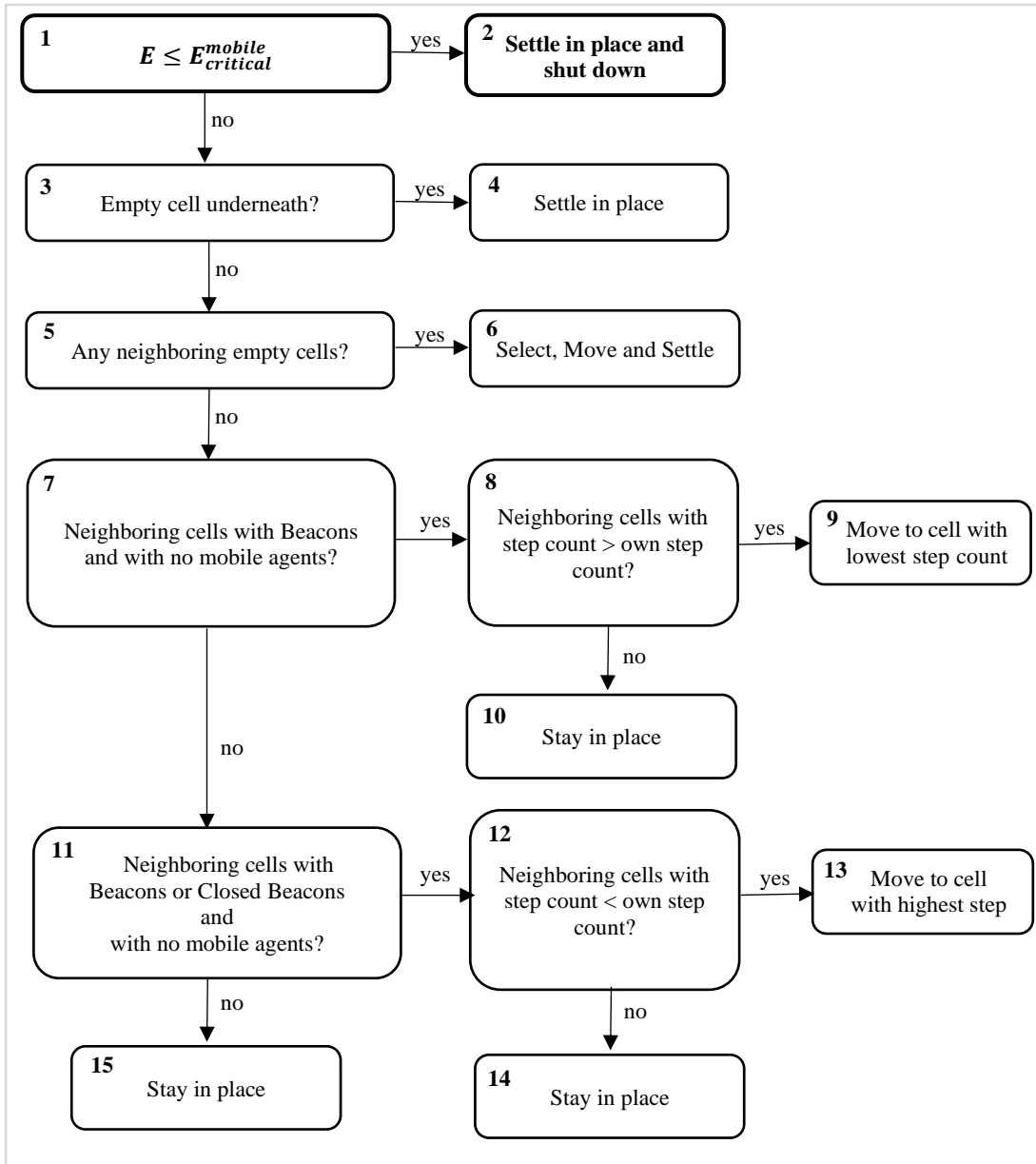

Figure 17          Single Layer, Unlimited Gradient – Energy Aware – mobile agent flowchart

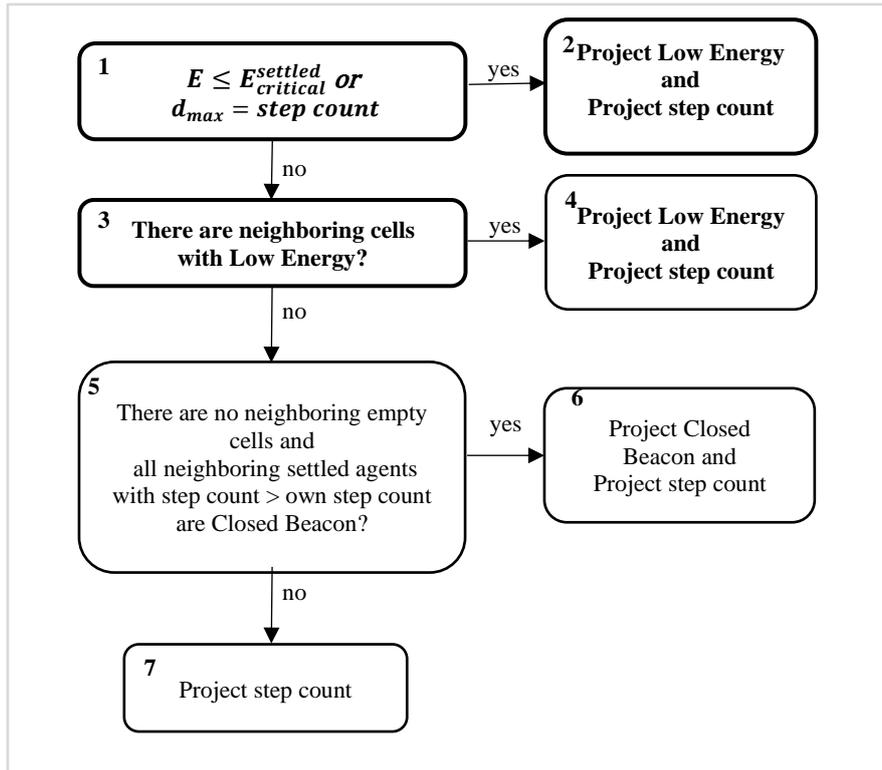

Figure 18  Single Layer, Unlimited Gradient – Energy Aware – Approach 1 - settled agent flowchart

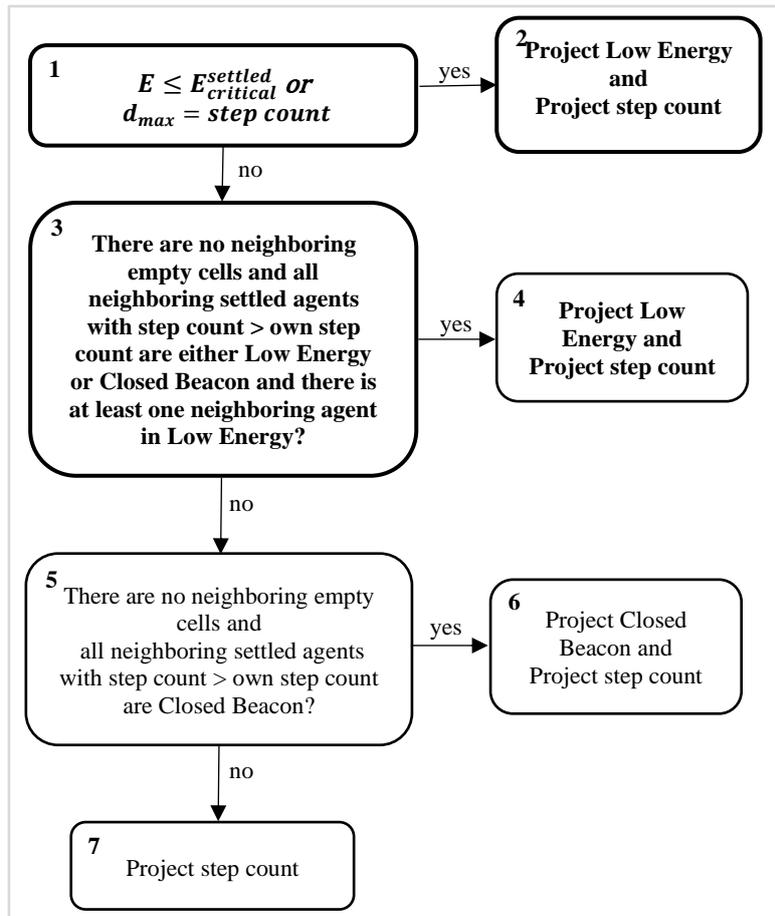

Figure 19  Single Layer, Unlimited Gradient – Energy Aware – Approach 2 - settled agent flowchart

**Initialization:** for all agents $a \in A$ located at t=0 at the agent source set $S_i = (s_1, s_2) = (\text{"mobile"}, 0)$ and $E_i = E_0$

---

```
If s_1 = "mobile"
    If E ≤ E_critical^mobile                                  /* if the current energy level is le. E_critical^mobile [1]
        u ← u;   shutdown                                     /* settle in place and shutdown [2]
    Elseif ∃(ξ^G) = 0                                         /* if there are empty, neighboring cells
        If ξ_1^G = 0                                          /* if the cell directly underneath is empty [3]
            u ← u;   s_1 ← "settled";   s_2 ← 1               /* set step count as 1 [4]
        Else                                                  /* if not [5]
            ξ̃^G = {ξ_j^G = 0: j = 1:4}                        /* define set ξ̃^G as all empty neighboring
                                                              /* cells
            v ← random(|ξ̃^G|)                                 /* select a cell from ξ̃^G
            u ← v;   s_1 ← "settled";   s_2 ← s_2 + 1         /* settle at selected empty cell, update state,
                                                              /* increase step count by 1 [6]
        End If
    Else
        possiblePos ← ∅                                       /* define possiblePos - the set of possible
                                                              /* locations
        destSC ← s_2 + 1                                      /* define the step count of the destination
                                                              /* cell
        relevantCells ← ξ_u^G with ξ_u^G.s_1 = "settled" and ξ_u^A = 0
                                                              /* subset of ξ_u^G that is filled with Beacons and
                                                              /* with no mobile agents [7]
        If |relevantCells| > 0                                /* if there are neighboring cells with Beacons
                                                              /* and no mobile agents
            destSC ← min(relevantCells.s_2 > s_2)             /* lowest step count in the set relevantCells
                                                              /* greater than current step count
            possiblePos ← relevantCells with relevantCells.s_2 = destSC
                                                              /* the subset of relevantCells with beacons
                                                              /* having a step count equal to destSC
            If |possiblePos| > 0                              /* if there are cells to move-to [8]
                v ← random(|possiblePos|)                     /* select a cell from possiblePos [9]
                u ← v;   s_1 ← "mobile";   s_2 ← destSC       /* move to selected cell and increase Step
                                                              /* Count by 1
            Else
                u ← u;   s_1 ← "mobile";   s_2 ← s_2          /* stay in place [10]
            End If
        Else
            destSC ← 0                                        /* initialize the step count of the destination
                                                              /* cell
            relevantCells ← ξ_u^G with (ξ_u^G.s_1 = "Beacons" or ξ_u^G.s_1
                          = "Closed Beacons") and ξ_u^A.s_1 = 0
                                                              /* search for cells that are filled by settled
                                                              /* agents and no mobile agents
```

Figure 20    Single Layer, Unlimited Gradient – Energy Aware - mobile agent pseudocode –- part 1

```
        If |relevantCells| > 0                                    /*[11]
           destSC ← max (relevantCells.s_2 < s_2)
           possiblePos ← relevantCells with relevantCells.s_2 = destSC
                                                    /* the subset of relevantCells with Step
                                                    /*count eq. to destSC no mobile agents

           If |possiblePos| > 0                     /*if there are cells to move-to [12]
              v ← random(|possiblePos|)             /*select a cell from possiblePos [13]
              u ← v;  s_1 ← "mobile";  s_2 ← destSC /*move to selected cell and set step count
                                                    /* to destSC
           Else
              u ← u;  s_1 ← "mobile";  s_2 ← s_2    /*stay in place [14]
           End If

        Else                                        /*[15]
           u ← u;  s_1 ← "mobile";  s_2 ← s_2       /*stay in place
        End If

     End If
  End If
Else                                                /* if the agent is not mobile
     Do SLUG-EA settled agent logic as defined in Figure 22 or Figure 23
End
```

Figure 21    Single Layer, Unlimited Gradient – Energy Aware mobile agent pseudocode - part 2

```
If s_1 ≠ "mobile"                                    /* if the agent is not mobile
    If E < E_{critical}^{settled} or d_{max} = step count   /* If the current energy < E_{critical}^{settled} or d_{max} =
                                                     /* step count [1]
        s_1 ← "Low Energy"                           /* state is "Low Energy" [2]
        Project (s_1)                                /* project state and step count
        Project (s_2)
    Elseif ∃ξ_j^G . s_1 = Low Energy, j = 1: 4      /* There are neighboring agents in state "Low Energy" [3]
        s_1 ← "Low Energy"                           /* state is "Low Energy" [4]
        Project (s_1)                                /* project state and step count
        Project (s_2)
    Else ∄ξ_j^G = 0, j = 1: 4                       /* If there are no empty cells [5]
        ξ̌_u^G ← ξ_u^G with ξ_u^G . s_2 > ξ_0^G . s_2   /* ξ̌_u^G - the set of neighboring settled agents with
                                                               step count > own step count
        ξ̂_u^G ← ξ_u^G with ξ_u^G . s_1 = "Closed Beacon"  /* ξ̂_u^G - the set of neighboring Closed Beacon
        If ξ̌_u^G = ξ̂_u^G                           /* [5]
            s_1 ← "Closed Beacon"                    /* state is "Closed Beacon" [6]
            Project (s_1)                            /* project state and step count
            Project (s_2)
        Else
            s_1 ← "Beacon"                           /* state is "Beacon"      [7]
            Project (s_2)
        End If
    End If
End If
```

Figure 22    Single Layer, Unlimited Gradient – Energy Aware pseudocode – Approach 1 - settled agent pseudocode

```
If s_1 ≠ "mobile"                                          /* if the agent is not mobile
    If E < E_critical^settled or d_max = step count        /* If the current energy < E_critical^settled or d_max =
                                                           /* step count [1]
        s_1 ← "Low Energy"                                 /* state is "Low Energy" [2]
        Project (s_1)                                      /* project state and step count
        Project (s_2)
    Else
        ξ̆^G ← ξ^G with ξ^G.s_2 > ξ_0^G.s_2                /* ξ̆^G - the set of neighboring settled agents with
                                                              step count > own step count
        ξ̂^G ← ξ^G with ξ^G.s_1 = Low Energy or Closed Beacon
                                                           /* ξ̂^G - the set of neighboring settled agents in Low
                                                           /* Energy or Closed Beacon
        If ξ̆^G = ξ̂^G and |ξ̂^G| > 0 and ∄ξ_j^G = 0, j = 1:4
                                                           /*[3]
            s_1 ← "Low Energy"                             /* state is "Low Energy" [4]
            Project (s_1)                                  /* project state and step count
            Project (s_2)
        Else
            ξ̂_u^G ← ξ_u^G with ξ_u^G.s_1 = "Closed Beacon"  /* ξ_u^G - the set of neighboring Closed Beacon
            If ξ̆_u^G = ξ̂_u^G
                s_1 ← "Closed Beacon"                      /* state is "Closed Beacon" [6]
                Project (s_1)                              /* project state and step count
                Project (s_2)
            Else
                s_1 ← "Beacon"                             /* state is "Beacon"  [7]
                Project (s_2)
            End If
        End If
    End If
End If
```

Figure 23    Single Layer, Unlimited Gradient – Energy Aware pseudocode – Approach 2 - settled agent pseudocode

## C. Single Layer, Tree Traversal, Energy Aware (SLTT-EA) algorithm

In SLTT-EA the settled agents project indication of the direction of movement rather than step counts and the mobile agents advance only to neighboring cells that are children of their current location. Since the mobile agents move on the tree implicitly generated by the settled agents, the scenario described in the previous section are impossible.

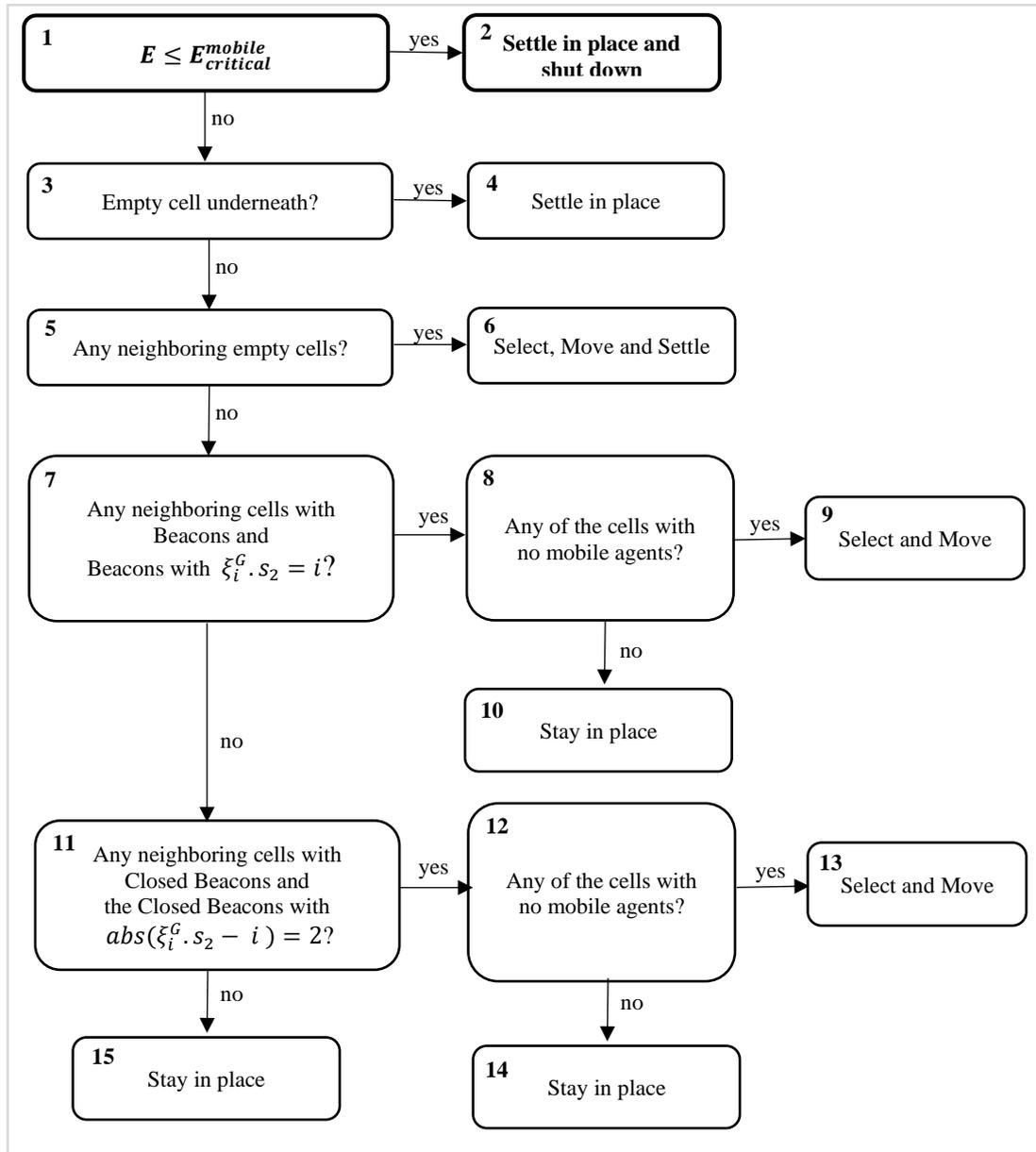

Figure 24        Single Layer, Tree Traversal, Energy Aware – mobile agent flowchart

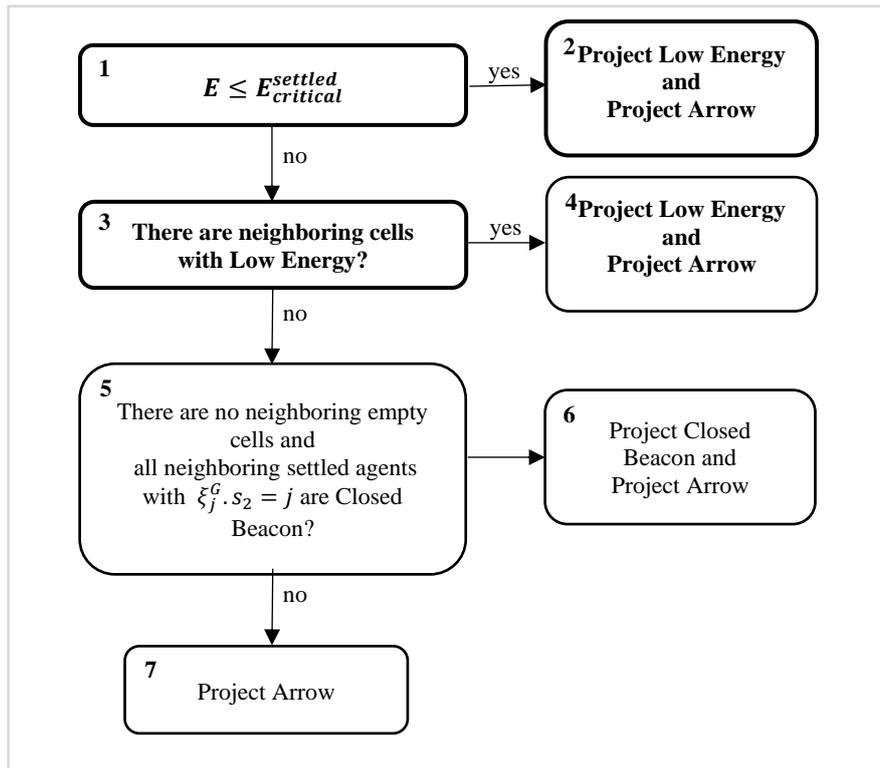

Figure 25                 Single Layer, Tree Traversal, Energy Aware – Approach 1 - settled agent flowchart

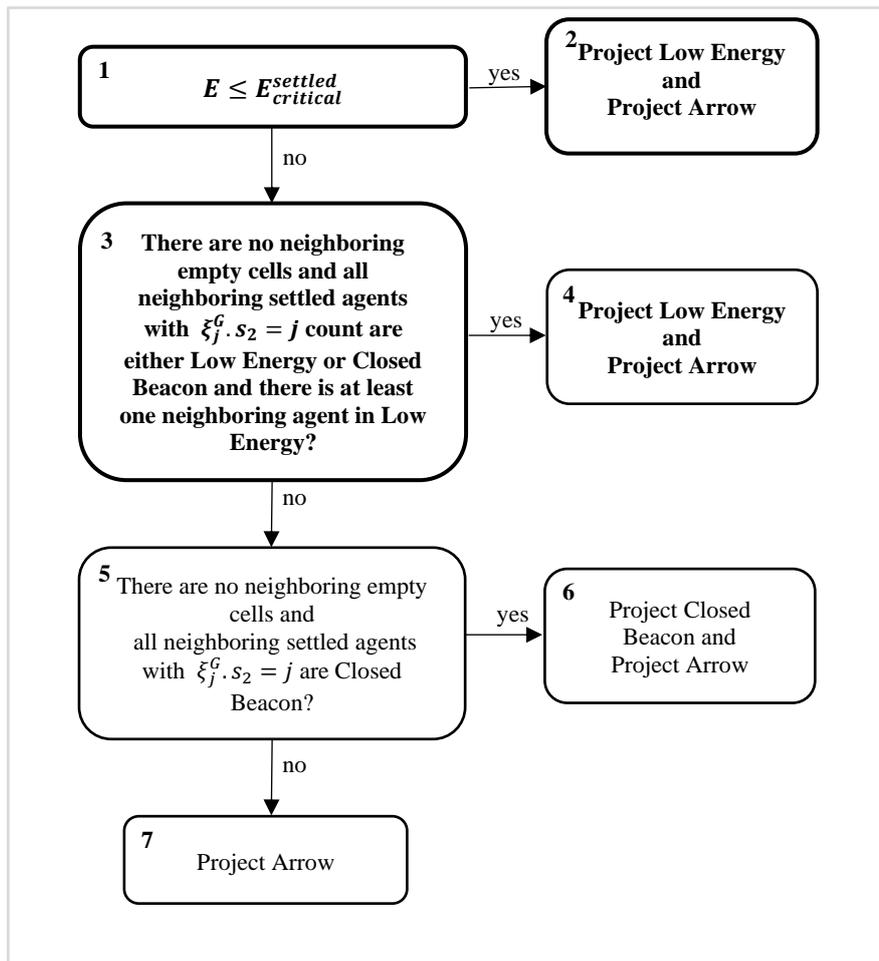

Figure 26                 Single Layer, Tree Traversal, Energy Aware – Approach 2 - settled agent flowchart

**Initialization**: for all agents $a \in A$ located at $t=0$ at the agent source set $S_i = (s_1, s_2) = ("mobile", 0)$ and $E_i = E_0$

```
If s₁ = "mobile"
    If E ≤ E_critical^mobile                                          /*if the current energy level is E_critical^mobile [1]
        u ← u;   shutdown                                             /*settle in place and shutdown [2]
    If ∃(ξ^G) = 0 then                                                /*if there are empty, neighboring cells
        If ξ_1^G = 0 then                                             /*if the cell directly underneath is empty [3]
            u ← u;   s₁ ← "settled";   s₂ ← 0                         /*[4]
        Else                                                          /*if not [5]
            ξ̃^G = {ξ_j^G = 0: j = 1:4}                                /* define set ξ̃^G as all empty neighboring
                                                                      /*cells
            v ← random(|ξ̃^G|)                                         /*select a cell from ξ̃^G
                                                                      /*settle at selected empty cell, update state,
                                                                      /*set A as the direction to move to v [6]
            u ← v;   s₁ ← "settled";   s₂ ← A(v)
        End If
    Else
        possiblePos ← ∅                                               /*initialize possiblePos - the set of possible
                                                                      /* locations
        relevantCells ← ξ^G with ξ^G.s₁ = Beacon and ξ_i^G.s₂ = i (for i = 1:4)
                                                                      /*the subset of ξ^G that is filled with Beacons
                                                                      /* and the direction they are signaling is eq.
                                                                      /* to the relevant position in the sensing
                                                                      /*neighborhood
        If |relevantCells| > 0                                        /*[7]
            possiblePos ← relevantCells with ξ^A = 0                  /*the subset of relevantCells with no mobile
                                                                      /*agents
            If |possiblePos| > 0                                      /*if there are cells to move-to [8]
                v = random(|possiblePos|)                             /*select a cell from possiblePos [9]
                u ← v;   s₁ ← "mobile";   s₂ ← A(v)                   /*move to selected cell and set A as the
                                                                      /*direction to move to v
            Else
                u ← u;   s₁ ← "mobile";   s₂ ← s₂                     /*stay in place [10]
            End If

        Else
            relevantCells ← ξ^G with ξ^G.s₁ = "Closed Beacons and abs(ξ_i^G.s₂ − i) = 2
                    (for i = 1:4)
                                                                      /*search for cells that are Closed Beacons
                                                                      /*and the direction they are signaling is
                                                                      /*opposite to the relevant position in the
                                                                      /*sensing neighborhood
```

Figure 27    Single Layer, Tree Traversal, Energy Aware algorithm – mobile agent pseudocode – part 1

```
        If |relevantCells| > 0                                    /* [11]
          possiblePos ← relevantCells with ξ^A = 0    /*the subset of relevantCells with no mobile
                                                                        /*agents
            If |possiblePos| > 0                                /*if there are cells to move-to [12]
              possiblePos ← possiblePos with ξ^A = 0
                                                                        /* find all the cells in possiblePos with no
                                                                        /* mobile agents
              v = random(|possiblePos|)                /*select a cell from possiblePos [13]
              u ← v;    s_1 ← "mobile";    s_2 ← 𝒜(v)
                                                                        /*move to selected cell and set 𝒜 as the
                                                                        /*direction to move to v
            Else
              u ← u;    s_1 ← "mobile";    s_2 ← s_2    /*stay in place [14]
            End If

        Else
              u ← u;    s_1 ← "mobile";    s_2 ← s_2    /*stay in place [15]
        End If

        End If
    End If
Else                                                                    /* if the agent is not mobile
    Do SLTT-EA settled agent logic as defined in Figure 29 or Figure 30
End
```

Figure 28       Single Layer, Tree Traversal, Energy Aware – mobile agent pseudocode – part 2

```
If s_1 ≠ "mobile"                              /* if the agent is not mobile
    If E < E_{critical}^{settled}              /* If there current energy < E_{critical}^{settled} [1]
        s_1 ← "Low Energy"                     /* state is "Low Energy" [2]
        Project (s_1)                          /* project state and Arrow
        Project (s_2)
    Elseif ∃ξ_j^G.s_1 = Low Energy, j = 1:4    /* There are neighboring agents in state "Low
                                               /* Energy" [3]
        s_1 ← "Low Energy"                     /* state is "Low Energy" [4]
        Project (s_1)                          /* project state and Arrow
        Project (s_2)
    Else ∄ξ_j^G = 0, j = 1:4                   /* If there are no empty cells [5]
        ξ̌^G ← ξ^G with ξ_j^G.s_2 = j          /* ξ̌^G - the set of neighboring settled agents with
                                               /* the direction they are signaling is eq. to the
                                               /* relevant position in the sensing neighborhood
        ξ̂^G ← ξ^G with ξ^G.s_1 = "Closed Beacon"  /* ξ̂^G - the set of neighboring Closed Beacon
        If ξ̌^G = ξ̂^G                          /* [5]
            s_1 ← "Closed Beacon"              /* state is "Closed Beacon" [6]
            Project (s_1)                      /* project state and Arrow
            Project (s_2)
        Else
            s_1 ← "Beacon"                     /* state is "Beacon" [7]
            Project (s_2)
        End If
    End If
End If
```

Figure 29     Single Layer, Tree Traversal, Energy Aware – Approach 1 - settled agent pseudocode

```
If s_1 ≠ "mobile"                                          /* if the agent is not mobile
  If E < E_critical^settled                                /* If there current energy < E_critical^settled [1]
    s_1 ← "Low Energy"                                     /* state is "Low Energy" [2]
    Project (s_1)                                          /* project state and step count
    Project (s_2)
  Else
    ξ̌^G ← ξ^G with ξ_j^G.s_2 = j                          /* ξ̌^G - the set of neighboring settled agents with
                                                           /* the direction they are signaling is eq. to the
                                                           /* relevant position in the sensing neighborhood
    ξ̂^G ← ξ^G with ξ^G.s_1 = Low Energy or Closed Beacon
                                                           /* ξ̂^G - the set of neighboring Low Energy or
                                                           /* Closed Beacon
    If ξ̌^G = ξ̂^G and |ξ̂^G| > 0 and ∄ξ_j^G = 0, j = 1:4
                                                           /* [3]
      s_1 ← "Low Energy"                                   /* state is "Low Energy" [4]
      Project (s_1)                                        /* project state and step count
      Project (s_2)
    Else
      ξ̂^G ← ξ^G with ξ^G.s_1 = "Closed Beacon"            /* ξ̂^G - the set of neighboring Closed Beacon
      If ξ̌^G = ξ̂^G                                        /* [5]
        s_1 ← "Closed Beacon"                              /* state is "Closed Beacon" [6]
        Project (s_1)                                      /* project state and Arrow
        Project (s_2)
      Else
        s_1 ← "Beacon"                                     /* state is "Beacon" [7]
        Project (s_2)
      End If
    End If
  End If
End If
```

Figure 30    Single Layer, Tree Traversal, Energy Aware – Approach 2 - settled agent pseudocode

### D. Algorithm Analysis & Performance

In this section we derive expressions for the upper bounds on the Covered Area when either of the approaches with the SLEAC algorithms. The agents are characterized by $E_0, E_{critical}^{mobile}, E_{critical}^{settled}, \alpha$ and $\Delta T$. The maximum time such agents can be mobile is given by $E_0 - E_{critical}^{mobile}$ and they can move to a maximal distance of $d_{max}$ cells from an entry point before settling given by Eq. (7). Consequently, the maximal area that can be covered by these agents is a skewed square – see Figure 31. This region can also be viewed as composed of concentric squares, one atop of the other (like pyramid) with the enveloping cells in each square having equal remaining energy. Using this interpretation all the outermost enveloping cells have an energy of $E_0 - E_{critical}^{mobile} - 1$ and successively inner envelopes have increasing values of (remaining) energy.

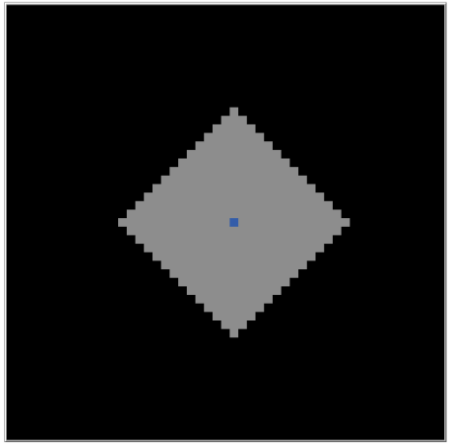

Figure 31   Maximal covered region of energy constrained agents using SLC algorithms. The covered region is colored gray, the empty cells in black and the entry point is in the middle in blue.

In Approach 1 the first agent to transition to Low Energy initiates the backward propagation that will halt the process once it reaches the entry point. Thus, the upper bound on the termination time is the sum of the time at which the first mobile agent settles at a distance of $d$ and the propagation time from that cell to the entry point.

From Figure 31 it is clear that the number of settled agents with a remaining energy equal to or greater than $E_0 - E_{critical}^{mobile} - 2$ is simply the sum of two arithmetic series and given by

$$N_{d-1} = (d_{max} - 2)^2 + (d_{max} - 1)^2 \qquad (8)$$

Hence the time the first mobile agent settles at a distance of $d_{max}$ is upper bounded by

$$(N_{d-1} + 1)\Delta T + d_{max} \qquad (9)$$

with the first term denoting the time until the agent $a_{N_{d-1}+1}$ enters the region and the second term the time until the agent $a_{N_{d-1}+1}$ settles at distance $d_{max}$. Based on [1] this expression is exact when $\Delta T \geq 2$ and SLLG-EC or SLTT-EC are used. Furthermore, it is an upper bound on the time when $\Delta T = 1$ and/or SLUG-EC is used. The propagation time of the information is upper bounded by the expression below when the adversarial scheduler is used:

$$T_{propogation}^{Upper\ bound} = d_{max} \qquad (10)$$

Hence the upper bound on the termination time is given by:

$$T_C^{Upper\ bound} = (N_{d-1} + 1)\Delta T + 2d_{max} \tag{11}$$

From which we can derive the upper bound on both the number of agents that will participate in the process and covered area simply by dividing the termination time by $\Delta T$ and get:

$$N^{Upper\ bound} = A_C^{Upper\ bound} = (N_{d-1} + 1) + \frac{2d_{max}}{\Delta T} \tag{12}$$

In approach 2 the objective is to explore the largest possible area and therefore continue the exploration and coverage process for as long as possible. The resulting upper bound on the covered area is when all the cells at a distance $d$ are occupied by settled agents. Calculating the area of the square is simply the sum of two arithmetic series and given by:

$$A_{covered}^{Upper\ bound} = d_{max}^2 + (d_{max} - 1)^2 \tag{13}$$

Deriving upper bounds for the termination time and the number of agents when using Approach 2 is not possible. The reason - the limited sensing range of the agents may result in an entering agent selecting a path that leads to a part of the region in which the Low Energy indication started propagating backwards but hasn't reached yet the entry point. This partial propagation is a direct result of the objective to maximize the covered area and why this cannot occur when using Approach 1. Once a mobile agent is surrounded by settled agents in Low Energy state it will never settle when SLLG-EC or SLTT-EC are used since the local action rules in these algorithms prohibit moving from Low Energy to Beacon. Thus, these agents will simply run out of energy. When using SLUG-EC such a move is possible but transition to settled agent depends on the remaining energy.

The above expressions were derived for $\alpha = 0$ meaning that an agent's energy does not decrease after settling. When $\alpha > 0$ some of the settled agents may run out of energy before termination and the gradient will become discontinuous. Whether or not this occurs depends on $E_{critical}^{settled}$. As long as the inequality below is true no settled agents will run out of energy prior to termination.

$$\frac{E_{critical}^{settled}}{\alpha} > T_C^{Upper\ bound} \tag{14}$$

The left side of the inequality describes the time a settled agent may operate before running out of energy whereas the right side is the upper bound on the termination time using energy constrained drones. While the left side depends solely on a user defined value $\left(E_{critical}^{settled}\right)$ and a characteristic of the drone ($\alpha$), the right side is also a function of the algorithm used and $\Delta T$. When using either approach an upper bound on the termination time can be derived above which agents may run out of energy. Additionally, when using Approach 1, an expression may be derived for the largest possible covered region for which no settled agents will fail.

## V. Results and Discussion

In this section the performance of the algorithms is experimentally evaluated using the metrics defined above. In the first sub-section, qualitative effects of both the termination approaches and the power factor are discussed using the complex region depicted in Figure 32a. In the following sub-section, the effect of the termination approach on the performance of the SLEAC algorithms is investigated for the case of $\alpha = 0$ whereas in the last sub-section, the effect of the power factor is discussed. All the quantitative investigations are done using various values of $E_0$ and $\Delta T$. Additionally, the region, shown in Figure 32b, is a square with the entry point located in the exact middle. This initial configuration negates the effects of region topology (such as rooms and corridors) on the results. Both $E_{critical}^{mobile}$ and $E_{critical}^{settled}$ are set to one. The results are derived from a simulator developed and implemented by the authors using the NetLogo programming language. Each data point in the following graphs is the average of fifty runs unless specifically noted.

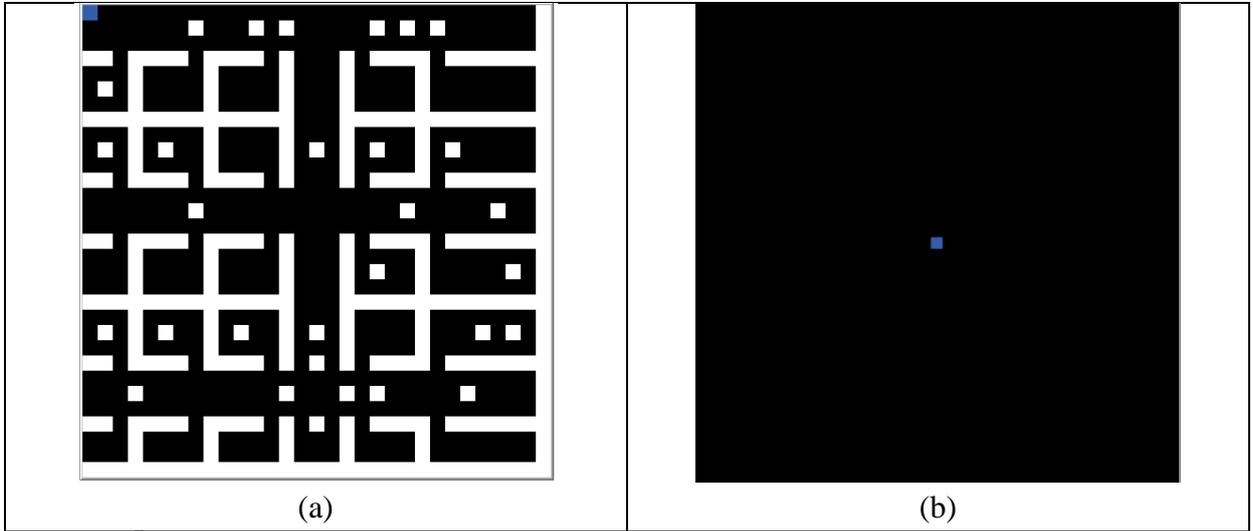

Figure 32  Regions used in the simulations. (a) A complex region (see [33]) with 636 cells; (b) a square with 2601 cells. The walls are marked in white, the empty cells in black and the entry point is marked in blue.

### A. Qualitative comparison

The qualitative effects of both the termination approaches and the power factor are shown in Figure 33. Each of the sub-figures shows the resulting coverage at the termination time for a different combination of power factor and termination approach. In addition, the number of agents and the covered area as a function of time are also shown along with the covered area, number of depleted agents (i.e., the number of agents that ran out of energy) and the termination time. Settled agents in "Low Energy" state are colored gray, in Beacon in yellow and in Closed Beacon in brown. The SLLG-EA algorithm is used with $E_0$=50, $\Delta T = 2$ and $\alpha = 0.025$ (when the power factor is not zero).

When the power factor is zero, the increased covered area when using Approach 2 compared to Approach 1 is readily apparent by comparing Figure 33(a) and (b) as are the accompanying increases in the number of depleted agents and termination time. The effect of the power factor is clearly evident when comparing Figure 33(b) and (d). The area covered is significantly smaller

(a) Approach 1 & $\alpha = 0$

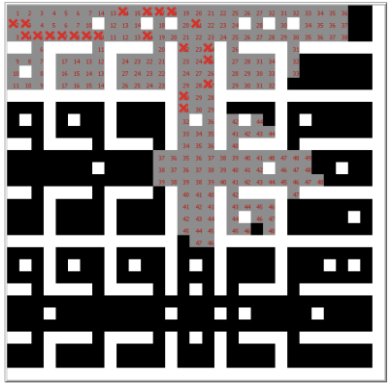

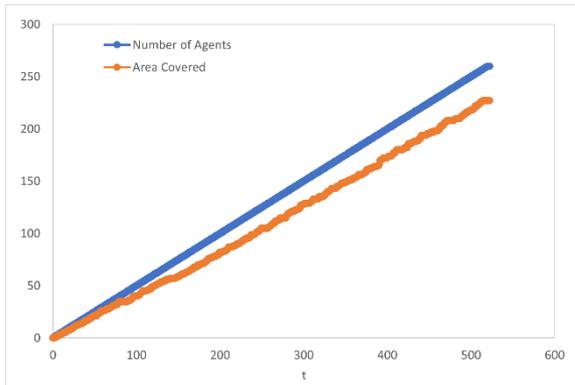

Tc=522, N=260, Ac= 227, depleted agents = 12

(b) Approach 2 & $\alpha = 0$

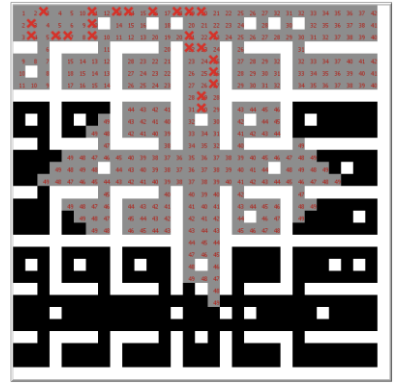

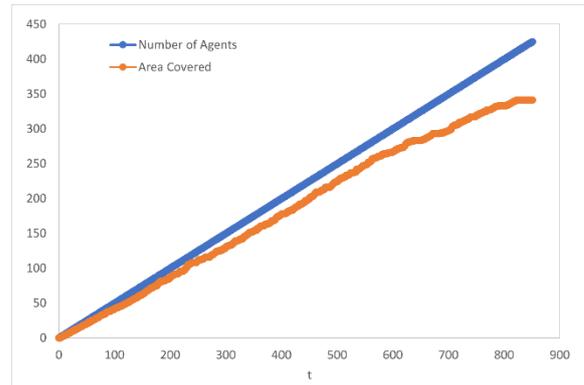

Tc=850, N= 425, Ac=341, depleted agents = 60

(c) Approach 1 & $\alpha > 0$

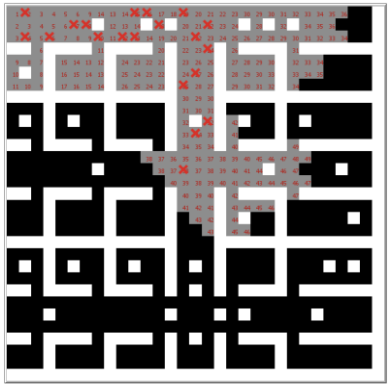

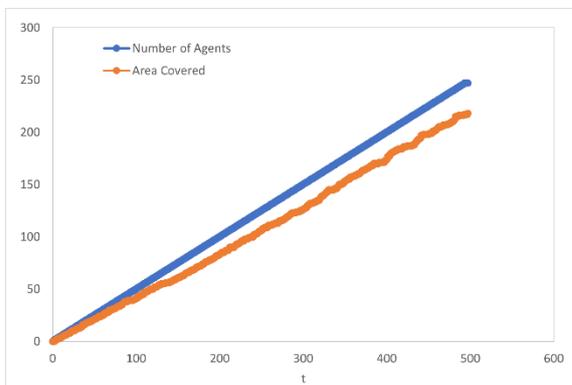

Tc= 497, N=247, Ac= 218, depleted agents = 9

(d) Approach 2 & $\alpha > 0$

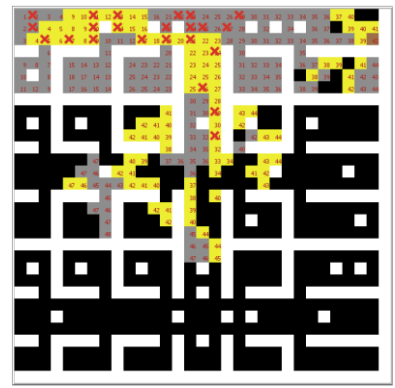

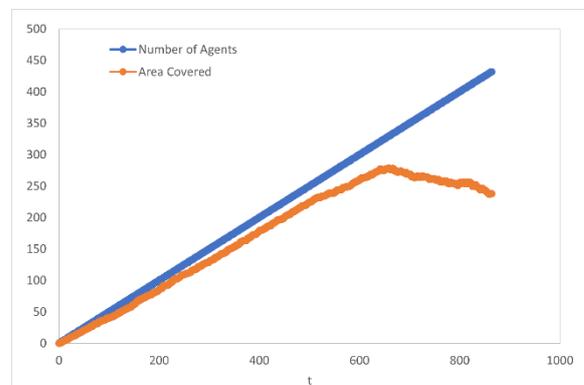

Tc=864, N= 432, Ac= 237, depleted agents = 171

Figure 33   Qualitative comparison of the termination approaches and power factor

and the number of depleted agents considerably higher when $\alpha$ is greater than zero. This is also portrayed in the area covered plot reaching a (momentary) maximum and then monotonously decreasing until termination time. Furthermore, the underlying graph is no longer connected because settled agents are running out of energy. In comparison, the effect of $\alpha > 0$ when Approach 1 is used is much smaller due to the rapid termination which means settled agents do not run out energy during the process.

### B. Algorithm performance when $\alpha = 0$

*1. Approach 1*

Figure 34 to Figure 36 show the area covered, termination time and total energy as a function of $\Delta T$ for $E_0 = 8, 15,$ and $23$ using SLUG-EA, SLLG-EA and SLTT-EA. Each algorithm is drawn in the figures with a different color and the three values of $E_0$ are denoted by solid, dashed, and dotted lines respectively. Several observations are readily apparent:

1) SLTT-EA out-performs both SLUG-EA and SLLG-EA. Moreover, the difference increases with $E_0$. This is due to the different conditions (that must be fulfilled in order to transition to Closed Beacon) used in SLTT-EA on the one hand compared to SLLG-EA and SLUG-EA on the other. The transition to Closed Beacon in SLUG-EA depends on at least as many neighboring cells as in SLLG-EA which in turn depends on at least as many cells as SLTT-EA. The result is that parts of the covered region transition to Closed Beacon (and by doing so funnel the mobile agents) earlier in SLTT-EA compared to SLLG-EA and SLUG-EA. As $E_0$ increases so does the covered area and the effect of the Closed Beacons on the movement of the mobile agents. This is clearly seen in Figure 37 that shows the covered area immediately after the first agent transitions to Low Energy. Using brown to denote the Closed Beacon cells, gray for the Low Energy cells and yellow for the Beacon cells the difference between the three algorithms is clear.
2) The performance of SLUG-EA and SLLG-EA is very similar regardless of $E_0$ with the sole exception of the total energy when $E_0 = 23$ and $\Delta T = 1$. The richer DAG generated by SLUG enables more efficient movement compared to SLLG.
3) All algorithms are indeed bounded by the theoretical upper bounds.
4) The covered area is not a function of $\Delta T$. This is because when $\alpha = 0$ the settled agents don't consume any energy and $\Delta T$ in itself does not affect the trajectories the individual agents take.
5) The termination time is linear with $\Delta T$ due to the linear dependency of the termination time on $\Delta T$ discussed in [1].
6) There is a weak dependency of the total energy on $\Delta T$. That is because the total energy is principally affected by the covered area which as discussed above is a function of $E_0$ rather than of $\Delta T$. Specifically, the total energy in SLTT-EA is higher since the covered area is greater. The effect of $\Delta T$ on the total energy is via the improved effectiveness of the agent funneling.

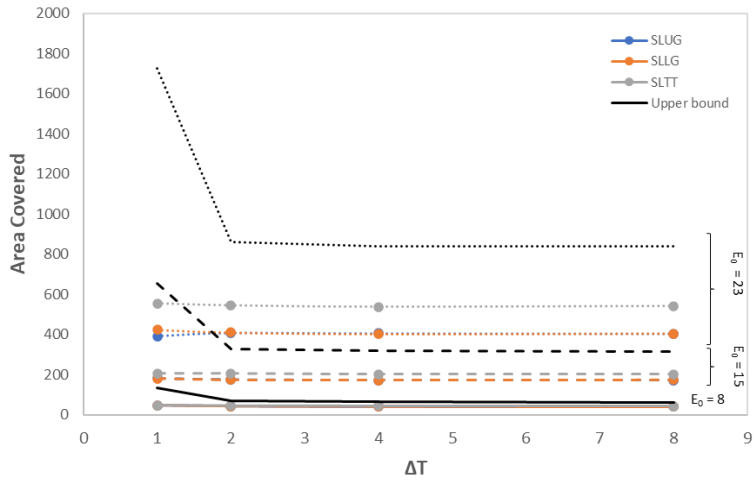

Figure 34     Area Covered as a function of the algorithm used and $E_0$ using Approach 1.
The theoretical upper bound is marked by a black line and the results for the different algorithms are colored as in the legend. The solid lines are for $E_0 = 8$, the dashed lines for $E_0 = 15$ and the dotted lines for $E_0 = 23$.

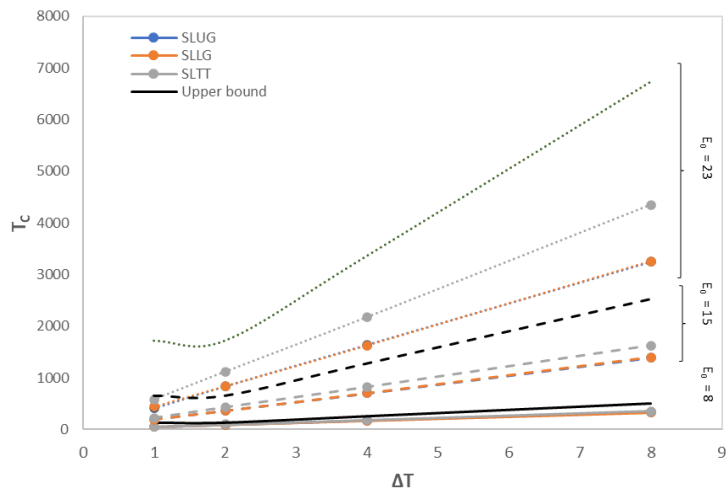

Figure 35     Termination time as a function of the algorithm used and $E_0$.
The theoretical upper bound is marked by a black line and the results for the different algorithms are colored as in the legend. The solid lines are for $E_0 = 8$, the dashed lines for $E_0 = 15$ and the dotted lines for $E_0 = 23$.

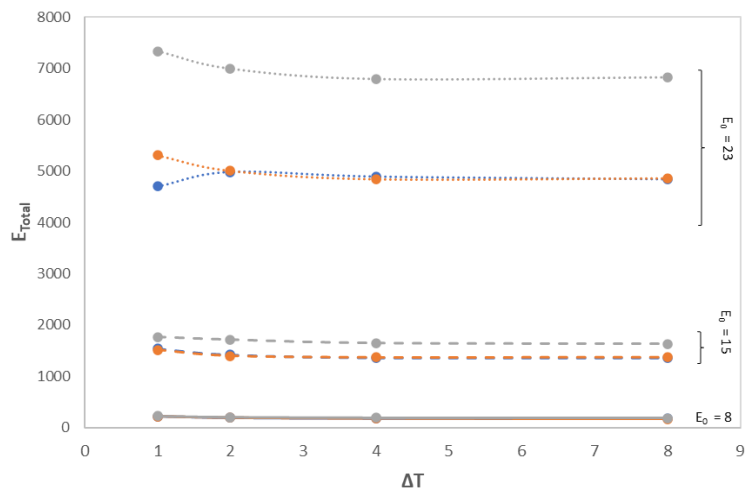

Figure 36     Total energy as a function of the algorithm used and $E_0$.
The solid lines are for $E_0 = 8$, the dashed lines for $E_0 = 15$ and the dotted lines for $E_0 = 23$.

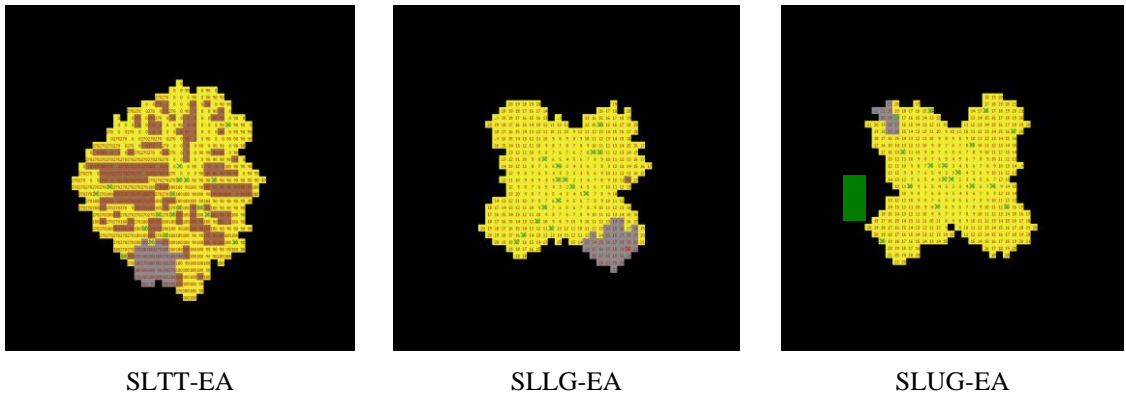

| SLTT-EA | SLLG-EA | SLUG-EA |

Figure 37 SLEAC algorithms – qualitative comparison of the transition to Closed Beacon using Approach 1

## 2. Approach 2

When Approach 2 is used to terminate the process ever greater parts of the covered region change to Low Energy until at the termination time the settled agent at the entry point changes to Low Energy. The limited energy capacity means that in many cases superfluous mobile agents run out of energy before the process terminates and thus their use as mobile explorers, done in SLUG-EC, is negligible at best. For the same reason, the benefit of guiding the superfluous agents to the entry point (shown in [1]) is negligible.

The effect of $E_0$ on the covered area and the number of depleted agents is shown in Figure 38. The lines corresponding to each algorithm are colored as shown in the legend with solid lines used to show the area covered and dashed lined the depleted agents. Several observations are obvious:

1) The covered area does not depend on $\Delta T$ unlike the number of depleted agents. Since $\alpha = 0$ the energy consumption of the settled agents does not affect the covered area and more importantly there is no cost (energy-wise) to increasing $\Delta T$. Moreover, increasing $\Delta T$ increases the effectiveness of the funneling of the mobile agents (due to the Backward Propagating Closure concept) thereby reducing the number of agents that run out energy. The transition of settled agents to Low Energy due to Approach 2 proceeds independently of the coverage process and the indication propagates faster than the rate additional agents enter the region. These additional agents then follow a more energy-efficient trajectory thereby reducing the number of agents that run out of energy. A similar effect of $\Delta T$ on the total energy was observed in [1].

2) The covered area using SLTT-EA is the largest regardless of $E_0$ and for the same reason described in the previous section.

3) SLTT-EA did not terminate in several cases when $\Delta T = 1$ and thus this data point is missing. That is because, regardless of the algorithm used, Approach 2 is problematic with $\Delta T = 1$ since mutual interference between mobile agents may result in an agent staying in place during one or more time step. This energy waste directly translates to a reduced maximum range from the entry point. The problem is especially severe with SLTT-EA in which the agents advance on a spanning tree. This means agent $a_i$ can cause following agent $a_j$ (i.e., $\delta(EP,i) \geq \delta(EP,j) + 1$) to stay in place if their relative wake-up order is $t_{i,k} < t_{j,k}$. When exploring on a tree, at any time step, there is always one agent furthest from the entry point and obviously the issue above does not hamper its advance. However, at some time step t, the agent at the head of the exploration on a branch will settle at a distance of $d_{max} - 1$ with an energy level of 2 (and not transition to Low Energy). Following agents that are blocked for at least one time step will reach a distance of $d_{max} - 1$ with an energy of 1 and consequently settle in place and shut down. The result is a branch that never transitions to Low Energy and prevents termination of the exploration process. Simulation runs show the probability of such a scenario is not negligible.

Figure 39 shows the effect of both the algorithm used and $\Delta T$ on the total energy. This decrease in the number of depleted agents (shown in Figure 38) is the reason for the drop in the energy consumption as $\Delta T$ increases.

The effect of $E_0$ on the termination time is shown in Figure 40. The termination time is linear with respect to $\Delta T$ as discussed at length in the previous chapters. SLTT-EA shows the longest termination time due to the larger covered area and this gap from SLUG-EA and SLTT-EA increases with $\Delta T$.

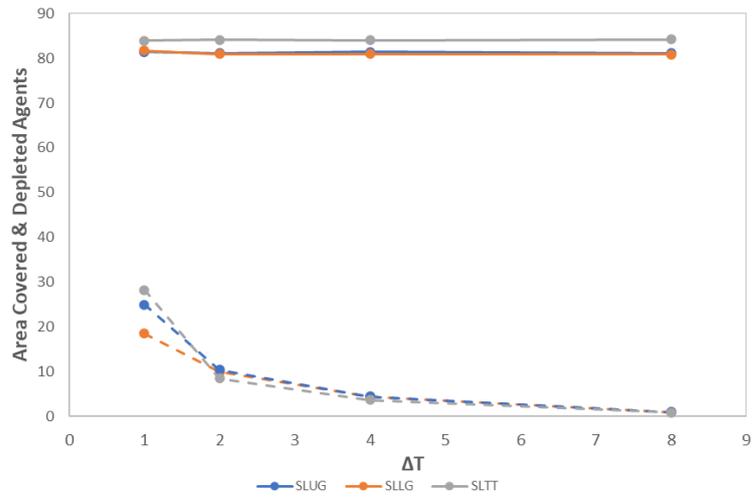

(a)

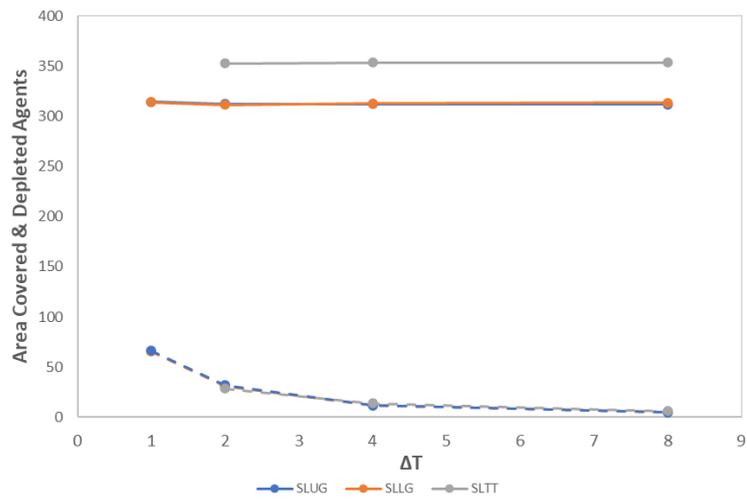

(b)

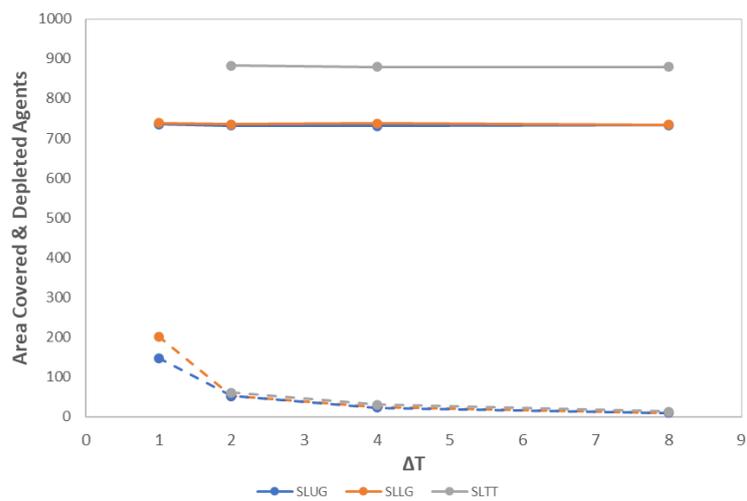

(c)

Figure 38     Area Covered and number of depleted agents as a function of the algorithm and $E_0$
(a) $E_0 = 8$; (b) $E_0 = 15$; (c) $E_0 = 23$
The solid lines denote the area covered and the dashed lines the depleted agents

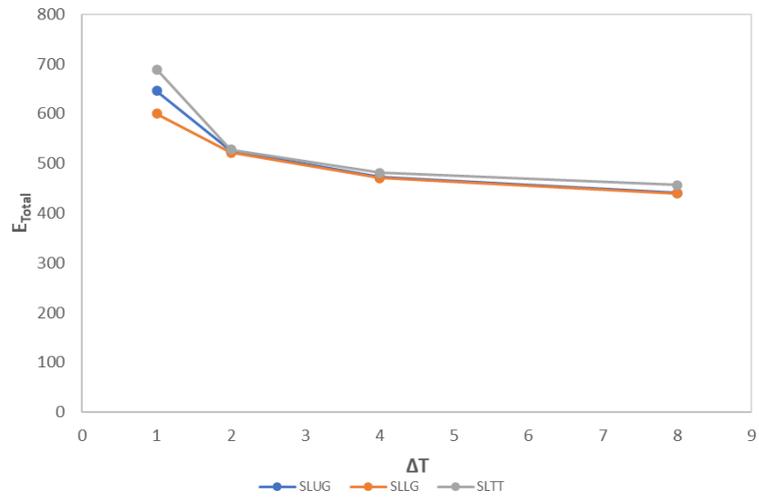

(a)

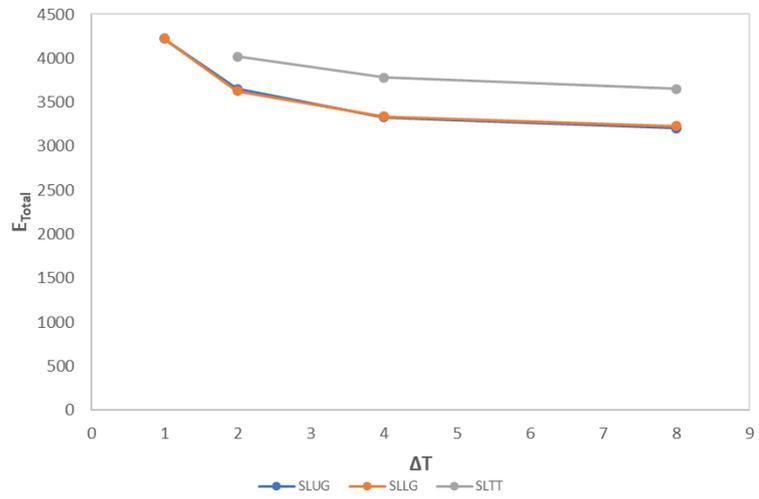

(b)

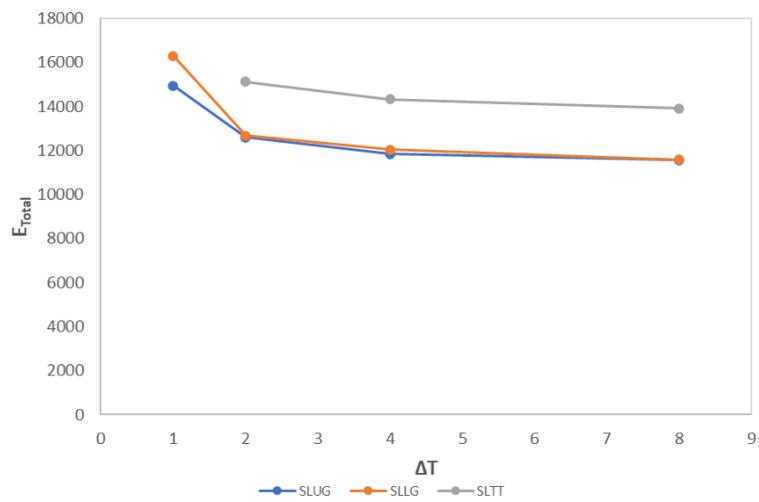

(c)

Figure 39   Total energy as a function of the algorithm and $E_0$
$E_0 = 8$; (b) $E_0 = 15$; (c) $E_0 = 23$

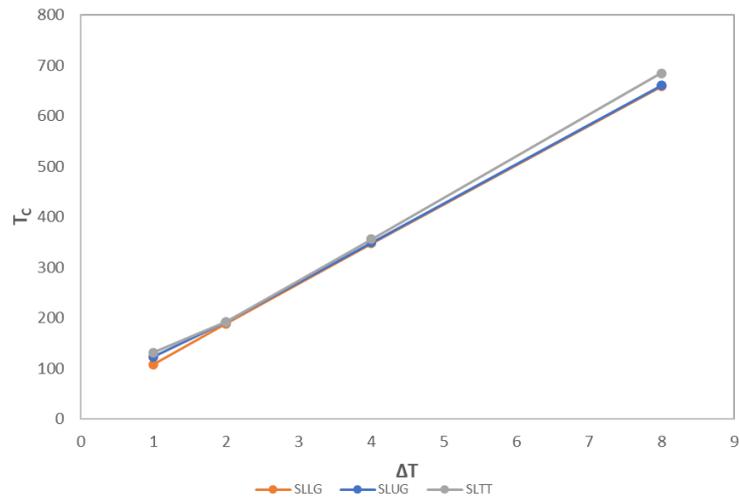

(a)

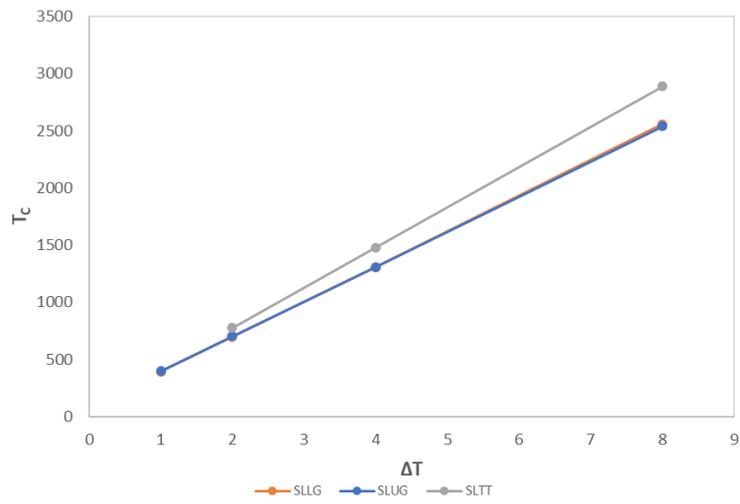

(b)

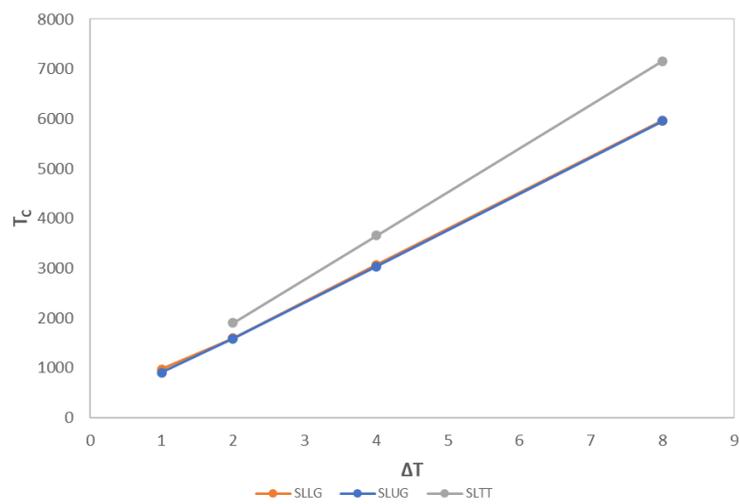

(c)

Figure 40  Termination time as a function of the algorithm and $E_0$
$E_0 = 8$; (b) $E_0 = 15$; (c) $E_0 = 23$

## C.   Algorithm performance when $\alpha > 0$

When the power consumption of a settled agent is greater than zero (i.e., $\alpha > 0$) the behavior of both the multi-agent system as well as that of the individual agents' changes. Regarding the agents, when $\alpha > 0$ the energy level of a settled agent continuously decreases after settling and the agent might run out of energy before process termination. Hence, the agent may transition to Low Energy at any time step after settling and not necessarily at the time step following its settling. The effect on the multi-agent system is much more complex. When a settled agent runs out of energy the result is the removal of a vertex in the underlying graph. If the underlying graph is a spanning tree the result is a spanning forest. If the underlying graph is a DAG, the outcome depends on the specific DAG. Moreover, the first settled agents to run out of energy will be those furthest from the entry point since most of their energy was spent being mobile. As the process continues, ever closer agents to the entry point will run out of energy until at some point the settled agent at the entry point will shut down. Entering agents will settle in the newly opened cells however since the agents have a sensing range of 1 and the action rules give preference to moving up the gradient rather than down the net result will be a gradual but monotonous decline in the covered area. This is exacerbated when $\Delta T$ increases since less agents enter the region in a given period of time while the energy consumed by the settled agents is constant. This phenomenon occurs when either termination approach is used however to a much greater extant with Approach 2.

In this paper we experimentally investigate the effect of $\alpha > 0$ and use $\alpha = 2.5\%$ based on data from [2] and [34]. While $\alpha = 2.5\%$ is small (40 time steps on the ground are energetically equivalent to a single time step of flying), neglecting this energy consumption can lead to significant error in the total energy and more importantly the termination time as shown below.

In Figure 41 snapshots of the coverage process of a square region using SLUG-EA with $\Delta T = 4$, termination according to Approach 1 and $\alpha = 2.5\%$ are shown. At $t = 441$ the first agent transitions to Low Energy as shown in Figure 41(c). The agent, marked in pink, is not located on the expansion frontier but rather inside the covered area. Since the indication propagated quicky neither the agent shut down nor any other agent transitioned to Low Energy in the time until termination. This example illustrates the complex interaction between the different parameters.

Comparisons of the area covered, termination time and total energy with $\alpha = 2.5\%$ and with $\alpha = 0$ and termination Approach 1 are shown in Figure 42 using the different SLEAC algorithms. The region is $41 \times 41$ square with the entry point in the middle, $E_{max} = 15, E_{critical}^{mobile} = 1$ and $E_{critical}^{settled} = 1$. The algorithms are colored differently as explained in the legend with solid lines for $\alpha = 2.5\%$ and dashed lines for $\alpha = 0$. In all three metrics, the performance is degraded when $\alpha = 2.5\%$ with the degradation becoming more significant as $\Delta T$ increases. When $\Delta T = 1$ the first agent to transition to Low Energy lies on the exploration frontier. When $\Delta T = 2$ or $4$ that first agent is located inside the covered area. And when $\Delta T = 8$ the energy of the settled agent at the entry point decreases to 1 after 480 timesteps at which time the process terminates, and entry of additional agents stops. This explains why as $\Delta T$ increases both the differences between the algorithms and the covered area become smaller.

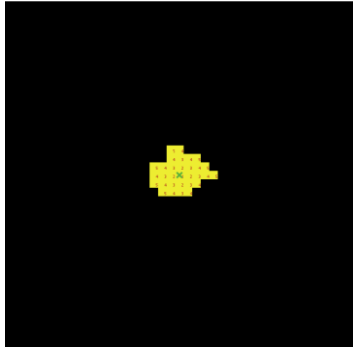

(a) t = 125

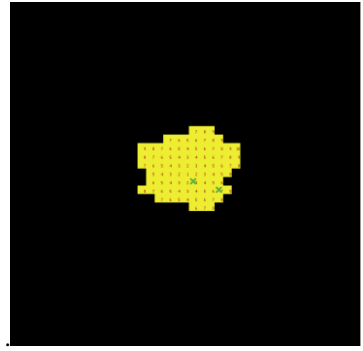

(b) t = 350

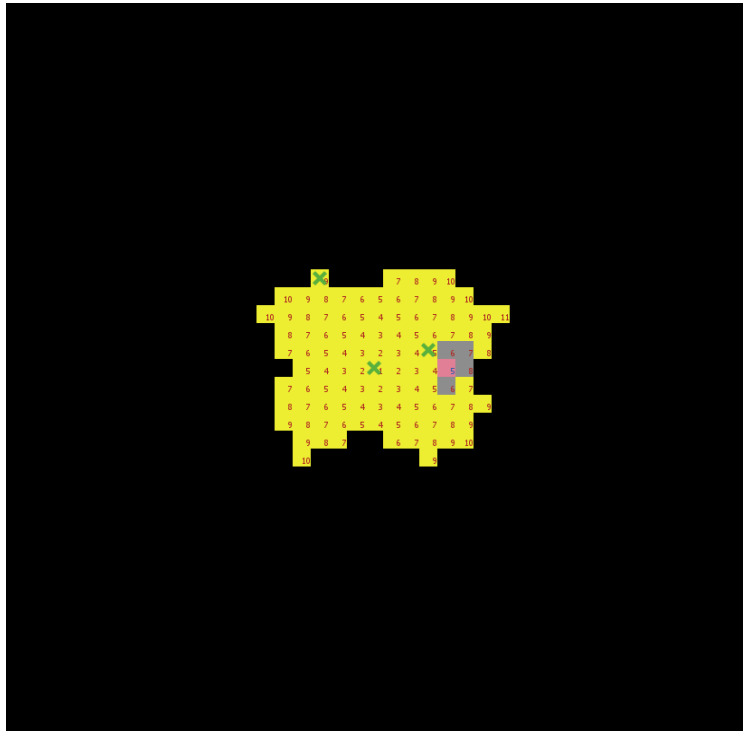

(c) t = 441

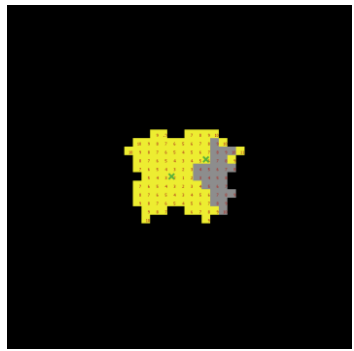

(d) t = 442

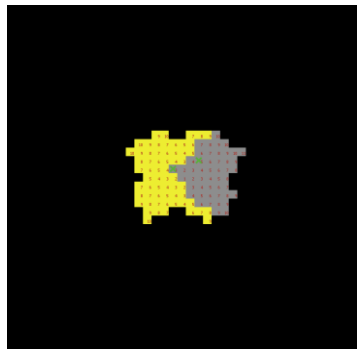

(e) t = 443

Figure 41 Single Layer, Unlimited Gradient – Energy Aware – Coverage process of a square region using $\alpha = 2.5\%$, Approach 1 and $\Delta T = 4$. The entry point is in the middle, $E_0 = 15$, $E_{critical}^{mobile} = 1$ and $E_{critical}^{settled} = 1$.

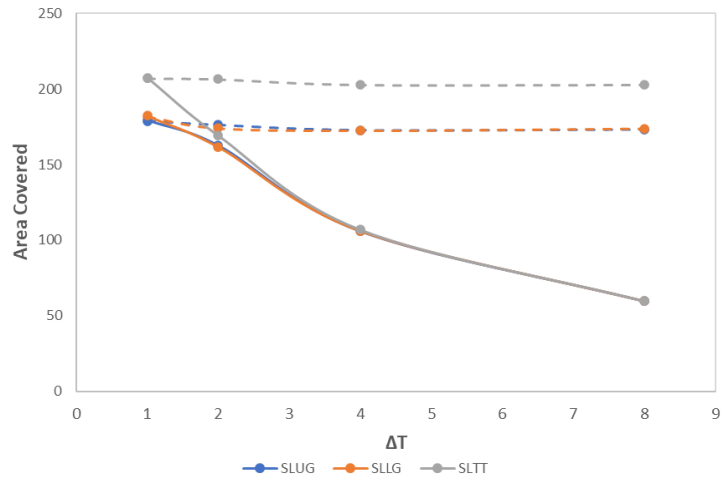

(a)

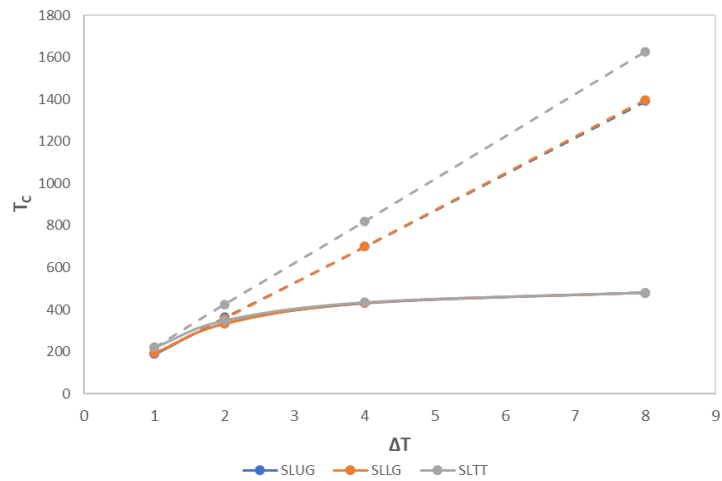

(b)

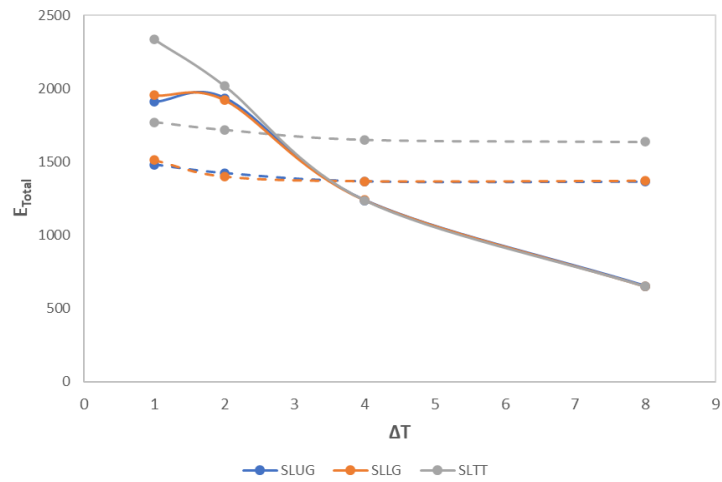

(c)

Figure 42 Effect of the power factor on the performance with $\alpha = 0$ and $\alpha = 2.5\%$ using Approach 1.
(a) Area covered; (b) Termination time; (c) Total energy
The solid lines are for $\alpha = 2.5\%$ the dashed lines for $\alpha = 0$
Each algorithm is marked by a different color per legend

In Figure 43 snapshots of the coverage process of a square region using SLUG-EA with $\Delta T = 4$, $\alpha = 2.5\%$ and termination approach 2 are shown. At $t = 433$ the first agent transitions to Low Energy as shown in Figure 43(a) and as the process progresses additional agents transition to Low Energy until at t=481 (see Figure 43(f)) the settled agent at the entry point transitions to Low Energy and the process terminates. By that time several agents depleted their energy completely and shut down resulting in the cells becoming empty again. Neither of the mobile agents in the region at that time can advance to the empty cells due to the existing gradient and consequently the two will either expand the region or settle and shut down. This example further illustrates the complexity of the exploration and uniform coverage process with energy limited agents.

In the figures below a comparison of the area covered, termination time and total energy with $\alpha = 2.5\%$ and with $\alpha = 0$ are shown using the different SLEAC algorithms in conjunction with termination Approach 2 for various values of $\Delta T$. The region is $41 \times 41$ square with the entry point in the middle, $E_0 = 15$, $E_{critical}^{mobile} = 1$ and $E_{critical}^{settled} = 1$. In all three metrics, the performance is degraded when $\alpha = 2.5\%$ with the degradation becoming more severe as $\Delta T$ increases. When $\Delta T = 1$ the first agent to transition to Low Energy lies on the exploration frontier. When $\Delta T = 2$ or 4 that first agent is located inside the covered area. When $\Delta T = 8$ the energy of the settled agent at the entry point decreases to 1 after 480 timesteps at which time the process terminates, and entry of additional agents stops. This explains why as $\Delta T$ increases the differences between the algorithms become smaller.

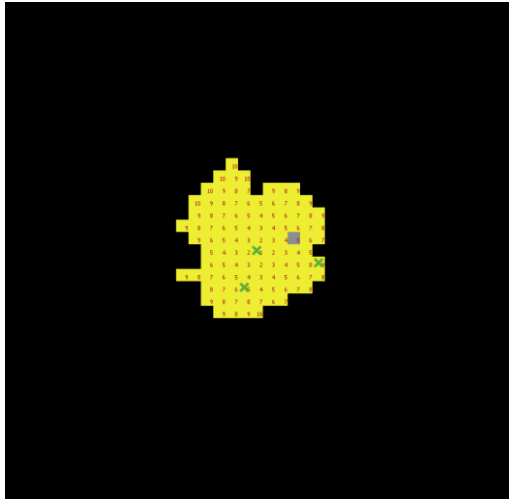
(a) t = 433

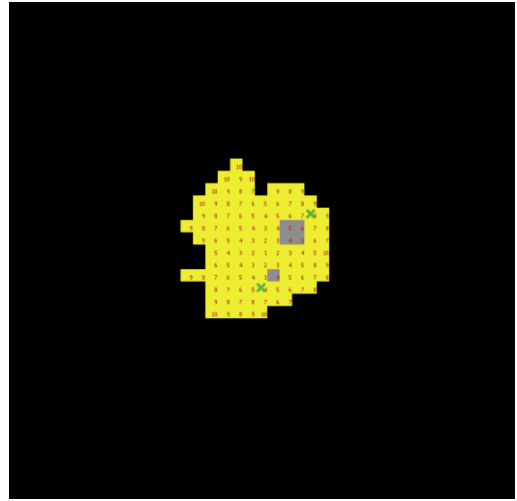
(b) t = 440

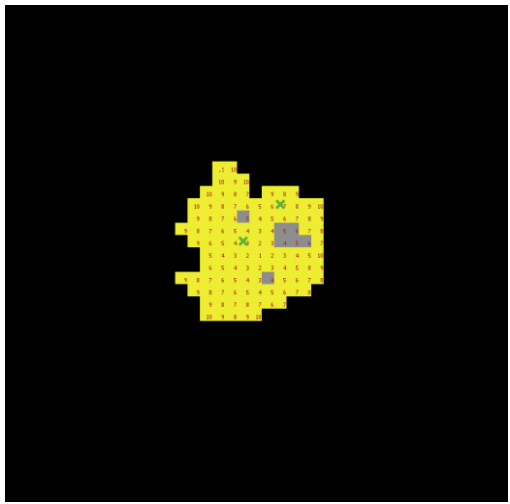
(c) t = 450

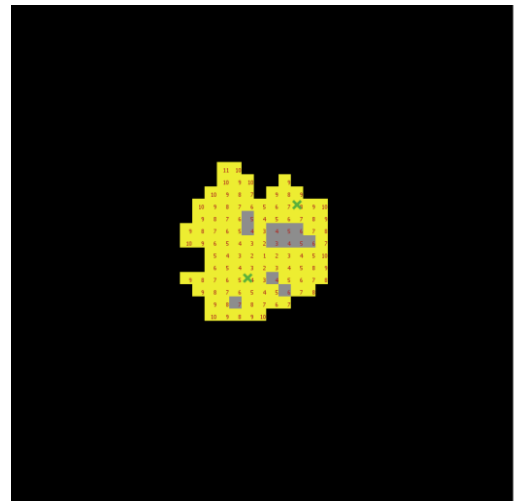
(d) t = 460

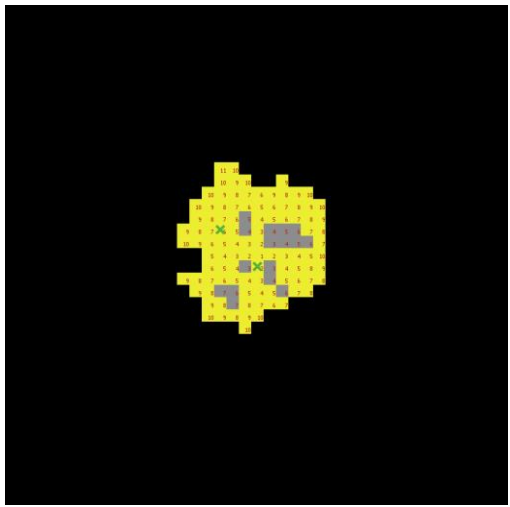
(e) t = 470

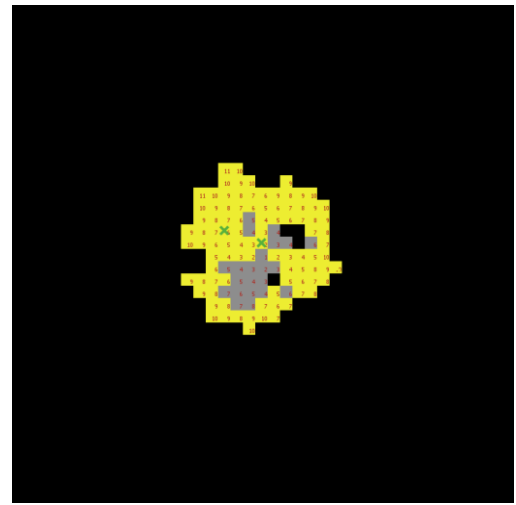
(f) t = 481

Figure 43 Single Layer, Unlimited Gradient – Energy Aware – Coverage process of a square region using Approach 2. The entry point is in the middle, $E_0 = 15$ and $E_{critical}^{mobile} = E_{critical}^{settled} = 1$.
The snapshots of the process are ordered left to right, top to bottom

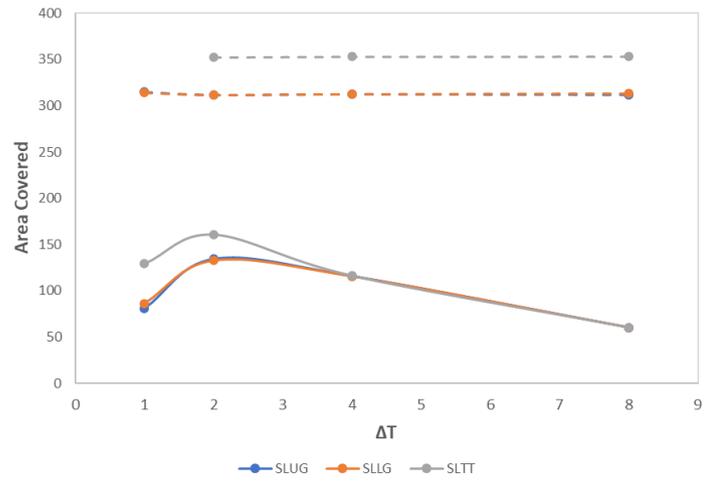

(a)

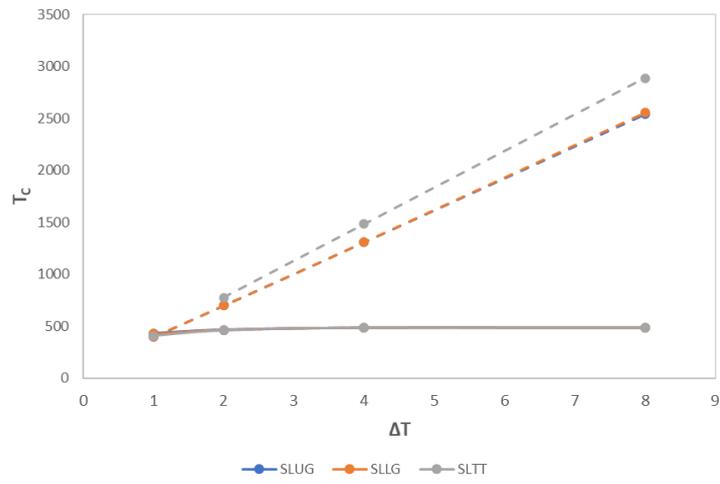

(b)

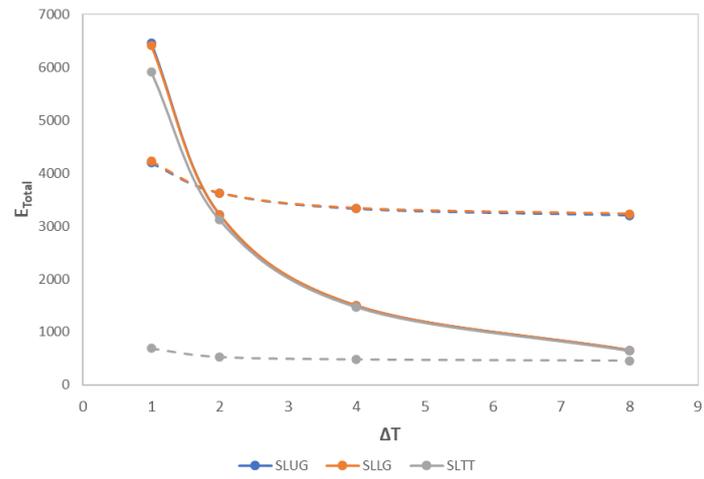

(c)

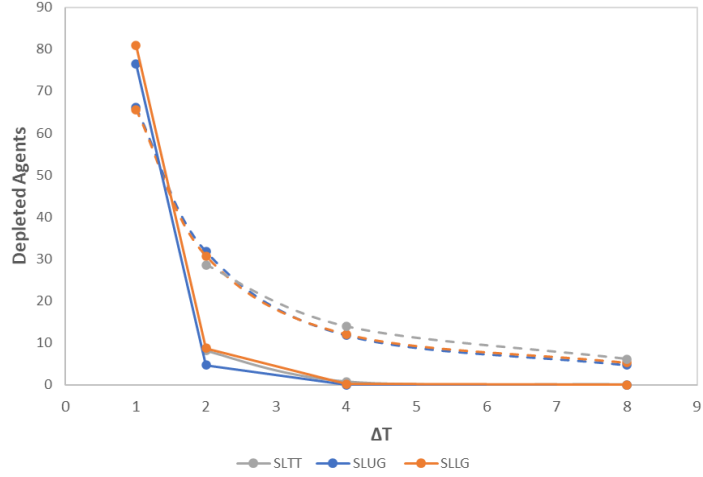

(d)

Figure 44 Effect of the power factor on the performance with $\alpha = 0$ and $\alpha = 2.5\%$ using Approach 2.
(a) Area covered; (b) Termination time; (c) Total energy; (d) Depleted agents
The solid lines are for $\alpha = 2.5\%$ the dashed lines for $\alpha = 0$
Each algorithm is marked by a different color per legend

Each of the two approaches is a tradeoff. Approach 1 can be viewed as the more robust approach in that it gives preference to assuring termination regardless of the resulting area coverage and for any the power factor. Consequently, as long as $E_{critical}^{settled}/\alpha > d_{max}$ there is no possibility settled agents will run out of energy during the process. Approach 2 is aimed at maximizing the covered area hence its termination criteria is more complicated. When α=0, using Approach 2 indeed results in a bigger covered area. The unlimited time settled agents can operate means the sole cost is an increase in the number of agents that run out of energy. However, when α>0, using Approach 2 may result in a smaller covered area since the settled agents start running out of energy and additional agents that come in cannot replace them at a sufficient rate as explained above[§]. Thus, in these cases, the covered areas using both approaches are similar however when using Approach 1 the process terminates faster resulting in far fewer depleted agents.

## VI. Energy Consumption Analysis

In this section upper bounds on both the energy consumption of a single agent and of the multi-agent system are derived for a linear region and compared to experimental results. Moreover, optimal values for $\Delta T$ that minimize the upper bounds are derived and their effect experimentally investigated. The first part of this section analyzes the energy consumption in a linear region with the entry point at one of the edges while in the second part the entry point is at neither of the edges. While the agents have limited energy, in this section it is assumed that the constraint is sufficient to achieve coverage. When $\Delta T \geq 2$, the sum of the time periods each agent $a_i$ spends in the "agent source", as a mobile agent and as a settled agent is exactly the termination time as described by the following relation:

---

[§] The macro-behavior in this case is reminiscent of the Cooperative Cleaners problem described [35].

$$T_c = t_0^i + t_m^i + t_s^i \tag{15}$$

Since the first agent enters the region at $t = 0$ and the i-th agent waits $(i - 1)\Delta T$ time steps before entering the region, Eq. (15) may be re-written as:

$$T_c = (i - 1)\Delta T + t_m^i + t_s^i \tag{16}$$

The energy consumption of agent $a_i$ (Eq. (5)) can thus be re-written as

$$E_i = t_m^i(1 - \alpha) + \alpha T_c - \alpha(i - 1)\Delta T \tag{17}$$

The total energy consumption (see Definition 7) is given by

$$E_{total} = \sum_{i=1}^{N} E_i \tag{18}$$

Substitution of Eq. (5) to Eq. (18) yields

$$E_{total} = \sum_{i=1}^{N} \left(t_m^i + \alpha t_s^i\right) \tag{19}$$

Or alternately by:

$$E_{total} = \sum_{i=1}^{N} \left(t_m^i(1 - \alpha) + \alpha T_c - \alpha(i - 1)\Delta T\right) \tag{20}$$

Since $T_c, N$ and $t_m^i, \forall i$ in Eq. (20) are functions of $\Delta T$ and there are no analytic expressions for them in a general region **R**, derivation of Eq. (20) with respect to $\Delta T$ will not yield a closed form expression for $\Delta T$. The sole exception is the linear graph in which it is possible to derive analytic expression for the time each agent is mobile and bounds, both upper and lower, on the termination time and number of agents used. These can then be substituted into Eq. (20). We start with the simplest case - the entry point located at either end of the region[**].

---

[**] The cells at either end are also known as the "terminal cells"

## A. Linear region with the entry point at either edge

In the SLEAC algorithms the termination time is the sum of the time until the last cell is filled and the time it takes the BPC signal to propagate as given by:

$$T_C(\mathbb{G}) = T_{settled} + T_{BPC} \tag{21}$$

with $T_{BPC}(\mathbb{G})$ defined as the time span from the time step an agent settles in the last empty cell in $\mathbb{G}(V, E)$ until the settled agent at the entry point transitions to "Closed Beacon". When deriving the upper bounds on $T_C$ and $N$ for a SLEAC algorithm with $\Delta T \geq 2$ an adversarial scheduler is used is that results in the longest $T_{BPC}$. Specifically, the adversarial scheduler defines the relative wake up order of any two neighboring agents in a time step. This is particularly useful in a linear graph in which movement of agents closer to the entry point is only affected by agents further away.

**Definition 12**: The distance of agent $a_i$ from the entry point is given by $\delta(EP, j)$.

**Definition 13**: The scheduler that will result in the largest termination time on a linear graph is defined as the wake-up order of the agents, in time step $t = k$, that begins with agents at the entry point and ends with the agents at the expansion frontier. We call this the *adversarial scheduler*. When zooming in on two adjacent, mobile agents, this means that at time step $t = k$ the agent closer to the entry point will wake up first.

$$t_{i,k} < t_{j,k} \quad \forall i, j \text{ such that } \delta(EP, j) \geq \delta(EP, i) + 1 \tag{22}$$

Using the above scheduler it is proven in [1] that $T_{BPC}(\mathbb{G}) \leq n$. in When $\Delta T \geq 2$ the entry process is non-blocking regardless of the scheduler. Hence, the $n - th$ agent will enter the region at $t = (n - 1)\Delta T$, begin moving up the gradient at $t = (n - 1)\Delta T + 1$ after which it will take the $n - th$ agent an additional $n - 1$ time steps to reach the last cell in the graph.

$$T_{covered} = (n - 1)\Delta T + 1 + n - 1 = n(\Delta T + 1) - \Delta T \tag{23}$$

Thus, the termination time in this case is hence upper bounded by:

$$T_C^{adversariel}(\mathbb{G}) = n(\Delta T + 2) - \Delta T \tag{24}$$

And the number of agents that participate in the coverage process using the SLC algorithms is given by:

$$N = \frac{T_C(\mathbb{G})}{\Delta T} \tag{25}$$

$$N = \frac{n(\Delta T + 2)}{\Delta T} - 1 \tag{26}$$

Similarly, we derive expressions for the upper bound on the time agent $a_i$ is mobile, $t_m^i$. Since $N > n$ and only the first $n$ agents will settle and become beacons there are different expressions for $i \in [1, n]$ and for $i \in [n + 1, N]$. Furthermore, since the exact wake-up moment in a time step is unknown and the objective is an upper bound on the total energy, any part of a time step is considered a full time-step. Thus:

$$t_m^i = \begin{cases} 2 & i = 1 \\ i & 2 \leq i \leq n \\ n(\Delta T + 2) - i\Delta T & n < i \end{cases} \quad (27)$$

Substituting Eq. (24), (26) and (27) into Eq. (20) we get the following expressions for the upper bound on the total energy as a function of $n, \Delta T$ and $\alpha$:

$$E_{Total} \leq \sum_{i=1}^{n} (i(1-\alpha) + \alpha(n(\Delta T + 2) - \Delta T) - \alpha(i-1)\Delta T) \\ + \sum_{i=n+1}^{\frac{n(\Delta T+2)-\Delta T}{\Delta T}} (n(\Delta T + 2) - i\Delta T) + 1 \quad (28)$$

After simplification, using symbolic Matlab, Eq. (28) becomes:

$$E_{Total} \leq \frac{\alpha n(n-1)}{2}\Delta T + \frac{2n^2}{\Delta T} + \frac{n(n-1)}{2} + \frac{\alpha n}{2}(3n-1) + 1 \quad (29)$$

Eq. (29) is the sum of three terms - a linear function of $\Delta T$, an inverse function of $\Delta T$ and a constant that depends solely on the region size and power ratio. When $\alpha = 0$, $E_{Total}$ is an inverse function of $\Delta T$ and given by

$$E_{Total} \leq \frac{2n^2}{\Delta T} + \frac{n(n-1)}{2} + 1 \quad (30)$$

from which it is clear that increasing $\Delta T$ results in a monotonous reduction in $E_{Total}$. Moreover, $E_{total} \to n^2/2$ as $\Delta T \to \infty$. The conclusion being that when $\alpha = 0$ the most energy efficient strategy is to enter the agents as slowly as possible there being no cost to keeping the agents on the ground. The results from [1] support this conclusion since the total energy behaves like a quadratic function of $n$. Furthermore, it is seen that the coefficients of the quadratic term decrease as $\Delta T$ increases as predicted by Eq. (30).

Figure 45 compares the upper bound on $E_{Total}$ given in Eq. (30) with results from [1] for a linear region. In order to show the results for the various linear regions on the same figure the dependency on region size is removed by showing $E_{Total}/n^2$. It is readily apparent that the total energy is upper bounded by Eq. (30) and that as $\Delta T$ increases the gap decreases.

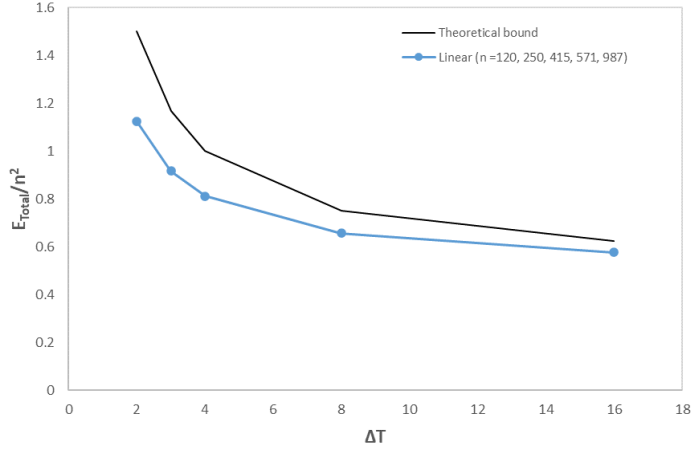

Figure 45 Comparison of theoretical bound and simulation results - $E_{Total}/n^2$ vs. $\Delta T$

When $\alpha > 0$, the energy used by the settled agents increases with time until at some point it is greater than the energy used by the mobile agents. To find the $\Delta T$ that will minimize $E_{total}$ when $\alpha > 0$, $\frac{dE_{total}}{d(\Delta T)} = 0$ is calculated resulting in:

$$\Delta T_{opt} = \sqrt{\frac{4n}{\alpha(n-1)}} \qquad (31)$$

For large regions ($n \gg 1$), Eq. (31) can be approximated by

$$\Delta T_{opt} \cong \frac{2}{\sqrt{\alpha}} \qquad (32)$$

It is clear from Eq. (31) that the optimal $\Delta T$ depends almost exclusively on $\alpha$. As described Eq. (5), the energy used by an agent is the sum of the energy used when mobile and the energy used when settled. Hence, as $\alpha$ increases so does the (relative) amount of energy used by the settled agents. Up to $\Delta T_{opt}$, the reduction in the energy consumption of the mobile agents outweighs the increase in the energy consumption of the settled agents. When $\Delta T > \Delta T_{opt}$, the increase in the energy consumption of the settled outweighs any decrease in the energy consumption of the mobile agents. Note that the upper bounds for $E_{total}$ are correct for all single layer coverage algorithms described in [1] since in linear region the coverage process is the same regardless of the traversal rules and/or the retracing logic. Substituting Eq. (31) to Eq. (29) and assuming large regions ($n \gg 1$) and typical MAV's ($\alpha \ll 1$) yields:

$$E_{Total} \leq n^2 \left( \frac{1}{2} + 2\sqrt{\alpha} + \frac{3\alpha}{2} \right) \qquad (33)$$

Figure 46 compares $E_{Total}/n^2$ from the numerical simulations with the upper bound on $E_{Total}/n^2$ given in Eq. (33) for $\alpha$=0.025. The theoretical $\Delta T_{opt}$ in this case is 12.6 while from the numerical simulations the $\Delta T$ at which the minimal value of $E_{Total}/n^2$ is reached is 12. While the values of $E_{Total}/n^2$ are different this good correlation in the value of $\Delta T_{opt}$ has a very practical and important application. It means, at least in linear-like regions, that rescue workers know the $\Delta T_{opt}$ to send the drones that will minimize $E_{Total}$ regardless of the region size.

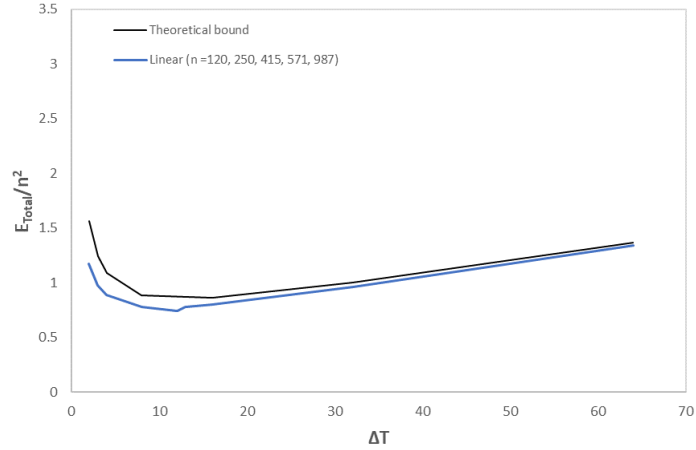

Figure 46 $E_{Total}/n^2$ as a function of $\Delta T$ for $\alpha=0.025$

While $E_{Total}$ gives the total cost of the coverage process, the energy constraint on the process comes from the agent with the highest energy consumption. Only when $max\,(E_i) < E_0$ will the swarm be able to complete the exploration and coverage of the region. If not, agents will run out of energy before complete exploration is achieved.

In the following paragraphs analytical expressions for the maximal energy consumption of each agent, $E_i^{max}$ are derived and compared with experimental results. Note, $max\,(E_i) \neq E_i^{max}$. The former is the maximum energy consumption by an agent belonging to the set $A$ (see Definition 6) whereas the latter is the maximum energy that agent $a_i$ can use during an exploration and coverage process of a linear region.

Based on Eq. (17) and since in a linear region only the first $n$ agents to enter will settle the expressions for the energy consumption of a single agent are:

$$E_i = \begin{cases} t_m^i(1-\alpha) + \alpha T_c - \alpha(i-1)\Delta T & i \in [1,n] \\ T_c - (i-1)\Delta T & i \in [n+1,N] \end{cases} \quad (34)$$

In order to get the maximal energy consumption of each of the agents we substitute Eq. (27) and Eq. (24) into Eq. (34) and get Eq. (35).

$$E_i^{max} = \begin{cases} i(1-\alpha-\alpha\Delta T) + \alpha n(\Delta T + 2) & i \in [1,n] & (a) \\ n(\Delta T + 2) - i\Delta T & i \in [n+1,N] & (b) \end{cases} \quad (35)$$

Since both Eq. (35) are linear functions of $i$, the maximal value lies at one of the bounds. Hence the maximal value of Eq. (35)a is reached when $i = n$ or in other words the last agent to settle will consume the highest energy of all the settled agents. Using the same rationale, the $n+1$ agent (i.e., the agent that will be mobile for the longest period) maximizes Eq. (35)b. Explicitly,

$$\begin{aligned} E_{settled}^{max} &= n(1+\alpha) & (a) \\ E_{mobile}^{max} &= 2n - \Delta T & (b) \end{aligned} \quad (36)$$

Eq. (36) supports the notion that when $\alpha \ll 1$, an energy limit will strongly affect the behavior of the mobile agents and that increasing $\Delta T$ will obviate that problem. Furthermore, typically $\Delta T < n$ hence $max(E_i^{max}) = E_{mobile}^{max}$. The markers in Figure 47 show $max(E_i), i \in [1:N]$ in various linear regions based on the numerical simulations along with the theoretical upper bound from Eq. (36) for $\alpha = 1$ and $\Delta T = 2$ and 4. It is clear that indeed the simulation results are smaller than the theoretical bounds.

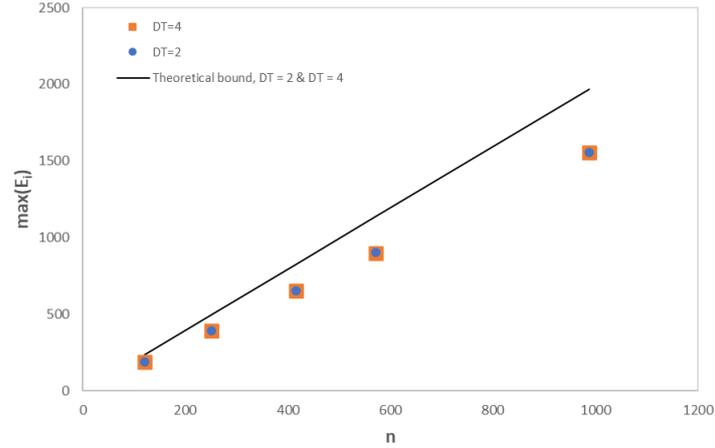

Figure 47 $max(E_i)$ in linear regions as a function of region size for $\Delta T = 2$ and 4

One of the main objectives of the SLEAC algorithms is to maximize the covered area. Intuitively, one would try to do so by equalizing the energy consumption of the settled agents. Since Eq. (35) is linear, setting $E_1 = E_n$ yields, for a given $\alpha$ and $n$, the $\Delta T$ that equalizes the energy consumption of all settled agents:

$$\Delta T = \frac{1 - \alpha}{\alpha} \qquad (37)$$

The $\Delta T$ given by Eq. (37) is significantly higher than the $\Delta T$ that minimizes the total energy. In other words, equalizing the energy use of the drones is possible but comes at the cost of increasing the energy consumption of the settled agents by a factor of $(1 - \alpha)/\sqrt{\alpha}$ found by substituting both expressions of $\Delta T$ to Eq. (35).

Figure 48 depicts $max(E_i)$ of the different agents for $\alpha = 0$ and 2.5% using Eq. (35). The agents are ordered according to their entry into the region and thus the first agent to enter the region is on the left with the first 100 agents being agents that settled and the rest being superfluous agents that remained mobile until the process terminated. A region with 100 cells is used and $\Delta T = 2$. The agent with the highest energy consumption is clearly the first superfluous agent to enter the region.

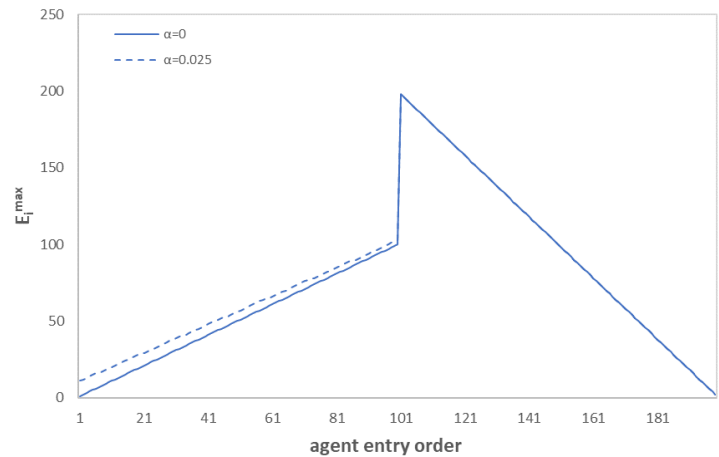

Figure 48 Maximal energy consumption of individual agents in a linear region with entry at one side

## B. Linear region with the entry point at neither edge

In the general case, shown in Figure 49, the entry point is not located at either of the edges. The exploration in this case is to both sides of the entry point and the underlaying graph is a tree with two branches – one to either side of the entry point which is the tree's root. In order to derive expressions for $E_{Total}$ and $E_i$ in the general case we first prove an upper bound on the termination time in this case. As described above, the upper bound on the termination time is the sum of the time until coverage by a single layer is achieved and the time until the BPC signal reaches the entry point. The upper bound is derived assuming local action rules that prevent superfluous, mobile agents on one branch from settling in empty cells in the other and the adversarial scheduler defined above. Such action rules increase the time until first coverage is reached by increasing the number of agents needed and are used in the SLLG-EA, SLTT-EA and SLDF algorithms (see [1]).

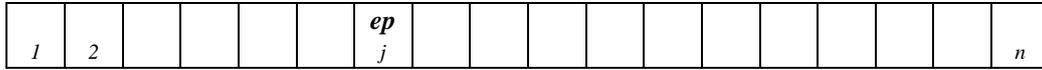

Figure 49 Linear graph with the entry point located at neither end. The graph has $n$ cells and the entry point is located at the $j - th$ cell from the left edge

When the entry point is located at an arbitrary cell $j \in [2, n-1]$ the coverage process is the superposition of two process – the coverage of the region to the right of the entry point and the coverage of the region to the left of the entry point.

**Lemma 1:** The termination time of the coverage process using either SLLG, SLDF or SLTT with $\Delta T \geq 2$ of a linear region with $n$ cells and the entry point at neither of the terminal points, is upper bounded by:

$$T_C(\mathbb{G}) \leq n(\Delta T + 2) - \Delta T \tag{38}$$

*Proof*: The deployment process is non-blocking for $\Delta T \geq 2$ as was proven in ?? [36]. In other words, once every $\Delta T$ time steps a mobile agents will enter the region. The coverage process will terminate when all the cells are at state Closed Beacon and specifically the two terminal cells. The worst case (i.e., maximal termination time) is when the two sub-regions are filled sequentially. The reason - in all other cases parts of coverage processes will overlap hence the termination time will be shorter. Suppose the entry point is located at $j \in [2, n-1]$ then the two branches going out of the entry point have $j$ and $n - j$ cells. Hence the $j - th$ agent will enter the region at $t = (j-1)\Delta T$, begin moving up the gradient at $t = (j-1)\Delta T + 1$ after which it will take the $j - th$ agent $j - 1$ additional time steps to reach the last cell in the graph followed by $j$ time steps until all the cells to the left of the entry point become Closed Beacons. Up to $\Delta T$ time steps may lapse between the time the left branch transitions to Closed Beacon and the entry of the next agent. The same process will occur in the other branch[††] resulting in the following equation:

---
[††] The entry point is not included in the second branch in order to prevent double counting.

$$\begin{aligned}
T_C(\mathbb{G}) \\
\leq\ & \underbrace{(j-1)\Delta T + 1 + (j-1) + j}_{time\ until\ the\ left\ brach\ is\ covered\ and\ transitions\ to\ Closed\ Beacon} \\
+\ & \underbrace{\Delta T}_{delay\ until\ next\ agent\ entersn} \\
+\ & \underbrace{(n-j-1)\Delta T + 1 + (n-j-1) + n-j}_{time\ until\ the\ right\ brach\ is\ covered\ and\ transitions\ to\ Closed\ Beacon}
\end{aligned} \quad (39)$$

Summing up yields

$$T_C(\mathbb{G}) \leq n(\Delta T + 2) - \Delta T \quad \blacksquare \quad (40)$$

The above sequence is possible when using SLDF however using SLLG-EA or SLTT-EA will result in one agent settling to the right of the entry point before the coverage of the left branch proceeds. The upper bound however doesn't change. The correctness of Lemma 1 was verified using numerical simulations with the various SLC algorithms.

As described by Eq. (35), the energy consumption of a settled agent is the sum of its energy consumption while mobile and energy consumption when settled whereas the energy consumption of a mobile agent (at the time of termination) is equal to the time the agent is in the region. Each entering agent makes an arbitrary selection of the branch it moves to. In both branches the mobile agents will first fill the empty cells and then start moving back towards the entry point. In other words, only the first $x$ agents to enter a branch with $x$ empty cells will settle while the rest of the agents that entered the branch will stay mobile until termination. As before, we call these agents the *superfluous agents*. While analysis of each of the large number of sequences of arbitrary decisions is impossible, it is possible to analyze the worst case from the energy perspective and derive the maximal energy consumption of each agent, $E_i^{max}$. This is done by successively, every $\Delta T$ time steps, deploying agents to the region and selecting for each agent the action that will result in its largest energy consumption. Due to symmetry and in order to simplify the analysis, whenever a movement to either side results in the same energy consumption the agent will move to the left. For the same reasons we set $j \leq n - j$.

The deployment sequence when using a greedy exploration strategy (namely either SLLG-EA or SLTT-EA) is described first followed by the sequence when using a depth-first exploration strategy (namely SLDF). The first agent to enter must settle down in the entry point. The second agent to enter will settle in the cell to the left of the entry point and the third agent will settle to the right of the entry point. While both will be mobile for the same length of time the 2$^{nd}$ agent will be settled for longer and thus have a higher energy consumption. The fourth will move two cells to the left and settle and so on until, and including, the $(j+1)th$ agent at which time the left branch is filled by settled agents. When the $(j+2)th$ agent enters the region, it may either move to the left and be a superfluous agent or move to the right and settle. Since the energy consumption of a mobile agent, in a single time step, is always greater than that of a settled agent, the agent's energy consumption will always be higher if it is constantly mobile as opposed to being settled part of the time. Consequently, the agent will move to the left. The same reasoning applies to the following agents until the BPC signal reaches the entry point. Thus agents $a_i$, $i \in [j+2, N_j + 1]$ will all move to the left branch and remain mobile throughout the coverage process. Since all the settled agents in the left branch will be in the Closed Beacon state all additional agents will move to the right branch. The following $n - j - 2$ agents to enter will all settle in the right branch and all the following agents to enter (during the propagation time of the Closed signal from the right terminal cell to the entry point) will stay mobile. Table *1* summarizes the behavior of each agent as discussed above with $N$ denoting, as above, the total number of agents participating in the coverage process. The resulting coverage process is hence the coverage of one branch followed by coverage of the other branch, as was discussed in the proof of Lemma 1.

| Group | Behavior | Agent index |
|---|---|---|
| a | Settle in place | $i = 1$ |
| b | Move to empty cell and then settle | $i = 2: j + 1$ |
| c | Remain mobile until process termination | $i = j + 2: N_j + 1$ |
| d | Move to empty cell and then settle | $i = N_j + 2: N_j + n - j$ |
| e | Remain mobile until process termination | $i = N_j + n - j + 1: N$ |

Table 1 Agent behavior as a function of its index using SLLG-EA or SLTT-EA

The maximum energy consumption for each agent and the total energy can now be derived using the expressions (derived above) for $T_c$, $N$, and $t_m^i$ using (either SLLG-EA or SLTT-EA).

Group a:
$$E_1^{max} = 2 + \alpha(n(\Delta T + 2) - 2\Delta T - 2) \tag{41}$$

Group b:
$$E_2^{max} = 2 + \alpha(n(\Delta T + 2) - 2\Delta T - 2) \tag{42}$$

$$E_3^{max} = 2 + \alpha(n(\Delta T + 2) - 3\Delta T - 2) \tag{43}$$

$$E_i^{max} = i(1 - \alpha - \alpha\Delta T) - (1 - \alpha) + \alpha n(\Delta T + 2) \quad \text{for} \quad i = 4: j + 1 \tag{44}$$

Group c:
$$E_i^{max} = n(\Delta T + 2) - i\Delta T \quad \text{for} \quad i = j + 2: N_j + 1$$
$$\text{and} \quad N_j = \frac{j(\Delta T + 2) - \Delta T}{\Delta T} \tag{45}$$

Group d:
$$E_i^{max} = (i - N_j + 1)(1 - \alpha) + \alpha(n(\Delta T + 2) - i\Delta T)$$
$$\text{for} \quad i = N_j + 2: N_j + n - j \tag{46}$$

Group e:
$$E_i^{max} = n(\Delta T + 2) - i\Delta T \quad \text{for} \quad i = N_j + n - j + 1: N \tag{47}$$

In all of the above equations, $E_i$ is a linear function of $j, \Delta T$ and $n$. Moreover, $E_i$ increases when $\Delta T$ and $n$ increase. The figure below shows $E_i^{max}$ for $n = 100$, $j = 20$, $\Delta T = 2$, $\alpha = 0$ and $\alpha = 2.5\%$ using the equations above. It is easily discernible that (1) the maximum energy consumption is by the first superfluous agent to enter the region and (2) the effect of $\alpha$ on the energy consumption is relatively minor.

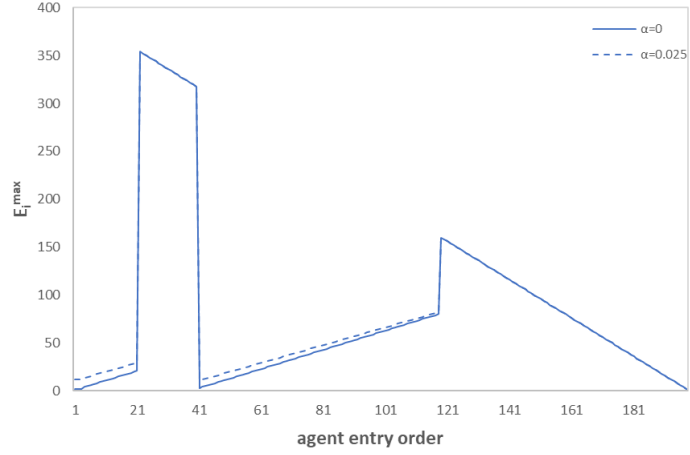

Figure 50 Energy consumption of individual agents in a linear region with entry 20 cells from the left terminal point

Since the above equations describe $E_i^{max}$ as a function of $i$, $E_{Total}$ of the coverage process is calculated by substituting the equations above to Eq. (18) and summing using symbolic Matlab:

$$E_{Total} = \frac{(j-n+1)(4\alpha + j - n + 3\alpha j - 3\alpha n + \Delta T\alpha j - \Delta T\alpha n - 4)}{2}$$
$$- \alpha(3\Delta T - \sigma_1 + 2) - \alpha(2\Delta T - \sigma_1 + 2)$$
$$+ (j-2)(\alpha + \alpha n(\Delta T + 2) - 1) - \alpha(\Delta T - \sigma_1 + 2)$$
$$- \frac{(j-2)(j+5)(\alpha + \Delta T\alpha - 1)}{2}$$
$$+ \frac{(\Delta T - 2j)(\Delta T + j - 2n + \Delta Tj - \Delta Tn)}{\Delta T}$$
$$- \frac{(j-n)(\Delta T - 2j + 2n)}{\Delta T} + 6 \tag{48}$$

with

$$\sigma_1 = n(\Delta T + 2) \tag{49}$$

After simplification (using symbolic Matlab) the result is:

$$E_{Total} = \left(1 - \alpha + j - n - \alpha j + \frac{\alpha n}{2} + \frac{\alpha n^2}{2}\right)\Delta T + \frac{2n^2}{\Delta T} - \alpha - 3j + \frac{n}{2}$$
$$+ 3\alpha j - \frac{3\alpha n}{2} + jn + \alpha j^2 + \frac{3\alpha n^2}{2} - j^2 + \frac{n^2}{2} - \alpha jn + 1 \tag{50}$$

It is easily seen that Eq. (50) is composed of a term linear with $\Delta T$, a term inverse with $\Delta T$ and a constant and thus similar to Eq. (29). When $\alpha = 0$ Eq. (50) becomes:

$$E_{Total} = (1 + j - n)\Delta T + \frac{2n^2}{\Delta T} - 3j + \frac{n}{2} + jn - j^2 + \frac{n^2}{2} + 1 \tag{51}$$

Parametric analysis of Eq. (51) reveals that (1) the linear term with $\Delta T$ is never positive if $j \leq n - 1$ which is always true since $j$ is not a terminal cell and (2) the constant term is positive if $n \geq 3$ which is also true this being the smallest possible region with propagation to both sides.

Thus, the total energy is a monotonically decreasing function of $\Delta T$. This is similar to the behavior seen when the entry point is at one of the terminal points (and expressed in Eq. (30)).

When $\alpha > 0$ the optimal $\Delta T$ (i.e., the $\Delta T$ that will minimize $E_{total}$) is found by equating to 0 the partial derivative of $E_{total}$ with respect $\Delta T$, hence:

$$\Delta T_{opt} = \frac{2n}{\sqrt{2(j-n) - 2j\alpha - 2\alpha + \alpha n + \alpha n^2 + 2}} \tag{52}$$

The right term is real only when the expression in the denominator is positive from which we get the following relationship between $n, j$ and $\alpha$ that must exist in order for there to be an optimal $\Delta T$ that minimizes the total energy:

$$\frac{n(2-\alpha n)}{2} - 1 < j \tag{53}$$

When SLDF is used the process is slightly different (since in this case the third agent to enter the region moves to the left) and is summarized in Table 2.

| Group | Behavior | Agent index |
|---|---|---|
| a | Settle in place | $i = 1$ |
| b | Move to empty cell and then settle | $i = 2:j$ |
| c | Remain mobile until process termination | $i = j+1:N_j$ |
| d | Move to empty cell and then settle | $i = N_j+1:N_j+n-j$ |
| e | Remain mobile until process termination | $i = N_j+n-j+1:N$ |

Table 2  Agent behavior as a function of its index using SLDF

The maximum energy consumption for each agent and the total energy can now be derived using the expressions (derived above) for $T_c$, $N$, and $t_m^i$ using SLDF.

Group a:
$$E_1^{max} = 2 + \alpha(n(\Delta T + 2) - 2\Delta T - 2) \tag{54}$$

Group b:
$$E_2^{max} = 2 + \alpha(n(\Delta T + 2) - 2\Delta T - 2) \tag{55}$$

$$E_3^{max} = 2 + \alpha(n(\Delta T + 2) - 3\Delta T - 2) \tag{56}$$

$$E_i^{max} = i(1 - \alpha - \alpha\Delta T) + \alpha n(\Delta T + 2) \quad \text{for} \quad i = 2:j \tag{57}$$

Group c:
$$N_j = \frac{j(\Delta T + 2) - \Delta T}{\Delta T} \tag{58}$$

$$E_i^{max} = n(\Delta T + 2) - i\Delta T \quad \text{for} \quad i = j+1:N_j$$

Group d:
$$t_m^i = (i - N_j) + 1 \tag{59}$$

$$E_i^{max} = (i - N_j + 1)(1 - \alpha) + \alpha(n(\Delta T + 2) - i\Delta T)$$
$$\text{for} \quad i = N_j + 1 : N_j + n - j$$

Group e:
$$E_i^{max} = n(\Delta T + 2) - i\Delta T \quad \text{for} \quad i = N_j + n - j + 1 : N \tag{60}$$

Using symbolic Matlab the result is:
$$E_{Total} = \left(j - n - \alpha j + \frac{\alpha n}{2} + \frac{\alpha n^2}{2}\right)\Delta T + \frac{2n^2}{\Delta T} - \alpha - j + \frac{n}{2} + \alpha j - \frac{3\alpha n}{2}$$
$$+ jn + \alpha j^2 + \frac{3\alpha n^2}{2} - j^2 + \frac{n^2}{2} - \alpha jn + 1 \tag{61}$$

It is easily seen that Eq. (61) is composed of a term linear with $\Delta T$, a term inverse with $\Delta T$ and a constant and thus similar to Eq. (29) and (50). When $\alpha = 0$ Eq. (61) becomes:
$$E_{Total} = (j - n)\Delta T + \frac{2n^2}{\Delta T} - j + \frac{n}{2} + jn - j^2 + \frac{n^2}{2} + 1 \tag{62}$$

Parametric analysis of Eq. (62) reveals that (1) the linear term with $\Delta T$ is never positive if $j \leq n$ which is by always true since $j$ is not a terminal cell and (2) the constant term is positive if $n \geq 3$ which is also true this being the smallest possible region. Thus, the total energy is a monotonically decreasing function of $\Delta T$. This is similar to the behavior seen when the entry point is at one of the terminal points or when using SLLG-EA or SLTT-EA.

When $\alpha > 0$ the optimal $\Delta T$ (i.e., the $\Delta T$ that will minimize $E_{total}$) is found by equating to 0 the partial derivative of $E_{total}$ with respect $\Delta T$, hence:
$$\Delta T_{opt} = \frac{2n}{\sqrt{2(j-n) - 2j\alpha + \alpha n + \alpha n^2}} \tag{63}$$

The right term is real only when the expression in the denominator is positive from which we get the following relationship between $n, j$ and $\alpha$ that must exist in order for there to be an optimal $\Delta T$ that minimizes the total energy:
$$\frac{n(2 - \alpha n)}{2} < j \tag{64}$$

It is easily discernable that the differences in the expressions for $E_{Total}$ and $\Delta T_{opt}$ are small. Moreover, the difference in the total energy between the two deployment sequences is bounded by $|(1 - \alpha)(\Delta T - 2j)|$ or in other words is less than $2n$ with SLDF having the higher total energy.

In summary, several interesting observations arise from the above analysis:
1. The maximal specific energy consumption by a settled agent is when the entry point at one of the edges.
2. The maximal specific energy consumption by a mobile agent is when the entry point is in the middle and it is reached by the first superfluous mobile agent that enters the region. Its energy consumption is significantly higher compared to the case the entry point is at the edge.
3. Typically, the maximum specific energy ($max(E_i)$) is by a mobile agent.
4. The total energy consumption is greater when the entry point is in the middle.

## VII. Conclusions

In this paper several distributed algorithms for the exploration and uniform coverage of unknown indoor regions by energy constrained agents are presented. Such scenarios raise the question of when, and even more importantly, how does the exploration stop when the region can't be fully explored by the agents. Consequently, two approaches were defined. In the first approach the objective is to update the user and stop the entry of additional agents as soon as possible. The second approach gives precedence to covering the maximal possible area and thus delays the termination as long as possible. In both approaches the Backward Propagating Closure meta-concept is essential to the propagation of information about the energy level in the swarm. Upper bounds on the covered area are derived for both approaches and for the termination time when the first approach is used. Good correlation is found with the results of the numerical simulations.

An analytical solution is derived for the case of a linear graph. Most importantly it is shown that the maximal specific energy consumption by a settled agent is when the entry point at one of the edges. Conversely, for a given region both the total energy and the specific energy are higher when the entry point is in the middle of the region.

The behavior of the multi-agent system is extremely complex when the energy is limited and the power factor (i.e. the power consumption on the ground) is greater than zero. The combined effect of these two parameters results in unpredictable macro-behavior in the sense that it is not possible to predict the behavior in the next step based on the macro-behavior in the current and previous time steps. To date the topic of energy constrained exploration and coverage by a swarm of airborne agents has received very limited research attention although it has many potential implications and applications. Hence, the combination of airborne agents, limited energy and simple agents merits further research in various scenarios and specifically in the context of multi-agent systems.

## References

[1] O. Rappel, J. Ben-Asher, and A. Bruckstein, "Exploration and Coverage with Swarms of Settling Agents," 2022, doi: 10.48550/ARXIV.2209.05512.

[2] T. Stirling, S. Wischmann, and D. Floreano, "Energy-efficient indoor search by swarms of simulated flying robots without global information," *Swarm Intelligence*, vol. 4, no. 2, pp. 117–143, 2010.

[3] P. Flocchini, G. Prencipe, and N. Santoro, Eds., *Distributed Computing by Mobile Entities: Current Research in Moving and Computing*, vol. 11340. in Lecture Notes in Computer Science, vol. 11340. Cham: Springer International Publishing, 2019. doi: 10.1007/978-3-030-11072-7.

[4] P. Flocchini, G. Prencipe, and N. Santoro, "Computing by Mobile Robotic Sensors," in *Theoretical Aspects of Distributed Computing in Sensor Networks*, S. Nikoletseas and J. D. P. Rolim, Eds., in Monographs in Theoretical Computer Science. An EATCS Series. Berlin, Heidelberg: Springer Berlin Heidelberg, 2011, pp. 655–693. doi: 10.1007/978-3-642-14849-1_21.

[5] T.-R. Hsiang, E. M. Arkin, M. A. Bender, S. P. Fekete, and J. S. B. Mitchell, "Algorithms for Rapidly Dispersing Robot Swarms in Unknown Environments," *SpringerLink*, pp. 77–93, 2004, doi: 10.1007/978-3-540-45058-0_6.

[6] L. BarrièRe, P. Flocchini, E. Mesa-Barrameda, and N. Santoro, "UNIFORM SCATTERING OF AUTONOMOUS MOBILE ROBOTS IN A GRID," *International Journal of Foundations of Computer Science*, vol. 22, no. 03, pp. 679–697, Apr. 2011, doi: 10.1142/S0129054111008295.

[7] A. Howard, M. J. Matarić, and G. S. Sukhatme, "An Incremental Self-Deployment Algorithm for Mobile Sensor Networks," *Autonomous Robots*, vol. 13, no. 2, pp. 113–126, Sep. 2002, doi: 10.1023/A:1019625207705.

[8] T. Stirling and D. Floreano, "Energy-Time Efficiency in Aerial Swarm Deployment," in *Distributed Autonomous Robotic Systems*, A. Martinoli, F. Mondada, N. Correll, G. Mermoud, M. Egerstedt, M. A. Hsieh, L. E. Parker, and K. Støy, Eds., Berlin, Heidelberg: Springer Berlin Heidelberg, 2013, pp. 5–18. doi: 10.1007/978-3-642-32723-0_1.

[9] F. Aznar Gregori, M. Pujol, and R. Rizo, "UAV Deployment Using Two Levels of Stigmergy for Unstructured Environments," Oct. 2020, doi: 10.3390/app10217696.

[10] H. J. Chang, C. S. George Lee, Yung-Hsiang Lu, and Y. Charlie Hu, "Energy-time-efficient adaptive dispatching algorithms for ant-like robot systems," in *IEEE International Conference on Robotics and Automation, 2004. Proceedings. ICRA '04. 2004*, Apr. 2004, pp. 3294-3299 Vol.4. doi: 10.1109/ROBOT.2004.1308762.

[11] A. Howard, M. J. Matarić, and G. S. Sukhatme, "Mobile Sensor Network Deployment using Potential Fields: A Distributed, Scalable Solution to the Area Coverage Problem," in *Distributed Autonomous Robotic Systems 5*, H. Asama, T. Arai, T. Fukuda, and T. Hasegawa, Eds., Tokyo: Springer Japan, 2002, pp. 299–308. doi: 10.1007/978-4-431-65941-9_30.

[12] Y. Mulgaonkar, A. Makineni, L. Guerrero-Bonilla, and V. Kumar, "Robust Aerial Robot Swarms Without Collision Avoidance," *IEEE Robotics and Automation Letters*, vol. 3, no. 1, pp. 596–603, Jan. 2018, doi: 10.1109/LRA.2017.2775699.

[13] E. M. Barrameda, S. Das, and N. Santoro, "Deployment of Asynchronous Robotic Sensors in Unknown Orthogonal Environments," in *Algorithmic Aspects of Wireless Sensor Networks*, S. P. Fekete, Ed., Berlin, Heidelberg: Springer Berlin Heidelberg, 2008, pp. 125–140. doi: 10.1007/978-3-540-92862-1_11.

[14] F. Blatt and H. Szczerbicka, "Combining the multi-agent flood algorithm with frontier-based exploration in search amp; rescue applications," in *2017 International Symposium on*

*Performance Evaluation of Computer and Telecommunication Systems (SPECTS)*, Jul. 2017, pp. 1–7. doi: 10.23919/SPECTS.2017.8046775.

[15] I. Rekleitis, A. P. New, E. S. Rankin, and H. Choset, "Efficient Boustrophedon Multi-Robot Coverage: an algorithmic approach," *Ann Math Artif Intell*, vol. 52, no. 2, pp. 109–142, 2008, doi: 10.1007/s10472-009-9120-2.

[16] J. Kim, "Topological Map Building with Multiple Agents Having Abilities of Dropping Indexed Markers," *J Intell Robot Syst*, vol. 103, no. 1, p. 18, Sep. 2021, doi: 10.1007/s10846-021-01473-4.

[17] M. Betke, R. L. Rivest, and M. Singh, "Piecemeal learning of an unknown environment," *Mach Learn*, vol. 18, no. 2, pp. 231–254, Feb. 1995, doi: 10.1007/BF00993411.

[18] S. Das, D. Dereniowski, and C. Karousatou, "Collaborative Exploration by Energy-Constrained Mobile Robots," in *Structural Information and Communication Complexity*, C. Scheideler, Ed., in Lecture Notes in Computer Science. Cham: Springer International Publishing, 2015, pp. 357–369. doi: 10.1007/978-3-319-25258-2_25.

[19] S. Das, D. Dereniowski, and C. Karousatou, "Collaborative Exploration of Trees by Energy-Constrained Mobile Robots," *Theory Comput Syst*, vol. 62, no. 5, pp. 1223–1240, Jul. 2018, doi: 10.1007/s00224-017-9816-3.

[20] S. Das, D. Dereniowski, and P. Uznański, "Energy Constrained Depth First Search," *arXiv:1709.10146 [cs]*, Feb. 2018, Accessed: Jul. 23, 2021. [Online]. Available: http://arxiv.org/abs/1709.10146

[21] M. Dynia, M. Korzeniowski, and C. Schindelhauer, "Power-Aware Collective Tree Exploration," in *Architecture of Computing Systems - ARCS 2006*, W. Grass, B. Sick, and K. Waldschmidt, Eds., in Lecture Notes in Computer Science. Berlin, Heidelberg: Springer, 2006, pp. 341–351. doi: 10.1007/11682127_24.

[22] E. Bampas, J. Chalopin, S. Das, J. Hackfeld, and C. Karousatou, "Maximal Exploration of Trees with Energy-Constrained Agents," Feb. 2018.

[23] A. Kwok and S. Martinez, "Energy-balancing cooperative strategies for sensor deployment," in *2007 46th IEEE Conference on Decision and Control*, New Orleans, LA, USA: IEEE, 2007, pp. 6136–6141. doi: 10.1109/CDC.2007.4434494.

[24] F. Aznar, M. Pujol, R. Rizo, and C. Rizo, "Modelling multi-rotor UAVs swarm deployment using virtual pheromones," *PLOS ONE*, vol. 13, no. 1, p. e0190692, Jan. 2018, doi: 10.1371/journal.pone.0190692.

[25] F. Aznar, M. Pujol, R. Rizo, F. A. Pujol, and C. Rizo, "Energy-Efficient Swarm Behavior for Indoor UAV Ad-Hoc Network Deployment," *Symmetry (20738994)*, vol. 10, no. 11, p. 632, Nov. 2018, doi: 10.3390/sym10110632.

[26] O. Rappel and J. Z. Ben-Asher, "Area coverage – A swarm based approach," in *59th Israel Annual Conference on Aerospace Sciences, IACAS 2019*, Israel Annual Conference on Aerospace Sciences, 2019, pp. 35–55.

[27] D. Peleg, "Distributed Coordination Algorithms for Mobile Robot Swarms: New Directions and Challenges," in *Distributed Computing – IWDC 2005*, Springer, Berlin, Heidelberg, Dec. 2005, pp. 1–12. doi: 10.1007/11603771_1.

[28] T. Peters, "Comparison of scheduler models for distributed systems of luminous robots."

[29] S. Das, P. Flocchini, G. Prencipe, N. Santoro, and M. Yamashita, "Autonomous mobile robots with lights," *Theoretical Computer Science*, vol. 609, pp. 171–184, Jan. 2016, doi: 10.1016/j.tcs.2015.09.018.

[30] P. Poudel and G. Sharma, "Time-optimal uniform scattering in a grid," in *Proceedings of the 20th International Conference on Distributed Computing and Networking*, in ICDCN '19. New York, NY, USA: Association for Computing Machinery, 2019, pp. 228–237. doi: 10.1145/3288599.3288622.

[31] X. Défago, M. Gradinariu, S. Messika, and P. Raipin-Parvédy, "Fault-Tolerant and Self-stabilizing Mobile Robots Gathering," in *Distributed Computing*, Springer, Berlin, Heidelberg, Sep. 2006, pp. 46–60. doi: 10.1007/11864219_4.

[32] G. Prencipe, "Instantaneous Actions vs. Full Asynchronicity: Controlling and Coordinating a Set of Autonomous Mobile Robots," *LECTURE NOTES IN COMPUTER SCIENCE*, no. 2202, p. 154, 2001.

[33] E. Ferranti, N. Trigoni, and M. Levene, "Brick Mortar: an on-line multi-agent exploration algorithm," in *Proceedings 2007 IEEE International Conference on Robotics and Automation*, Apr. 2007, pp. 761–767. doi: 10.1109/ROBOT.2007.363078.

[34] H. V. Abeywickrama, B. A. Jayawickrama, Y. He, and E. Dutkiewicz, "Comprehensive Energy Consumption Model for Unmanned Aerial Vehicles, Based on Empirical Studies of Battery Performance," *IEEE Access*, vol. 6, pp. 58383–58394, 2018, doi: 10.1109/ACCESS.2018.2875040.

[35] I. A. Wagner and A. M. Bruckstein, "Cooperative Cleaners: A Study in Ant Robotics," in *Communications, Computation, Control, and Signal Processing: a tribute to Thomas Kailath*, A. Paulraj, V. Roychowdhury, and C. D. Schaper, Eds., Boston, MA: Springer US, 1997, pp. 289–308. doi: 10.1007/978-1-4615-6281-8_16.

[36] O. Rappel, M. Amir, and A. M. Bruckstein, "Stigmergy-based, Dual-Layer Coverage of Unknown Indoor Regions." arXiv, Sep. 18, 2022. doi: 10.48550/arXiv.2209.08573.